\documentclass[draft,phd,12pt]{psuthesis}
\usepackage{amsmath}
\usepackage{amssymb}
\usepackage{amsthm}
\usepackage{exscale}
\usepackage[mathscr]{eucal}
\usepackage{bm}
\usepackage{eqlist} % Makes for a nice list of symbols.
\usepackage[final]{graphicx}
\usepackage[dvipsnames]{color}
\usepackage{url}
\DeclareGraphicsExtensions{.pdf, .jpg}

\usepackage[Lenny]{fncychap}
\ChTitleVar{\Huge\sffamily\bfseries}

\title{Transverse Single Spin Asymmetries and Cross-Sections for Forward $\pi^0$ and $\eta$ Mesons at Large $x_F$ in $\sqrt{s}=200$ GeV $p^{\uparrow}+p$ Collisions at STAR}

\author{Len K. Eun}
\dept{Physics}
\degreedate{August 2011}
\copyrightyear{2011}

\honorsdegreeinfo{for a baccalaureate degree \\ in Engineering Science \\ with honors in Engineering Science}

\documenttype{Dissertation}

\submittedto{The Graduate School}

\numberofreaders{4}

\honorsadviser{Honors P. Adviser}

\secondthesissupervisor{Second T. Supervisor}

\honorsdepthead{Department Q. Head}

\advisor[Dissertation Advisor, Chair of Committee]
        {Steve F. Heppelmann}
        {Professor of Physics}

\readerone
          {John C. Collins}
          {Professor of Physics}

\readertwo
          {St\'{e}phane Coutu}
          {Professor of Physics}

\readerthree
            {Horacio Perez-Blanco}
            {Professor of Mechanical Engineering}

\readerfour[Associate Head of the Department of Physics]
			{Nitin Samarth}
			{Professor of Physics}

\begin{document}

%%%%%%%%%%%%%%%%%%%%%%%%
% Preliminary Material %
%%%%%%%%%%%%%%%%%%%%%%%%
% This command is needed to properly set up the frontmatter.
\frontmatter

%%%%%%%%%%%%%%%%%%%%%%%%%%%%%%%%%%%%%%%%%%%%%%%%%%%%%%%%%%%%%%
% IMPORTANT
%
% The following commands allow you to include all the
% frontmatter in your thesis. If you don't need one or more of
% these items, you can comment it out. Most of these items are
% actually required by the Grad School -- see the Thesis Guide
% for details regarding what is and what is not required for
% your particular degree.
%%%%%%%%%%%%%%%%%%%%%%%%%%%%%%%%%%%%%%%%%%%%%%%%%%%%%%%%%%%%%%
% !!! DO NOT CHANGE THE SEQUENCE OF THESE ITEMS !!!
%%%%%%%%%%%%%%%%%%%%%%%%%%%%%%%%%%%%%%%%%%%%%%%%%%%%%%%%%%%%%%

% Generates the signature page. This is not bound with your
% thesis.
%\psusigpage

% Generates the title page based on info you have provided
% above.
\psutitlepage

% Generates the committee page -- this is bound with your
% thesis. If this is an baccalaureate honors thesis, then
% comment out this line.
\psucommitteepage

% Generates the abstract. The argument should point to the
% file containing your abstract. 
%
\pagestyle{plain}
\chapter*{Abstract}
    \begin{singlespace}
        The STAR collaboration has previously reported a large transverse single spin asymmetry, $A_{N}$, for forward $\pi^{0}$ meson production. The cross-section in this region was measured up to $x_{F}$ of 0.55, and found to be consistent with perturbative Quantum Chromo-Dynamics predictions. During RHIC running in the year 2006 (6.8 pb$^{-1}$, 56 \% average polarization), an even larger $A_{N}$ was observed for the $\eta$ meson at forward rapidity for $x_{F}>$ 0.5. All data were collected by the STAR Forward Pion Detector during $\sqrt{s}$ = 200 GeV polarized p+p collisions at RHIC. Understanding these large spin asymmetries requires information on the production cross-section. In this thesis, we present the results of the cross-section as well as the $A_{N}$ measurements for both $\pi^0$ and $\eta$ mesons at average pseudo-rapidity of 3.7, for $0.4 < x_F < 0.75$. The details of the calibration, and the analysis methods are also discussed. Finally, we present selected aspects of the high voltage system for the STAR Forward Meson Spectrometer inner calorimeter, which was designed primarily by the author and his thesis advisor.

    \end{singlespace}
\newpage

% Generates the Table of Contents
\thesistableofcontents

% Generates the List of Figures
\thesislistoffigures

% Generates the List of Tables
\thesislistoftables

% Generates the List of Symbols. The argument should point to
% the file containing your List of Symbols. 
%\thesislistofsymbols{SupplementaryMaterial/ListOfSymbols}

% Generates the Acknowledgments. The argument should point to
% the file containing your Acknowledgments. 
%
\chapter{Acknowledgments}
%\addcontentsline{toc}{chapter}{Acknowledgments}
\begin{singlespace}
    % Place acknowledgments below.
First of all, I want to thank my parents. Of course to thank them for everything they've done for me in one paragraph would be an impossible task, so here I'll limit my scope to the topic at hand, and thank them specifically for cultivating my interest in science from an early age. Despite being scientists themselves, they never gave me any pressure to become one, and that probably had a lot to do with me getting a Ph.D. in physics. I also have to admit that the path I took to get here wasn't exactly free of major potholes, and I know that they always believed in me through that process. Thank you, mom and dad.\\

I really cannot thank Steve enough, whom I consider the best teacher I have ever had. It is literally true that I would not be here doing what I'm doing, if he hadn't put faith in me when I first met him as an undergraduate student. His insights and his unique way of thinking has been both an influence and a challenge to me, and the countless discussions we've had over lunch have been invaluable in forming my own perspective. He has been my mentor not only professionally, but also personally, and I feel very fortunate to have had a wonderful advisor like him. \\

I also want to thank my colleagues. I had the good fortune to work with some very good people who have provided me with guidance and support through the years. Carl Gagliardi has been an invaluable source of ideas and advices for me, who has helped me with all stages of my analysis detailed in this thesis. I want to thank Akio Ogawa, who always took the time to help me through the inner workings of the experiment, which often felt to me like a gigantic maze. I also want to thank Enrst Sichterman, Stephen Trentalange, and everyone else in the spin physics working group for their support. In addition, I want to thank Ming Liu from LANL for the always insightful conversations. Finally, I want to thank my good friends at STAR, Jim Drachenberg, Ermes Braidot, and Pibero Djawotho, who have made my experience in the collaboration not only fulfilling, but also fun. \\

More personally, I have to thank my best friend Ivan Iordanov, who, despite his affinity to density functional theory and ignorance of the Standard Model, has been the most loyal friend to me since college. Even though most of what he says is wrong, I have to admit I've always enjoyed the daily arguments we had over diverse subjects ranging from religion to politics, in which I was right and he was wrong. For that, and all the help he gave me when I moved, I thank him reluctantly.\\ 

Finally, I want to thank my wife, Doeun. There are undoubtedly more glorious things in this world than being a wife of a physics graduate student, and yet, she has been nothing but supportive and understanding throughout my time in graduate school, all the while being such a wonderful mother to our daughter. So here I leave in writing, that I believe she is the best thing that ever happened to me.\\

\end{singlespace}

% Generates the Epigraph/Dedication. The first argument should
% point to the file containing your Epigraph/Dedication and
% the second argument should be the title of this page. 
\thesisdedication{SupplementaryMaterial/Dedication}{Dedication}

%%%%%%%%%%%%%%%%%%%%%%%%%%%%%%%%%%%%%%%%%%%%%%%%%%%%%%
% This command is needed to get the main part of the %
% document going.                                    %
%%%%%%%%%%%%%%%%%%%%%%%%%%%%%%%%%%%%%%%%%%%%%%%%%%%%%%
\thesismainmatter

%%%%%%%%%%%%%%%%%%%%%%%%%%%%%%%%%%%%%%%%%%%%%%%%%%
% This is an AMS-LaTeX command to allow breaking %
% of displayed equations across pages. Note the  %
% closing the "}" just before the bibliography.  %
%%%%%%%%%%%%%%%%%%%%%%%%%%%%%%%%%%%%%%%%%%%%%%%%%%
\allowdisplaybreaks{
%
%%%%%%%%%%%%%%%%%%%%%%
% THE ACTUAL CONTENT %
%%%%%%%%%%%%%%%%%%%%%%
% Chapters
\chapter{Motivation}\label{ch:AN}
\setlength{\parindent}{1cm}
\begin{doublespace}

\section{Introduction}
According to our current understanding of the universe, there are four types of fundamental interactions that govern all physical processes: electromagnetism, strong interaction, weak interaction, and gravitation. The Standard Model of particle physics describes the first three of these interactions in the language of quantum field theory, providing us with a coherent framework to understand the fundamental processes.  The strong interaction is the force between quarks and gluons, which are the building blocks of the nucleons such as protons and neutrons. It is also responsible for binding together protons and neutrons inside a nucleus, overcoming the electromagnetic repulsion between protons. The part of the standard model that deals with the strong interaction is called Quantum Chromo-Dynamics (QCD), and the experiment described in this thesis is primarily aimed at studying an aspect of QCD that is related to the intrinsic spin of the nucleons.\\

Despite the formal similarities of the theories, QCD works very differently from electromagnetism (Quantum Electro-Dynamics, QED). In electromagnetism, the force between two charged particles decrease rapidly as the distance between them is increased. In QCD, however, the attractive force between two quarks does not diminish as they are separated. Once the separation between two quarks becomes large enough (but still microscopic), it becomes energetically favorable to create their anti-quarks from vacuum to ``neutralize" the long range color force. This phenomenon is called the ``confinement", and it is the reason that no one has ever observed a free, isolated quark or gluon. On the other hand, when the separation between two quarks become extremely small, the strength of the strong interaction becomes very weak, to a point where quarks and gluons inside a nucleon can be viewed as weakly interacting, almost free particles. This phenomenon is referred to as the ``asymptotic freedom" \cite{asympt}.\\

In general, strongly interacting systems do not lend themselves to perturbation theory, an essential analytical techniques that physicists employ to understand quantum systems. In fact, the only part of the nucleon scattering event for which perturbative QCD can be applied is the brief instance during which the quarks and gluons come extremely close together and scatter off each other. But due to confinement, the participants of this "hard scattering" are hidden from us both before and after the scattering. Consequently, what we actually observe are the indirect results of the hard scattering event, which involves processes that cannot be calculated analytically. \\

The QCD factorization deals with these difficulties, by separating the non-perturbative parts in the initial and final states of the scattering from the hard scattering cross-section that can be calculated precisely. While the non-perturbative parts cannot be calculated, they are nonetheless believed to be universal, so they can be obtained by global analyses of various different types of experiments. This factorized perturbative approach allows for a precise, analytic application of QCD to nuclear scattering, but only for a certain class of processes; namely, one that includes a scattering between quarks and/or gluons with a high enough momentum transfer (so the quarks and gluons are close enough together during the interaction, making it ``hard"), and the non-perturbative parts that are universal. This limitation of applicability implies that any process that can be understood by the perturbative approach gives us a valuable opportunity to put our understanding of QCD to a rigorous test. \\  

One example of such processes is the production cross-sections of jets and hadrons in proton-proton collisions, where the perturbative QCD has been successful in predicting these observables in variety of experiments. From here, one logical step forward, at least theoretically, is to explain the dependence of the cross-section on nucleon spin. While this involves yet more non-perturbative quantities that need to be obtained experimentally, the spin dependence of the hard scattering between quarks and/or gluons is something many believe the perturbative QCD framework should be able to handle as well as the spin averaged counterpart. Experimentally, however, it is considerably more difficult to have polarized experiments that can probe the spin dependent processes. The experiment described in this thesis is one such example, in which we study the effects of the proton spin on the particle production in polarized proton collisions. It is a part of a much broader effort to understand a diverse range of QCD related physics at the Relativistic Heavy Ion Collider (RHIC), the world's only polarized particle collider located at Brookhaven National Laboratory (BNL). \\  

The main subject of this thesis is the measurements of the forward production cross-section, and transverse single spin asymmetry for two neutral mesons, $\pi^0$ and $\eta$. The data for this analysis were taken with the Forward Pion Detector (FPD), which was a part of the first iteration of STAR (Solenoidal Tracker At RHIC \cite{star}) forward calorimetry. The descriptions of the detector setup, and more broadly of the RHIC environment, can be found in chapter \ref{ch:exp}. The theoretical background for the transverse spin physics can be found later in this chapter.  \\

The first analysis topic that we cover is the study of electromagnetic shower shape in the FPD, which is the subject of chapter \ref{ch:shower}. This is an important keystone to the rest of the analysis, as everything from off-line calibration to unfolding the energy smearing depends crucially on our ability to accurately simulate the shower development. The next topic, covered in chapter \ref{ch:calib}, is the off-line calibration. While this very data set was previously used for a published spin asymmetry measurement, \cite{STARAN} the fact that cross-section measurements require a much more rigorous calibration of the detector, and that we are measuring the spin asymmetry at higher energies than what was done before, lead us to work on improving the calibration methods. The details of the data analysis are covered in chapter \ref{ch:anal}, including background corrections, detection efficiency, and the unfolding of energy smearing. Finally, the physics results for the cross-section and the spin asymmetry can be found in chapter \ref{ch:result}.\\

The last part of this thesis (Appendix \ref{ch:CW}) covers a topic that is somewhat disjointed from the rest of this document. As a member of the STAR collaboration since 2004, I have participated in the STAR forward physics effort both in data analysis, and in hardware design. The STAR forward calorimetry has undergone continual upgrades and changes during my participation, and as a result, I had an invaluable opportunity to work on the design of the Cockroft-Walton high voltage system for the STAR Forward Meson Spectrometer (FMS). Appendix \ref{ch:CW} includes the study of a simple theoretical model of the Cockroft-Walton voltage multiplier, and covers selected elements of the system design.\\ 

\section{Transverse Single Spin Asymmetry}
Spin is a fundamental degree of freedom in quantum field theory formalism, in which an elementary particle is assigned an internal vector space representing the state of its intrinsic angular momentum. Particles that carry half integer unit of spin are called fermions, whose collective behavior is governed by Fermi-Dirac statistics. Particles that carry integer unit of spin are called bosons, which follow Bose-Einstein statistics. \\

Nucleons (protons and neutrons) are composite fermions that carry one half unit of spin, $\lvert s_z \rvert=\frac{\hbar}{2}$. The constituent particles that make up the nucleons are quarks and gluons, both of which are elementary particles often collectively referred to as partons. Of these, the quarks are fermions with $\lvert s_z \rvert =\frac{\hbar}{2}$, while the gluons are gauge bosons (force carriers) with $\lvert s_z \rvert =\hbar$. The spins of the partons are correlated with the spin of the parent nucleon. This dependence is often described by polarized parton distribution functions, which measure the probability difference between finding a parton with a particular momentum fraction whose spin is aligned with that of the parent nucleon, and the one whose spin is anti-aligned.\\

In a common nuclear/particle scattering experiment, the nucleons are statistically unpolarized, in that the measurements of their spins are equally likely to yield one direction versus the other. In such an experiment, effects due to the nucleon spin tend to average out, and we are left with observables that do not depend on the spin of the nucleon. However, it is also possible to have a polarized experiment, in which the spin of the nucleon is aligned in a chosen direction. In such an experiment, we may hope to observe effects due to the spin of nucleons and/or partons.    \\

The aspect of nucleon spin that we focus on in this analysis is called the transverse single spin asymmetry, which is the effect of transverse (to the direction of motion) polarization on the particle production cross-section in polarized scattering experiments. Our interest is in the particles produced in the forward region (the region close to the beam line, where small angle scattering is measured), where the largest transverse spin effects have been observed. The polarization of the colliding nucleons may point either upward or downward relative to the beam direction. The effects that we observe are found to be correlated with the polarization of the beam heading towards our detector, but not with the polarization of the other beam. This type of phenomena are known as ``single spin" effects, in contrast to ``double spin" effects that depend on the polarizations of both beams. What we actually measure is a left-right asymmetry in the forward production cross-section, as a function of the up-down polarization of the incoming beam. For instance, we may observe more final state particles on the left side of the beam if the beam was polarized upward, and more on the right side if it was polarized downward. A simple schematic of the experiment geometry is shown in figure \ref{ANscheme}.\\

\begin{figure} [t]
\centering
\includegraphics[width=1.\textwidth]{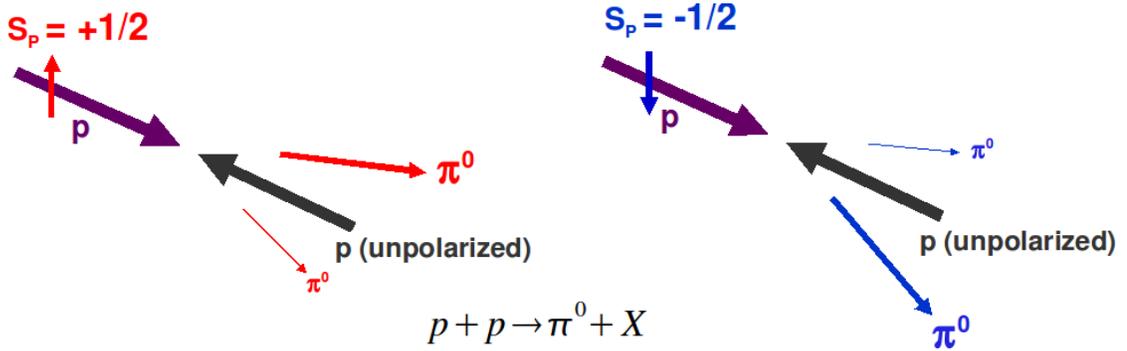}
\caption[The basic geometry of transverse single spin asymmetry]{The basic geometry of transverse single spin asymmetry in case of forward $\pi^0$ production}
\label{ANscheme}
\end{figure}  	

The quantity that is commonly used to measure the transverse single spin asymmetry is called the analyzing power, denoted $A_N$.  It is defined as the following when measured on the left side of the beam.
\begin{equation}\label{eq:AN}
A_N = \frac{d \sigma^{\uparrow} - d \sigma^{\downarrow}}{d \sigma^{\uparrow} + d \sigma^{\downarrow}} 
\end{equation}
Here, $d \sigma^{\uparrow}$ ($d \sigma^{\downarrow}$) indicates the differential cross-section of the final state particle (such as $\pi^0$ and $\eta$) when the incoming beam polarization was up (down). If the measurement is made on the right side of the beam, the signs on the numerator need to be reversed.\\ 

\begin{figure} [t]
\centering
\includegraphics[width=1.\textwidth]{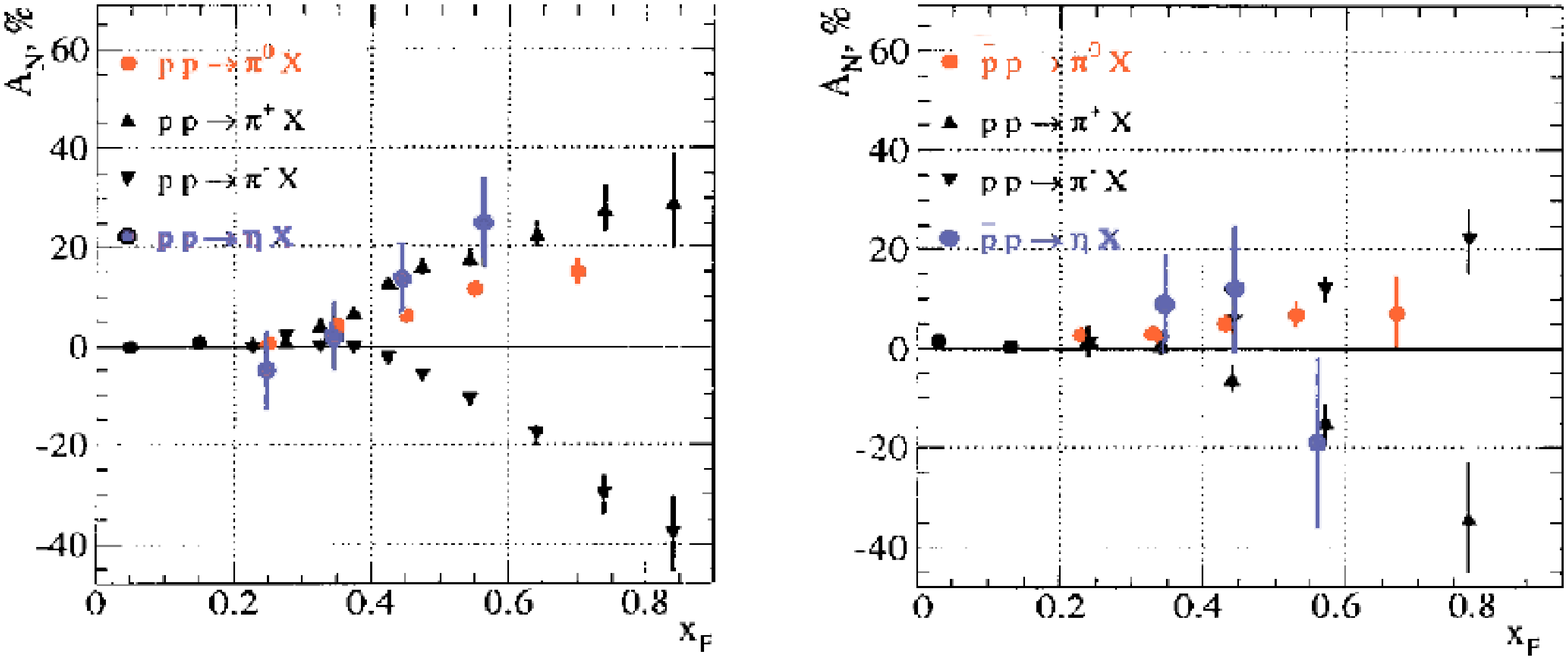}
\caption[FNAL E704 transverse single spin asymmetry results, $A_N$ vs. $x_F$]{FNAL E704 transverse single spin asymmetry results, $A_N$ vs. $x_F$. \cite{paper704} \cite{paper704b} LEFT: $p^{\uparrow}+p \rightarrow h+X$, RIGHT: $\bar{p}^{\uparrow}+p \rightarrow h+X$. Average $p_T \sim 1$ GeV. The $x_F$ denotes ``Feynman x", defined as the fraction of the total hadron energy carried by a parton.}
\label{e704}
\end{figure}  	

The theoretical expectation, which dates back to 1978, \cite{KPR} was that in the regime where perturbative Quantum Chromo-Dynamics is valid, such transverse spin effects should be suppressed at leading twist (simplest parton topology). Nevertheless, large analyzing power ($A_N$) in forward meson production was observed at various energies, \cite{oldAN1} \cite{oldAN2} \cite{oldAN3} most notably by the E704 collaboration at Fermi National Accelerator Laboratory. \cite{paper704} \cite{paper704b} \cite{paper704c} \cite{paper704d} Their results are shown in figure \ref{e704}, in which they saw sizable analyzing powers for all three species of $\pi$ mesons, as well as the $\eta$ meson. The measurement was made at center of mass energy of 19.4 GeV, with an average transverse momentum ($p_T$) of around 1 GeV. It is important to note that at this magnitude of transverse momentum, cross-section measurements usually could not be explained by perturbative QCD, raising concerns as to the applicability of the available theoretical techniques to the observed spin effects.\\

The STAR collaboration, along with other experiments at RHIC, found the large transverse spin effects to persist up to RHIC energy ($\sqrt{s}=200$ GeV). \cite{STARX2} The right-hand panel of figure \ref{starAN} shows the analyzing power ($A_N$) for forward neutral pion ($\pi^0$) production at a center of mass energy of 200 GeV. The transverse momentum of these data points ranges from 1 GeV to 3.5 GeV. The significance of this result lies in the fact that in the very same kinematic region, the cross-section for $\pi^0$ was measured to be in good agreement with perturbative QCD (pQCD) predictions, as shown on the left-hand panel of figure \ref{starAN}. It implies not only that the pQCD based theories should be able to explain the observed spin effects, but also that the transverse spin asymmetry measurements can be considered as useful tests of our current understanding of QCD.    

\begin{figure}[t]
\begin{minipage}{15pc}
\includegraphics[width=15pc]{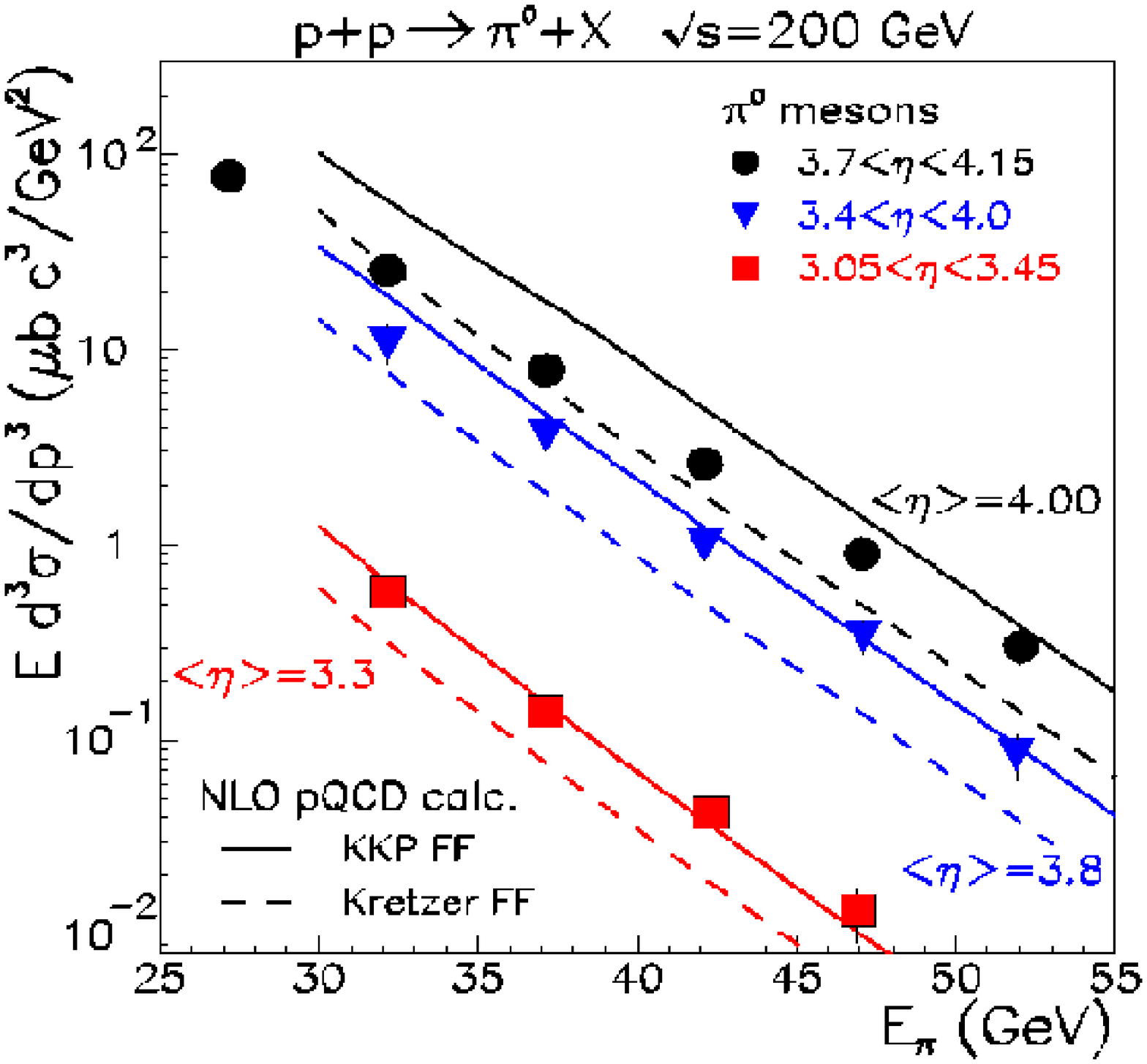}
\end{minipage}\hspace{2pc}%
\begin{minipage}{17pc}
\includegraphics[width=17pc]{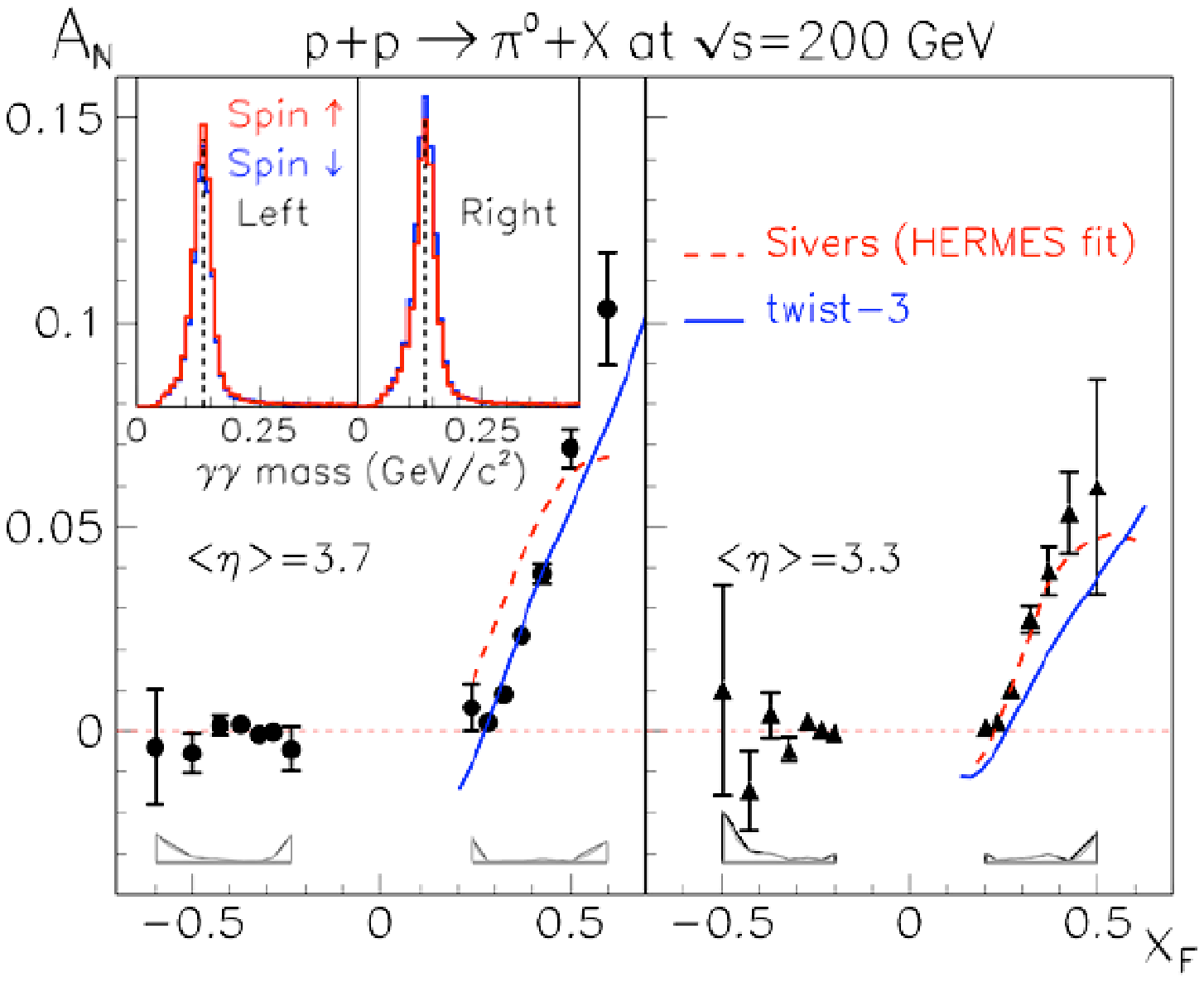}
\end{minipage} 
\caption[STAR results: cross-section and $A_N$]{STAR results. LEFT: $\pi^0$ invariant cross-section at $\sqrt{s}$=200 GeV \cite{STARX}. The cross-section is measured at three pseudo-rapidity ($\eta$) points. The curves represent next-to-leading-order (NLO) pQCD calculations, based on two different fragmentation functions. RIGHT: Transverse single spin asymmetry ($A_N$) vs. $x_F$ for forward $\pi^0$ production. \cite{STARAN} $A_N$ is measured at pseudo-rapidity ($\eta$) points of 3.3 and 3.7. The inset shows the invariant mass spectrum for left and right detector sorted by the spin states.}
\label{starAN}
\end{figure}

\section{Theoretical Background}\label{sec:theory}
\begin{figure} [t]
\centering
\includegraphics[width=1.\textwidth]{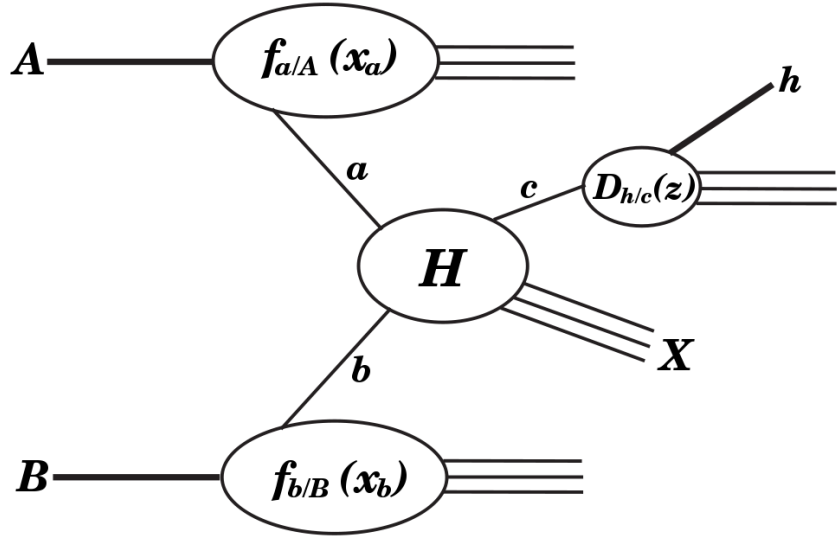}
\caption{Schematic of inclusive hadron production in hadron-hadron collisions, $A + B \rightarrow h + X$}
\label{feyn}
\end{figure}  	

We consider inclusive hadron production in hadron-hadron (in our case, protons) collisions, $A + B \rightarrow h + X$, as shown in figure \ref{feyn}. The QCD factorization theorem \cite{facto} allows us to write the unpolarized differential cross-section as a product of probability functions from the two initial states ($A$ and $B$), hard-scattering ($H$), and the final state ($h$).\\
\begin{equation}\label{eq:x1}
E_h \frac{d \sigma}{d^3 p_h} = \sum_{abc} \int dx_a \: dx_b \: \frac{dz}{z} \; f_{a/A}(x_a) \: f_{b/B}(x_b) \: H(a + b \rightarrow c + X) \: \mathcal{D}_{h/c}(z)
\end{equation}\\
Here, $E_h$ is the energy of the observed hadron, and $f_{a/A}(x_a)$ and $f_{b/B}(x_b)$ are the parton distribution functions for the two colliding hadrons. $f_{a/A}$ ($f_{b/B}$) returns the probability of finding a parton $a$ ($b$) with the momentum fraction $x_a$ ($x_b$) (fraction of the hadron momentum that a parton carries) inside the hadron $A$ ($B$). $\mathcal{D}_{h/c}(z)$ is the fragmentation function that returns the probability of the parton $c$ fragmenting to a hadron $h$ with the momentum fraction $z$ (fraction of the parton momentum that a hadron carries). The sum runs over all flavors of quarks, anti-quarks, and gluons. All of the functions are leading twist. $H$ is the elementary hard scattering cross-section for $a + b \rightarrow c + X$. The hard-scattering term is the only part of this equation that can be calculated using pQCD. All other terms need to be constructed based on experimental data.\\

To deal with the singly polarized collisions in which the hadron $A$ is polarized and the hadron $B$ is unpolarized, we must equip the parton $a$ with a spin density matrix $\rho^a$ to account for the polarization of the parton $a$ inside the hadron $A$. Similarly, we attach a ``decay" matrix $\rho^c$ to the fragmentation function $\mathcal{D}_{h/c}(z)$ to account for the spin dependence of the fragmentation, as the parton $c$ is now polarized. Both $\rho^a$ and $\rho^c$ are $2 \times 2$ matrices whose indices take values of $+$ and $-$. Then equation \ref{eq:x1} is replaced by the following.\\ 
\begin{equation}\label{eq:onespin}
E_h \frac{d \sigma}{d^3 p_h} = \sum_{abc} \sum_{\alpha \alpha^{\prime} \gamma \gamma^{\prime}} \int dx_a \: dx_b \: \frac{dz}{z} \; f_{a/A}(x_a) \: \rho_{\alpha \alpha^{\prime}}^{a} \: f_{b/B}(x_b) \: H_{\alpha \alpha^{\prime} ; \gamma \gamma^{\prime}} \: \rho_{\gamma \gamma^{\prime}}^{c} \: \mathcal{D}_{h/c}(z)
\end{equation}
The sum over $a$ and $c$ only goes over quark and anti-quark flavors, excluding gluons that are unpolarized inside a transversely polarized hadron. The sum over $b$ may still include gluons.\\

When constructing the spin density matrix $\rho^a_{\alpha \alpha^{\prime}}$, it is more natural to use the helicity (spin along the direction of the motion) basis even though we are interested in the transverse spin effects. In the limit where the masses are small compared to the transverse momentum of the scattering, QCD vertices conserve helicities. Explicitly, $\rho^a$ is given by,
\begin{equation}\label{eq:spinden}
\rho^a=
\begin{pmatrix}
1+\lambda & S_x-iS_y \\
S_x+iS_y & 1-\lambda
\end{pmatrix}
\end{equation}

Using spin operators, it is easy to see that $S_x$, $S_y$, and $\lambda$ correspond to the expectation values of the parton spin in x, y, and z (helicity) directions, respectively. Clearly, the transverse spin components belong to the off-diagonal ($+-$ and $-+$) elements of the density matrix. (From now on, we assume that the up-down polarization is in the Y-direction)\\

The decay matrix $\rho^c$ is normalized so that if the hadron $h$ is unpolarized, or spin-less, it becomes an identity matrix. (This is also true if the polarization of hadron $h$ is unmeasured.) In our experiment, the observed final states are $\pi^0$ and $\eta$ mesons, both of which are spin-less. \\

The hard scattering cross-section ($H_{\alpha \alpha^{\prime} ; \gamma \gamma^{\prime}}$) now has four spin indices to accommodate the spin of the initial and final state partons. We integrate over the spin of all other particles, which are unpolarized. Since the transverse spin information is stored in the off-diagonal terms of the spin density matrix $\rho^a_{\alpha \alpha^{\prime}}$, the hard scattering terms that are relevant to transverse spin effects are the ones with $\alpha \neq \alpha^{\prime}$ ($H_{+ - ; \times \times}$ and $H_{- +; \times \times}$).\\

If we let $\mathcal{M}_{\alpha \gamma x}$ be the helicity dependent scattering amplitude, where $\alpha$ and $\gamma$ are the helicities of quarks $a$ and $c$, and $x$ is the index for the spin states of all other partons, we have,
\begin{equation}
H_{\alpha \alpha^{\prime} ; \gamma \gamma^{\prime}} \propto \sum_x \mathcal{M}_{\alpha \gamma x} \: \mathcal{M}_{\alpha^{\prime} \gamma^{\prime} x}^{*}
\end{equation}
Confining ourselves to the case where the partons $a$ and $c$ are both quarks of the same flavor, the helicity conservation in the mass-less limit implies that the hard scattering term is non-zero only when $\alpha = \alpha^{\prime}$ and $\gamma = \gamma^{\prime}$, or when $\alpha = \gamma$ and $\alpha^{\prime} = \gamma^{\prime}$. (If the partons $a$ and $c$ are the quark and its corresponding anti-quark, then the latter condition becomes $\alpha \neq \gamma$ and $\alpha^{\prime} \neq \gamma^{\prime}$.) So the only non-zero terms are $H_{++ ; ++}$, $H_{++ ; --}$, $H_{+- ; +-}$, and since QCD is parity invariant, their parity transformed ($+ \mapsto -$ and $- \mapsto +$) counterparts. \cite{collins}\\

Combining the above two requirements, we find that the hard scattering terms necessary to generate the transverse spin effects are $H_{+- ; +-}$ and $H_{-+ ; -+}$. However, both of these terms, with $\gamma \neq \gamma^{\prime}$, couple to the off-diagonal elements of the decay matrix $\rho_{\gamma \gamma^{\prime}}^{c}$, which are zero since $\rho_{\gamma \gamma^{\prime}}^{c} = \delta_{\gamma \gamma^{\prime}}$. Clearly, the hard scattering terms we need are the ones that couple to the off-diagonal elements of two spin matrices. Consequently, we need another spin matrix with non-zero off diagonal elements in the system, or in other words, another polarized quark whose polarization can be measured. Since our final state is spin-less, the only object with known polarization in the system is the parton $a$. Therefore, we may conclude that at leading twist, the production cross-section is insensitive to the transverse spin of parton $a$. A very good overview of the subject can be found in \cite{barone}.\\  

Before we look at how we can get around this conclusion to obtain the transverse spin effect, it is necessary to introduce transverse momentum dependent (TMD) factorization. The analyzing power is an asymmetry in the azimuthal distribution of particles, and for a single final state like $\pi^0$ and $\eta$, we need to introduce its transverse momentum $k_{\perp}$ into the problem. Equation \ref{eq:onespin} uses collinear factorization, which has no room to accommodate unbalanced $k_{\perp}$. Therefore we rewrite equation \ref{eq:onespin} using TMD factorization as the following.\\
\begin{equation}\label{eq:tmdspin}
\begin{split}
E_h \frac{d \sigma}{d^3 p_h} =& \sum_{abc} \sum_{\alpha \alpha^{\prime} \gamma \gamma^{\prime}} \int dx_a \: d^2k_{\perp a} \: dx_b \: d^2k_{\perp b} \: \frac{dz}{z} \: d^2k_{\perp h} \; \\
& \hat{f}_{a/A}(x_a, k_{\perp a}) \: \rho_{\alpha \alpha^{\prime}}^{a} \: \hat{f}_{b/B}(x_b, k_{\perp b}) \: H_{\alpha \alpha^{\prime} ; \gamma \gamma^{\prime}}(k_{\perp a},k_{\perp b},k_{\perp h}) \: \rho_{\gamma \gamma^{\prime}}^{c} \: \mathcal{\hat{D}}_{h/c}(z, k_{\perp h})
\end{split}
\end{equation}\\
The hats on the parton density and fragmentation functions (PDF and FF) indicate that they are transverse momentum dependent, and they are defined so that once the transverse momentum is integrated out, we recover the collinear PDF and FF.
\begin{equation}
f_{a/A}(x_a) = \int d^2 k_{\perp a}  \:  \: \hat{f}_{a/A}(x_a, k_{\perp a})
\end{equation}
\begin{equation}
\mathcal{D}_{h/C}(z) = \int d^2 k_{\perp h}  \:  \: \mathcal{\hat{D}}_{h/C}(z, k_{\perp h})
\end{equation}

We note that the QCD factorization theorem is proven only for the collinear case, and the validity of the TMD approach is conjectural. In fact, universality of the TMD factorization has been found to be violated in some cases. \cite{nofacto2} \cite{nofacto} These results do not invalidate the TMD approach in general, but limits the scope of its applicability.  

\subsection{Collins Effect}

Even if the final state is spin-less, the second polarized object can still be the parton $c$ if the fragmentation into hadron $h$ depends on its spin. \cite{collins} \cite{collins2} The spin dependence may be observed in the form of an analyzing power, generated by the fragmentation process itself. This effectively provides the information on the spin state of the parton $c$. Formally, this is equivalent to allowing the decay matrix $\rho^c$ to be non-diagonal. The product of the hard scattering terms necessary for the transverse spin effects and the two spin matrices, $\rho^a_{+-} \cdot H_{+- ; +-} \cdot \rho^c_{+-}$ and $\rho^a_{-+} \cdot H_{-+ ; -+} \cdot  \rho^c_{-+}$, are now non-zero. As a result, we obtain an analyzing power that is proportional to the following: \\
\begin{equation}\label{eq:collins}
\begin{split}
A_N &\propto d\sigma^{\uparrow} - d\sigma^{\downarrow} \\
&\propto \sum_{abc} \int dx_a \: dx_b \: \frac{1}{z} \: d^2k_{\perp h} \; f_{a/A}(x_a) \: f_{b/B}(x_b) \: H_{+- ; +-}(k_{\perp h}) \: \Delta^N \mathcal{D}_{h/c^{\uparrow}}(z,k_{\perp h})
\end{split}
\end{equation}\\
Notice that the transverse momentum dependences for all initial states have been integrated out. Only fragmentation remains $k_T$ dependent. $\Delta^N \mathcal{D}_{h/c^{\uparrow}}$ is called the Collins function, and defined as below.
\begin{equation}
\Delta^N \mathcal{D}_{h/c^{\uparrow}}(z, k_{\perp h}) \equiv \mathcal{\hat{D}}_{h/c^{\uparrow}}(z, k_{\perp h}) - \mathcal{\hat{D}}_{h/c^{\downarrow}}(z, k_{\perp h})
\end{equation}

With the Collins effect, the hadron $h$ is produced with ``intrinsic" transverse momentum (one that does not come from hard scattering) $k_T$, the direction of which is correlated with the transverse spin of the quark $c$. The spin of quark $c$ is the same as the spin of quark $a$, because the hard scattering term is $H_{+- ; +-}$. In other words, the quark maintains its spin through the hard scattering. Finally, if parton $a$ is a large $x_F$ quark, it is likely that its spin is aligned with that of the hadron $A$. This way, the $k_T$ of the observed hadron $h$ can be correlated with the spin of the incoming polarized hadron $A$.\\

The $k_T$ of the hadron $h$ is orthogonal to the momentum and spin of parton $c$. That is, the up-down spin of parton $c$ would produce a left-right asymmetry. It also means that the asymmetry is defined with respect to the direction of parton $c$'s momentum, not the direction of the beam. In other words, the asymmetry is confined within the jet produced by parton $c$, and the jet axis itself, which is in the same direction as parton $c$, does not have an analyzing power. This can be verified in an experiment by performing a full jet reconstruction and measuring the jet asymmetry. If the observed final state is not coming from jet fragmentation, such as the case with prompt photons, then the Collins effect yields zero asymmetry. In addition, as mentioned before, the lack of gluon transverse polarization means that the jets that originate from gluons have zero Collins effect as well.  \\

\subsection{Sivers Effect}
Another approach relies on the intrinsic transverse momentum ($k_T$) in the initial state to generate the observed asymmetry. In this model, called the Sivers effect, \cite{sivers} \cite{sivers2} the transverse spin of the partons play no role, and the previous discussions regarding the off-diagonal elements of the spin matrices do not apply. Instead, the asymmetry comes from the $k_T$ of parton $a$, which is correlated with the transverse spin of the polarized hadron $A$. The orbital angular motion of quarks and gluons inside a proton would provide such a mechanism. If a parton with left-going $k_T$ contributes a different scattering amplitude from a parton with right-going $k_T$ to the same final state, there can be a net spin asymmetry. This situation could arise if the remaining parts of the hadrons that do not participate in hard scattering provide an environment with which the scattering partons interact, and if the extra interactions depend on the $k_T$ of the parton. The resulting analyzing power is proportional to the following:\\
\begin{equation}\label{eq:sivers}
A_N \propto \sum_{abc} \int dx_a \: dx_b \: \frac{1}{z} \: d^2k_{\perp a} \; \Delta^N f_{a/A^{\uparrow}}(x_a,k_{\perp a}) \: f_{b/B}(x_b) \: H(k_{\perp a}) \: \mathcal{D}_{h/c}(z)
\end{equation}\\
The transverse momentum dependences of the unpolarized parton distribution function and the fragmentation function have been integrated out. $\Delta^N f_{a/A^{\uparrow}}$ is given by,
\begin{equation}
\Delta^N f_{a/A^{\uparrow}}(x_a, k_{T a}) \equiv \hat{f}_{a/A^{\uparrow}}(x_a, k_{T a}) - \hat{f}_{a/A^{\downarrow}}(x_a, k_{T a})
\end{equation}
This function is called the Sivers function. It is clear that even if the difference between the hard scattering cross-section for spin up and spin down is zero, a non-zero Sivers function will generate the analyzing power. \cite{sivers3}\\ 

As a heuristic example, consider the following case. The orbital angular motion causes a correlation between the direction of $k_T$ and the location of the parton inside the polarized hadron. (Partons in the ``front" of the hadron are going one way, and the ones in the ``back" are going the other way.) If the amplitude for scattering off a ``front" parton is different from scattering off a ``back" parton because the incoming scatterer has to go through more soft interactions to get to the back of the proton, it can lead to a bias in $k_T$ that is correlated with the spin of the polarized hadron. In the Collins effect, the bias in $k_T$ was on the unpolarized final state hadron, and it was correlated with the spin of the polarized quark from which it fragments. In the Sivers effect, the bias is on the unpolarized parton, and it is correlated with the spin of the polarized initial state hadron from which it is pulled.\\

Unlike the Collins effect, the Sivers function needs not be zero for gluons. The gluons are not transversely polarized, but they may carry orbital angular momentum. The Sivers effect is also applicable to a wider range of final states, as all the action occurs before the hard-scattering, and the fragmentation plays no role. It is in principle possible to generate the Sivers asymmetry for prompt photons. Finally, the two-body final states are not precisely back-to-back in azimuth, because the $k_T$ bias from the initial state feeds into the hard scattering. A good way to observe this effect is the full reconstructions of di-jets and photon-jets, from which we can measure their relative azimuthal angle distribution. \\

\subsection{Boer-Mulders Effect}
The Boer-Mulders effect \cite{boer} \cite{boer2} has formal similarities to the Collins effect. Firstly, it introduces the second polarized quark to the system to allow the off-diagonal hard-scattering terms to survive. Secondly, the bias in $k_T$ that generates the asymmetry is related to a polarized quark and an unpolarized hadron. In the Collins effect, the unpolarized hadron fragmented from a polarized quark. In the Boer-Mulders effect, the unpolarized hadron is the incoming hadron $B$, and the polarized quark is the parton $b$ that is pulled from it. Physically, however, it is very similar to the Sivers effect, in that it requires a correlation between spin and orbit, and soft interactions from the environment that distinguish different parts of the orbit. In the Sivers effect, the orbital direction of the quarks was correlated with the spin of the proton. In Boer-Mulders, it is correlated with the spin of the quarks themselves. We rewrite the TMD factorized formula in a form more specific to this effect.\\
\begin{equation}\label{eq:boer}
\begin{split}
E_h \frac{d \sigma}{d^3 p_h} =& \sum_{abc} \sum_{\alpha \alpha^{\prime} \beta \beta^{\prime}} \int dx_a \: dx_b \: d^2k_{\perp b} \: \frac{dz}{z} \; \\
& f_{a/A}(x_a) \: \rho_{\alpha \alpha^{\prime}}^{a} \: \hat{f}_{b/B}(x_b, k_{\perp b}) \:\rho_{\beta \beta^{\prime}}^{b} \: H_{\alpha \alpha^{\prime} ; \beta \beta^{\prime}}(k_{\perp b}) \: \mathcal{D}_{h/c}(z)
\end{split}
\end{equation}\\
Notice that instead of the decay matrix $\rho^c_{\gamma \gamma^{\prime}}$, we now have a spin density matrix $\rho^b_{\beta \beta^{\prime}}$ to describe the spin state of parton $b$. We have also integrated out the transverse momentum dependences of the polarized hadron and the fragmentation process.\\

For the Collins effect, the spin density matrix $\rho^b_{\beta \beta^{\prime}}$ was considered effectively to be unity, similar to the decay matrix for the ``unpolarized" fragmentation process. Unlike $\rho^a$, $\rho^b$ can be proportional to an identity matrix because an unpolarized quark is necessarily in a mixed state. (A pure state has only one eigenvalue.) But once we allow the quark $b$ to be polarized, $\rho^b$ picks up off-diagonal elements just like $\rho^a$. The hard scattering terms that are relevant here are again $H_{+- ; +-}$ and $H_{-+ ; -+}$, despite the change of indices from $\gamma$ to $\beta$ compared to the Collins effect. It is clear that these terms will now survive by coupling to the non-zero off-diagonal elements of $\rho^a$ and $\rho^b$. \\

The reason that the quark $b$, which comes from the unpolarized hadron $B$, can be polarized is because the only hard scattering terms that go into the analyzing power are the ones with $\alpha=\beta$. In the Collins effect, the quark ``remembered" ($\alpha=\gamma$) its spin through the hard scattering. In Boer-Mulders, the polarized quark ``selects" a quark with the same spin from the unpolarized hadron. Since the spin of the quark from the polarized hadron is correlated with the spin of the hadron itself, we get a correlation between the polarization of the beam, and the spin of the quark from the unpolarized beam. From here, we just need the intrinsic $k_T$ of a quark inside an unpolarized hadron to be correlated with the spin of the quark to generate the asymmetry. \\

Again as a heuristic example, consider a case in which there is a spin-orbit coupling inside an unpolarized hadron, which causes spin up quarks to orbit in one way, and the spin down quarks in the other way. If there is an interaction with the environment that produces different scattering amplitudes for different points in the orbit, (like the Sivers effect) then we may find that spin up quarks from an unpolarized hadron are more likely to be going one way than the other.\\

\subsection{Twist-3 Effect}
So far, the three mechanisms that we've discussed have all been leading twist (twist-2) effects built on transverse-momentum dependent (TMD) factorization. However, it is also possible to generate transverse spin effects based on higher twist phenomena within the framework of the proven collinear factorization. \cite{twist3} The twist count can be thought of as the degree of suppression by the hard-scattering transverse momentum ($p_T$), where twist-2 has no $p_T$ suppression, and twist-3 is suppressed by one power of $p_T$ ($\propto \frac{1}{p_T}$). (A good review on this topic can be found in \cite{Jaffe}) However, in reality the above discussed ``twist-2" effects have their origins in the non-perturbative parts of the problem (initial and final state effects), and may be suppressed by powers of $p_T$. \\

\begin{figure} [t]
\centering
\includegraphics[width=1.\textwidth]{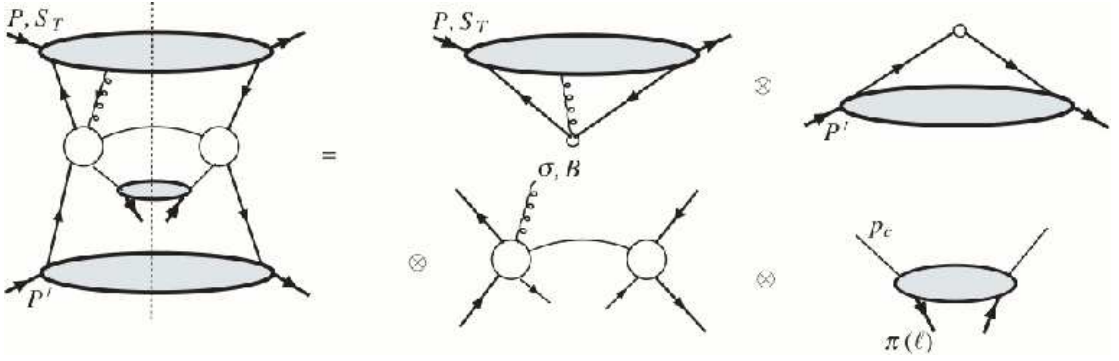}
\caption[Generic Feynman diagram for twist-3 contribution to $A_N$]{Generic Feynman diagram for twist-3 contribution to the inclusive pion single spin asymmetry in singly polarized proton-proton collision. \cite{twist3a} RHS, clockwise from top left, twist-3 quark-gluon correlation function for the polarized proton, PDF for the unpolarized proton, fragmentation function, and the hard-scattering function with interference between two-particle and three particle initial states.}
\label{t3}
\end{figure}  	

The twist-3 function, which has an additional gluon propagator, can be associated with the polarized initial state hadron, the unpolarized initial state hadron, and the fragmentation function. Again for $A + B \rightarrow h + X$ scattering, we may write the transverse spin dependent cross-section as the following: \cite{twist3a}
\begin{equation}
\begin{split}\label{eq:t3}
d \sigma (p_T, \vec{s_T}) &= \sum_{abc} f_{a/A}^{(3)}(x_{a1}, x_{a2}, \vec{s_T}) \otimes f_{b/B}(x_b) \otimes H^{\prime}(p_T, \vec{s_T}) \otimes \mathcal{D}_{h/c}(z)\\
&+ \sum_{abc} f_{a/A}(x_a, \vec{s_T}) \otimes f_{b/B}^{(3)}(x_{b1}, x_{b2}) \otimes H^{\prime\prime}(p_T, \vec{s_T}) \otimes \mathcal{D}_{h/c}(z)\\
&+ \sum_{abc} f_{a/A}(x_a, \vec{s_T}) \otimes f_{b/B}(x_b) \otimes H^{\prime\prime\prime}(p_T, \vec{s_T}) \otimes \mathcal{D}_{h/c}^{(3)}(z_1, z_2)
\end{split}
\end{equation}

Since we are using collinear factorization, the intrinsic $k_T$ does not appear in the formula. As before, the sum runs over all flavors of quarks, anti-quarks, and gluons. Functions with the symbol (3) on top indicate the twist-3 function, which has an additional independent variable for the extra gluon. Notice that the elementary scattering cross-section $H$ (for $a + b \rightarrow c + X$) is different for each line, due to the three different initial and final state configurations. Figure \ref{t3} depicts the Feynman diagrams for the first line of the equation \ref{eq:t3}. The modifications from all-leading-twist formula are visible both in polarized initial state hadron, and hard scattering term through the extra gluon propagator.\\   

It turns out that of the three lines, only the first one has a sizable contribution to the transverse single spin asymmetry. \cite{twist3a} The key ingredients that produce the spin effects are the twist-3 quark-gluon correlation function associated with the polarized hadron, and the modified elementary scattering cross-section that includes interference between the two scattering amplitudes; one with a two parton initial state, and another with a three parton initial state. \\

The twist-3 effect can be thought of as the simplest perturbative approximation of the interaction between the scattering partons and the environment. In fact, the first line of equation \ref{eq:t3} is closely related to the Sivers effect, \cite{sivt3} which also needs the scatterer-environment interaction to generate the asymmetry.  \\

\section{Measurement Description}
Our measurement is an extension of the results shown in figure \ref{starAN}. From singly polarized proton-proton collisions ($p^{\uparrow}+p \rightarrow h+X$), we make inclusive measurements of $\pi^0$ and $\eta$ mesons produced at very forward region, with average pseudo-rapidity of 3.7. Pseudo-rapidity, often denoted $\eta$, is a measure of the scattering angle, commonly used in relativistic scattering experiments because the difference in pseudo-rapidity is independent of beam energy (or Lorentz boost in the beam direction). It is defined in terms of the scattering angle $\theta$ (angle from the beam axis) by,
\begin{equation}
\eta=-ln \lbrace tan(\frac{\theta}{2}) \rbrace
\end{equation}

In this forward region, the dominant scattering channel is the one between a large momentum quark from the polarized proton, and a small momentum gluon from the unpolarized proton. The previous STAR forward cross-section and asymmetry measurements (for $\pi^0$ only) were made at $x_F$ of up to 0.55. \cite{STARX} \cite{STARAN} In this measurement, we extend the $x_F$ coverage significantly by measuring both final states from $x_F=0.4$ to $x_F=0.8$. We measure and compare the analyzing powers of the two neutral mesons. In addition, we measure their cross-sections in the same $x_F$ range in which the asymmetries are measured, which is crucial in understanding the nature of the spin effects. \\

\end{doublespace}
\chapter{Experimental Setup}\label{ch:exp}
\setlength{\parindent}{1cm}
\begin{doublespace}

\section{Introduction}
Out experiment is performed at the Relativistic Heavy Ion Collider (RHIC), as a part of the STAR (Solenoidal Tracker At RHIC) collaboration. In this chapter, we will cover the basics of the RHIC environment during polarized proton collisions, as well as the details of the detector used for this analysis, the Forward Pion Detector (FPD). The discussion includes a brief description of the machine called the Siberian Snake, a crucial device that allows RHIC to successfully deliver the world's only polarized particle collisions.  \\

\section{Relativistic Heavy Ion Collider}
Relativistic Heavy Ion Collider (RHIC) is a high energy particle collider located at Brookhaven National Laboratory in Long Island, New York. In operation since the year 2000, RHIC has unique physics capabilities that make it one of the premier particle colliders in the world. RHIC was the first machine of its kind capable of colliding two beams of heavy ions, such as gold nuclei, at relativistic energy. RHIC can accelerate the heavy ion beams up to 250 GeV of energy per nucleon, or 99.999 \% of the speed of light. This high energy heavy ion collisions are used to create an extremely hot and dense state of nuclear matter, thought to have existed during a brief period in the early universe following the Big Bang. \cite{white} The precise nature of this exotic state of matter, called quark gluon plasma, has fundamental ramifications in diverse fields of physics, from nuclear and particle physics to cosmology. In addition, RHIC remains the first and the only particle collider in the world capable of colliding polarized beams at relativistic energy, providing a vastly higher center of mass collision energy than any previous spin physics experiments. At RHIC, beams of protons are polarized either longitudinally, where the spins of the protons are parallel/anti-parallel to the beam direction, or transversely, where they are perpendicular to the beam direction. The longitudinal polarization is used to study the spin structure of the proton, whose one half unit of spin must originate from the spin and the orbital motion of its constituents. \cite{deltaG} \cite{deltaG2} \cite{deltaG3} Despite the progress made in our understanding of the unpolarized structure of the proton, a large portion of its spin structure still remains unknown. The RHIC spin program probes the role of gluon polarization in constructing the proton spin. In addition, the transverse polarization provides a unique opportunity to test the inner workings of Quantum Chromo-Dynamics (QCD), an extremely successful theory of nuclear interaction that nevertheless has struggled to explain the role of transverse spin in relativistic scatterings of quarks and gluons. \cite{STARAN} By studying transverse spin dependent observables in processes for which the unpolarized observables are well described by QCD, we can test the current theoretical framework at its frontier.\\
\begin{figure} [t]
\centering
\includegraphics[width=1.\textwidth]{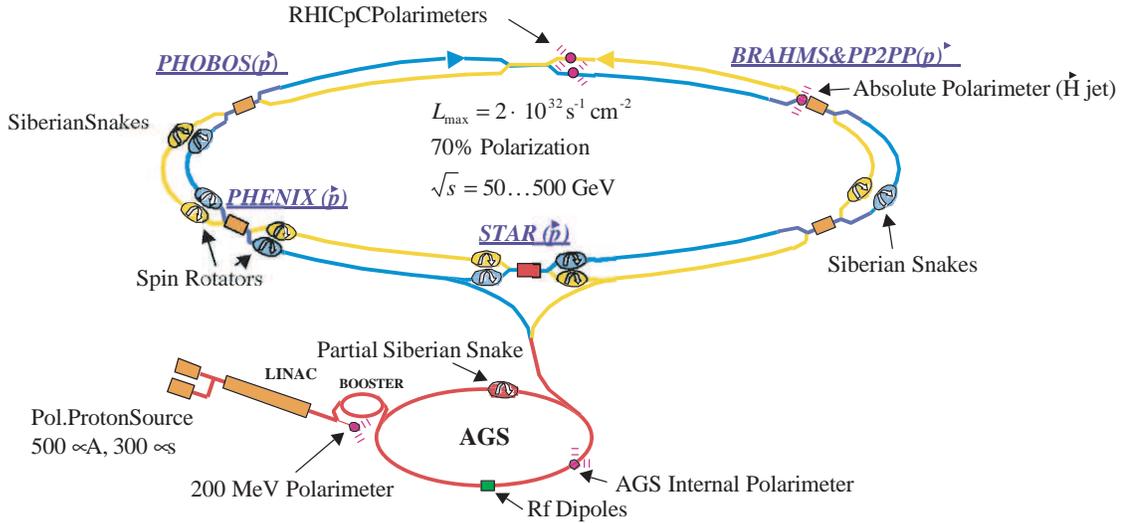}
\caption[Schematic of RHIC complex]{Schematic of the RHIC complex as of 2006, emphasizing the facilities and functions related to the polarized proton collisions. The two smaller experiments, BRAHMS and PHOBOS, are no longer active. Figure from \cite{nuclinst}.}
\label{rhic2}
\end{figure}  

Figure \ref{rhic2} shows the schematic of the RHIC complex. (The tandem Van de Graff, which is used as a heavy ion source, is not shown in the schematic.) In addition to the two main rings of RHIC, we have the Alternating Gradient Synchrotron (AGS) complex, which serves as both the particle source and pre-accelerator for RHIC. The linear accelerator (LINAC) includes a polarized ion source that produces beams of polarized protons. The LINAC accelerates the proton beam up to 200 MeV, \cite{nuclinst} and injects it into the AGS booster. AGS booster in turn forms high intensity bunches by accumulating protons from the Linac. (The ``beam" of particles in a particle collider consists of many bunches of highly concentrated particles. At RHIC, each beam is typically made up of around 100 bunches.) Once the bunches are formed, they are sent to the AGS, which accelerates the beam up to 25 GeV in preparation for the injection into the main rings of RHIC. \cite{nuclinst} The AGS itself used to be the main high energy physics facility at Brookhaven until the construction of RHIC, and it was used for three Nobel prize winning researches (76', 80' and 88').\\

\subsection{Spin Precession in an External Magnetic Field}

Until the RHIC era, virtually all the experiments that involved polarization were performed in fixed target environments, where only one beam is accelerated and collided into a stationary target. The advantage of this type of experiment is that since only the target is polarized, one does not have to deal with accelerating a polarized beam to a high energy. The disadvantage is that the center of mass energy in a fixed target experiment is substantially lower than what can be achieved in a collider environment, where two high energy beams are collided with each other. However, the difficulty in maintaining polarization during acceleration and storage of high energy beams is so significant, that RHIC will remain the only polarized collider in the world in the foreseeable future. To examine the challenge associated with polarizing high energy beams, we first look at the Thomas-BMT equation, which classically describes the motion of the spin vector of a particle in an external magnetic field: \cite{rhicprl}
\begin{equation}\label{thomas}
\frac{d\vec{S}}{dt} = \frac{e}{\gamma m} \: \vec{S} \times \lbrace (1+G\gamma)\vec{B}_{\perp}+(1+G)\vec{B}_{\parallel} \rbrace
\end{equation}  
Here, $\vec{S}$ is the spin vector of the particle, $\gamma=1/\sqrt{1-\beta^2}$ is the Lorentz factor, and ${B}_{\perp}$ and ${B}_{\parallel}$ are the transverse and longitudinal components of the external magnetic field, respectively. The $e$ and $m$ are the charge and mass of the particle, and $G$ is its anomalous magnetic moment, defined as:
\begin{equation}
G=\frac{1}{2}(g-2)
\end{equation} 
It is very close to zero for an ideal, structureless fermion like the electron, which has $G=0.001159652$. For a proton, which is a composite particle, $G=1.792947$.\\

For a nearly ideal particle collider, the longitudinal component of the magnetic field should be negligible compared to the transverse field produced by the orbit guiding dipoles. Under that assumption, we can compare equation \ref{thomas} with the Lorentz force equation for the orbit of a charged particle in an external magnetic field: 
\begin{equation}\label{orbit}
\frac{d\vec{v}}{dt} = \frac{e}{\gamma m} \: \vec{v} \times \vec{B}_{\perp}
\end{equation} 
Here, $\vec{v}$ is the velocity vector of the charged particle. It is easy to see that in the absence of ${B}_{\parallel}$, the two equations are nearly identical up to the factor of $G\gamma$. In fact, one of the clearest manifestations of the anomalous magnetic moment is that if $G=0$, the spin of a particle orbiting under a vertical magnetic field rotates along the vertical axis at the same rate as the orbital motion. That is, the spin vector makes one full rotation in the same time that it takes the particle to complete a full revolution around its orbit. Consequently, the spin vector is stationary with respect to the momentum vector. \\

In reality, because real particles have non-zero $G$, the spin will precess along the direction of the transverse magnetic field. Again comparing the equations \ref{thomas} and \ref{orbit}, we see that the number of times the spin vector precesses along the vertical axis is equal to $G\gamma$. For instance, if $G\gamma=1$, then the spin will rotate twice as fast as the orbital motion, making one extra rotation relative to the momentum vector per every revolution. Figure \ref{precess} shows an example of a spin precession, in the case of $G\gamma=6$. We call this quantity, $G\gamma$, the ``spin tune". Consequently, the stable spin direction for a polarized beam in a particle accelerator is vertical, the same as the guiding dipole field. If the direction of polarization was precisely identical to the direction of the magnetic field, there will be no precession. In reality, due to various imperfections, there will always be some level of precession along the vertical axis. \\

\begin{figure} [t]
\centering
\includegraphics[width=.5\textwidth]{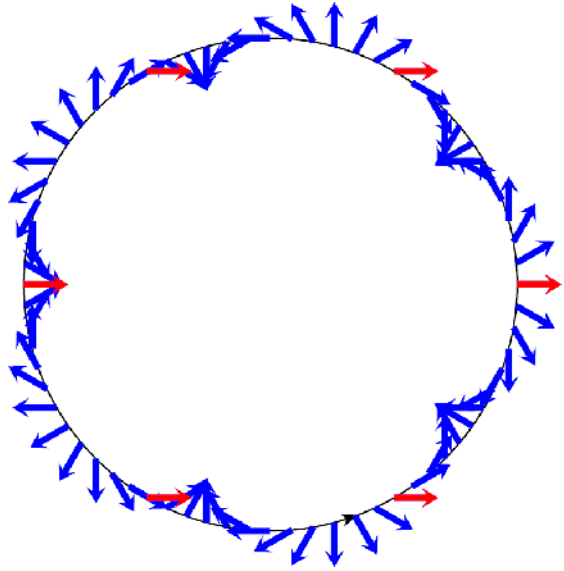}
\caption[Schematic of the spin vector precessing in an orbit]{Schematic of the spin vector precessing around the vertical axis, for the case of $G\gamma=6$. The frequency of the spin precession is exactly 6 times that of the orbital motion, locking the spin direction to the position in the orbit. Figure from RHIC presentation by W. MacKay.}
\label{precess}
\end{figure}  

The difficulty in accelerating a polarized beam without losing its polarization comes from the following. Both the imperfections of a real-life accelerator, and the need to have focusing magnetic fields for the beam, introduce depolarizing effects that drive the spin direction away from the stable vertical direction. These effects are generally localized in the ring, and the direction of the perturbation is determined by the direction of the spin at that location. This means that if the spin tune has an integer value, so that the spin direction is fixed at a given point in the orbit, we could encounter a resonance condition in which the spin perturbing field pushes the spin vector away from the vertical in the same direction every time the particle completes a revolution. This is referred to as a depolarizing resonance. For a proton beam with $G=1.792947$, the energy interval between the resonances is only 523.34 MeV. \cite{snake} It is clear that this poses an enormous challenge when we want to accelerate the beam up to 100 GeV and higher. \\

\subsection{Siberian Snakes}
One way to avoid the depolarizing resonances is to flip the direction of the spin once every revolution. This way, the effects of the depolarizing field will cancel out over two complete revolutions, and the polarization can be retained. In effect, this makes the spin tune always half-integer, never satisfying the condition for depolarizing resonances. \cite{snake} The device designed specifically to perform this function is called a Siberian Snake, named for the two Soviet scientists (Derbenev and Kondratenko) from the Institute of Nuclear Physics at Novosibirsk in Siberia. The original idea was proposed by them in 1976. \cite{originsnake}\\

The simplest form of a Siberian Snake would be a solenoid, which provides a longitudinal magnetic field. The advantage of a longitudinal field is that it does not alter the orbit of the particle, and only rotates its spin. However, looking back at equation \ref{thomas}, we see that the strength of the $\vec{B}_{\parallel}$ term is proportional to $1/\gamma$. This means that, assuming we cannot make an arbitrarily long solenoid, the field strength has to scale with the Lorentz factor $\gamma$ to produce a constant degree of spin rotation. This is clearly very difficult, as $\gamma$ scales with the beam energy. Therefore a solenoid Snake is only suitable for low energy applications.\\

On the other hand, the $\vec{B}_{\perp}$ term in equation \ref{thomas} is proportional to $(1+G\gamma)/\gamma$. This means that for a large value of $\gamma$, the spin rotating effect of a transverse field becomes constant independent of the beam energy, making it much more suitable for high energy applications. The clear downside to this approach is that $\vec{B}_{\perp}$ changes the orbit along with the spin, potentially disturbing the orbital motion of the beam. However, it is important to note that the rates of change for the spin and the orbit are different. Consequently, it is in principle possible to design a device that reverses the spin direction, while inducing a change in the orbital motion that cancels itself out.  \\

\begin{figure} [t]
\centering
\includegraphics[width=1.\textwidth]{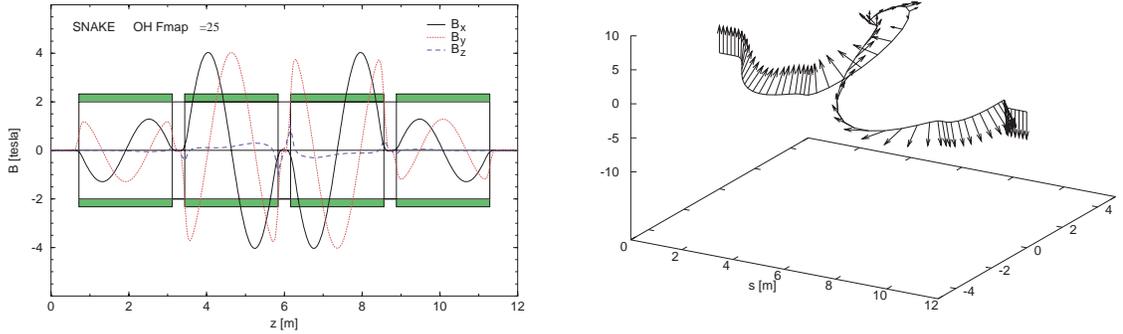}
\caption[Magnetic field and spin orbit tracking for Siberian Snakes]{LEFT: X, Y, and Z components of the rotating magnetic field through the four helical magnets of a Siberian Snake. RIGHT: Spin and orbit tracking of a particle going through a full Snake. The outgoing particle has the same momentum vector as the incoming one without any transverse offset. Only the spin direction has been reversed. Figures from \cite{nuclinst}}
\label{snake}
\end{figure}  

At RHIC, a full Siberian Snake employs four helical dipole magnets to produce a rotating transverse magnetic field that flips the spin vector, without altering the orbit in any direction. The changes in orbital motion introduced by the first two magnets are exactly canceled out by the last two magnets. Figure \ref{snake} shows field configuration of the four helical magnets, and the particle trajectory and its spin direction inside a Snake. We see that the outgoing beam is on the same path as the incoming beam with no change in the direction of the motion or the transverse offset, but with a reversed polarization. \\

\begin{figure} [t]
\centering
\includegraphics[width=1.\textwidth]{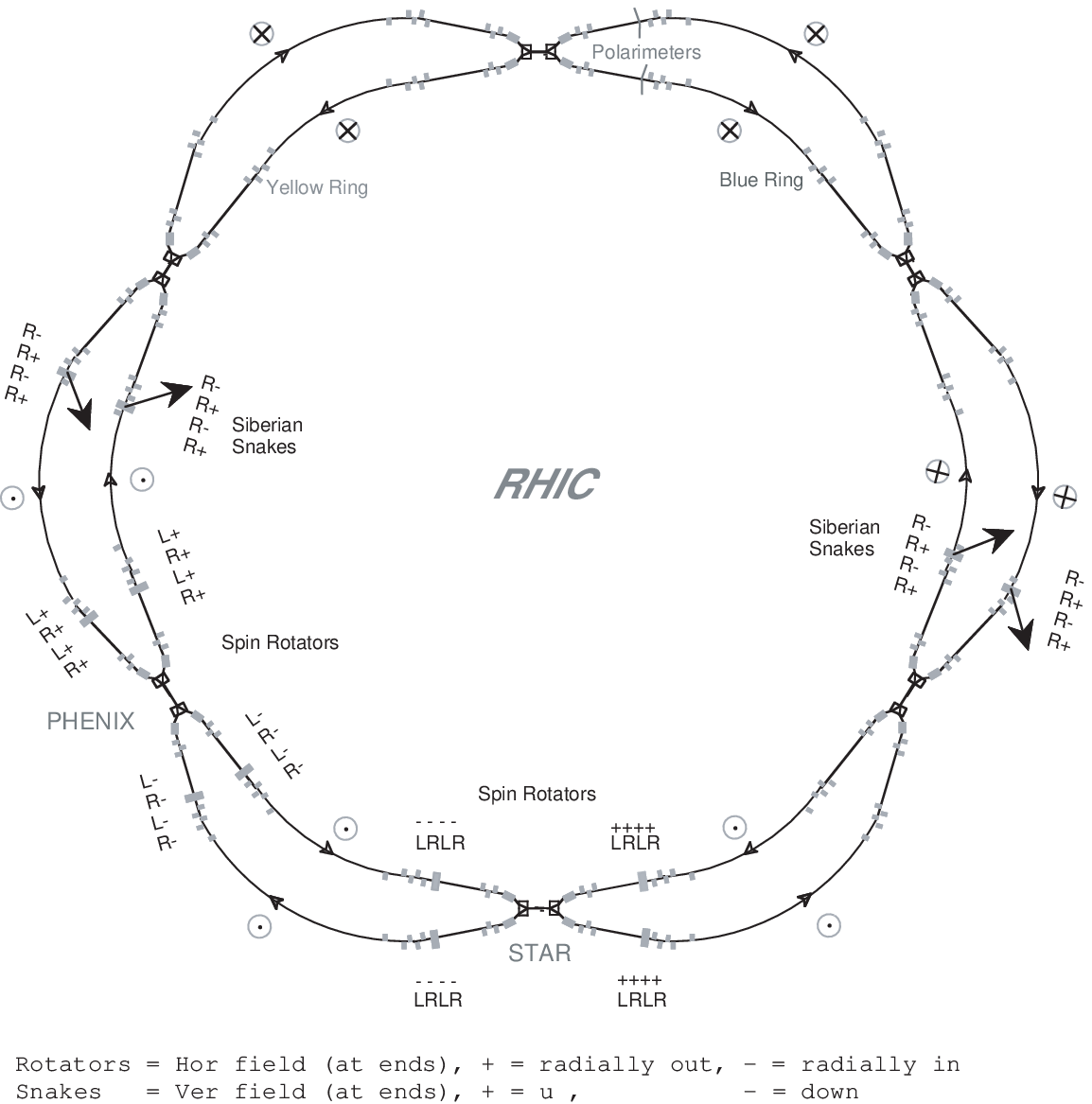}
\caption[Siberian Snake and spin rotator configuration at RHIC]{Locations of the Siberian Snakes and spin rotators. Also shown are the rotational axis for the Snakes, and the spin directions along the ring. Figure from \cite{nuclinst}}
\label{rhic}
\end{figure}  

There are two full Siberian Snakes for each of the two main rings, installed at the 3 o'clock and 9 o'clock sections. \cite{nuclinst} In addition, RHIC requires spin rotators for the collision points in order to provide longitudinal polarization for the experiments. The spin rotators are essentially Siberian Snakes that rotate the spin half-way through, turning a vertical spin state into a helicity state. There are a total of four rotators, two for each of the two ``large" experiments, STAR and PHENIX. Figure \ref{rhic} shows the overall layout of the Snakes and Rotators at RHIC. Also shown in the figure are the axis around which the spin is flipped for each of the Snake. Through the use of Siberian Snakes, RHIC is capable of delivering over 60 \% polarization for the proton beams.\\

\section{STAR Forward Pion Detector}
The Solenoidal Tracker At RHIC (STAR) is one of the two large experiments at RHIC. The collaboration consists of fifty five institutions from twelve countries, with more than 500 collaborators. The complex detector system centers around the massive Time Projection Chamber, capable of simultaneously tracking thousands of particles created by a heavy ion collision. The subsystem of interest for this analysis is the Forward Pion Detector (FPD), located in the very forward region of STAR. It is a set of small electromagnetic calorimeters, covering pseudo-rapidity range from 3.3 to 4.0. \cite{fpd} First commissioned during RHIC run 3 (in 2003), the FPD calorimeter modules were initially placed on both east and west sides of the STAR wide angle hall, in front of the DX magnets that steer the beams for the collision. All of the data for the current analysis were taken with the east FPD during RHIC run 6 (2006). At the time, the west side of the forward calorimetry was undergoing an upgrade, with a transitional detector called the FPD$^{++}$ replacing the west FPD. The data from the FPD$^{++}$ are not included in this analysis.\\

\begin{figure} [t]
\centering
\includegraphics[width=.5\textwidth]{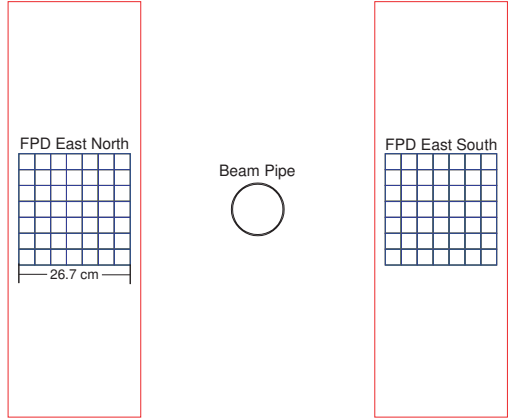}
\caption[Front view of the east Forward Pion Detector]{Front view of the east Foward Pion Detector (FPD). Each cell is a $3.81 \times 3.81 \times 45$ cm lead glass column. The distance from the beam line to the detector is about 30 cm. }
\label{face}
\end{figure}  

The FPD is a modular lead (Pb) glass calorimeter. A main module contains 49 Pb glass cells, forming a $7 \times 7$ square array. Figure \ref{face} shows a simplified schematic of the FPD in its run 6 configuration. The glasses were previously used in FermiLab E704, \cite{paper704d} and have been donated to the FPD project by IHEP Protvino. The Moliere radius of the glass is 3.32 cm, and the radiation length is around 2.5 cm. Each cell has a cross-sectional shape of a square, with the nominal side length of 3.81 cm. The length of the cell is 45 cm, making it about 18 radiation lengths long. The optical property of the cells are also available, which can be found in \cite{NETshower}, and chapter \ref{ch:shower}. Each main FPD module is about 26.7 cm wide, and there are two such modules placed on either sides of the beam pipe. As will be discussed later, the size and granularity of the FPD is best suited to detect $\pi^0$'s in the energy range from around 15 GeV to 50 GeV. During RHIC run 6, the distance between the beam line and the nearest edge of the FPD was set at about 30 cm. The modules are horizontally movable, and during other runs the distance to the beam line was varied to provide a wider pseudo-rapidity coverage. \\

\begin{figure} [t]
\centering
\includegraphics[width=.8\textwidth]{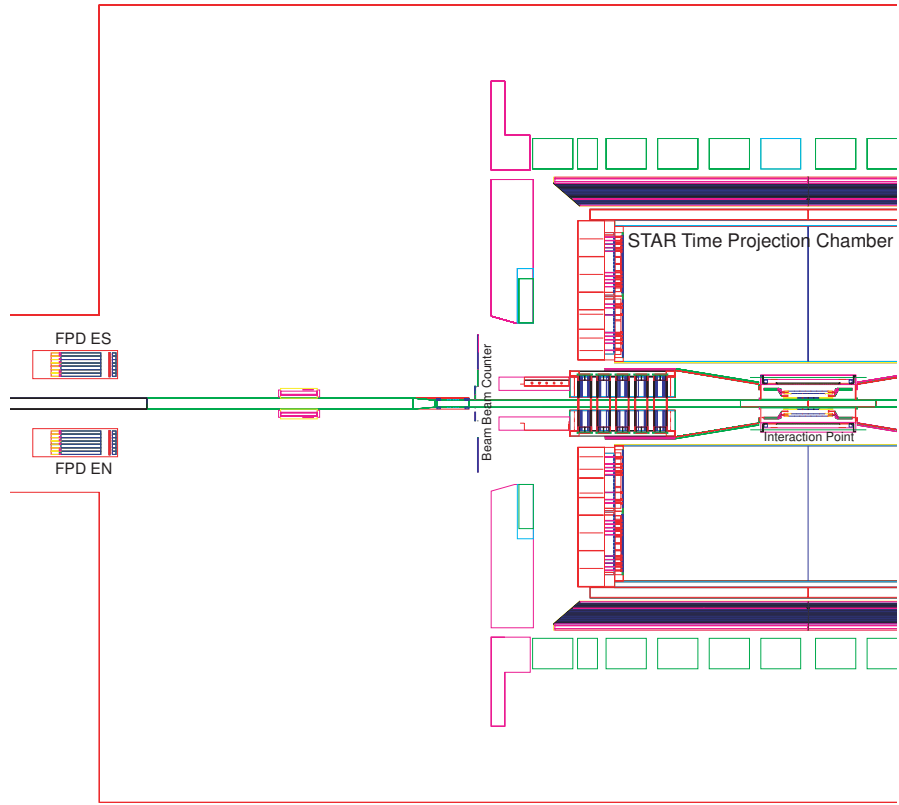}
\caption[Top view of the east half of the STAR wide angle hall]{Top view of the east half of the STAR wide angle hall. The distance between the FPD and the interaction point is about 8 m.}
\label{top}
\end{figure}  

Figure \ref{top} shows the top view of the east side of the STAR wide angle hall. The two FPD modules can be seen on the left corner, placed roughly 8 m away from the interaction point. The long blue rectangles inside the FPD enclosure represent the Pb glass cells. Also visible are blue squares at the front of the enclosure, which represent the pre-shower detector. It is essentially 7 FPD Pb glass cells placed vertically, providing an additional $\sim 1.5$ radiation length. Overall, there are about 2 radiation lengths worth of material in front of the main Pb glass array. (In other runs, a 1.27 cm thick Pb plate was often inserted in front of the pre-shower, adding 2 more radiation lengths. This was absent during run 6.) \cite{NETshower} \\  

A 12 stage, Russian design photo-multiplier tube (FEU-84) was attached to the end of each glass column to collect the light from the electromagnetic shower. The nominal phototube gain for the FPD main modules was set at 0.2 GeV per count, based on on-line analysis of $\pi^0$ events. Further corrections to the gain are done off-line by more detailed analysis utilizing full simulation, as described in chapter \ref{ch:calib}. All of the events for our analysis come from the FPD sum trigger, for which the sum of ADC for any single module was required to be greater than or equal to 150 counts, nominally equivalent to 30 GeV. \\   

The FPD has been taking data continuously since RHIC run 3 (2003). Since then, the STAR forward calorimetry has received significant upgrades, leading to the commissioning of the Forward Meson Spectrometer (FMS) in 2008, a substantially larger detector that replaced the west FPD \cite{fms}. Nevertheless, the east FPD remains an important part of the forward physics effort, complementing the FMS with a finer spatial resolution enabled by the shower maximum detector placed in front of the FPD \cite{yiqun}. \\

\end{doublespace}
\chapter{Electromagnetic Shower in FPD}\label{ch:shower}
\setlength{\parindent}{1cm}
\begin{doublespace}

\section{Introduction}
The first step in our analysis is to understand how an electromagnetic shower develops in Pb-glass cells of the FPD. There is a previous test beam measurement \cite{NETshower} on this subject, where electron beams with energy between 3 GeV and 23 GeV were used to illuminate a calorimeter that consisted of the same Pb-glass and photomultiplier tubes as the FPD. However, there are many differences between the STAR environment and the test beam set up, and it is necessary to verify both in simulation and data what the shape of the shower actually is in the FPD. The purpose of this chapter is to compare and study three objects: the shower function used in the reconstruction, the shower shape in the simulation, and the shower shape in the data.  \\

\section{FPD Shower Function}
We present the transverse profile of the shower in the following way (For now, we integrate over the longitudinal direction, z). For every event, we record the fraction of the cluster energy deposited in each cell. For each cell, we calculate the X and Y coordinates of the cell center relative to the true position of the photon.\\ 
\begin{equation}\label{eq:31}
\begin{split}
x &= x_{photon} - x_{cell}\\
y &= y_{photon} - y_{cell}
\end{split}
\end{equation}
The resulting distribution tells us the expected amount of energy deposited in a given cell, as a two dimensional function of the relative coordinates between the cell center and the photon location. We can think of this function as a definite integral of the ``true" shower shape over the transverse dimension of a cell.\\

To model this ``apparent" shower shape (which contains the information about the geometry), we use the following functional form.
\begin{equation}\label{eq:Gxy}
\begin{split}
G(x,y) = &F(x+\frac{d}{2},y+\frac{d}{2})-F(x-\frac{d}{2},y+\frac{d}{2}) \\
&-F(x+\frac{d}{2},y-\frac{d}{2})+F(x-\frac{d}{2},y-\frac{d}{2})
\end{split}
\end{equation}
\begin{equation}\label{eq:ShowerF}
F(x,y) = \frac{1}{2\pi} \sum_{i}^{N} a_i \cdot \lbrace arctan(\frac{x}{b_i}) + arctan(\frac{y}{b_i}) + arctan (\frac{xy}{b_i \sqrt{b_i^2+x^2+y^2}}) \rbrace 
\end{equation}
Here, $d$ is the width of an FPD cell, which is equal to 3.81 cm. $F(x,y)$ is the indefinite integral of the ``true" shower shape, and $G(x,y)$ the definite integral that folds in the detector geometry. We also note that the first two $arctan$ terms in equation \ref{eq:ShowerF} cancel when substituted into $G$, and only the third term that contains both $x$ and $y$ survives. Often, we add multiple copies of this function to describe the shower shape, in which case $N$ is the number of copies. $a_i$ and $b_i$ are free parameters for the height and the width of each copy, respectively. When using multiple copies, we can impose a normalization condition that the sum of $a_i$'s should be equal to one. 
The ``true" shower shape is given by,
\begin{equation}
f_i(x,y) = \frac{1}{2\pi} \cdot \lbrace \frac{1}{ 1 + (\frac{x}{b_i})^2 + (\frac{y}{b_i})^2 } \rbrace ^{\frac{3}{2}}
\end{equation}

So we have a function that falls as $\frac{1}{1+r^2}^{\frac{3}{2}}$. The choice of this functional form, as far as we know, is purely empirical. It is introduced in the test beam measurement \cite{NETshower}, although in that paper only one copy was used to describe the shower. For the FPD analyses, we have traditionally used three copies of the shower function with the ``default" parameters shown in table \ref{Tdefault}. A much more detailed discussion on the FPD reconstruction algorithm in general can be found in Yiqun Wang's PhD thesis \cite{yiqun}.\\

\begin{table}
\centering
\begin{tabular}{ | c || c | c | c |}
    \hline
    i & 1 & 2 & 3 \\ \hline  
    \hline
    $a_i$ & 0.8 & 0.3 & -0.1 \\ \hline 
    $b_i$ & 0.8 & 0.2 & 7.6 \\ 
    \hline
\end{tabular}
\caption[Default parameters for the shower function]{Default parameters for the shower function. $b_i$'s are in units of cm.}
\label{Tdefault}
\end{table}

\section{Photon Shower Shape in Geant}

As the FPD has been in use since 2003, it has been incorporated into the STAR specific version of the Geant3 package, called GSTAR. (\textbf{GE}ometry \textbf{AN}d \textbf{T}racking is a detector response simulator developed by CERN. The newer, C$^{++}$ based Geant4, which addresses at least one issue of Geant3 discussed later, has not been implemented within the STAR software framework.) The natural place to start then is to look at the photon showers in GSTAR as it was configured at the beginning of the current analysis, which was the same set up used by all previous FPD analyses.\\

In the ``default" setting, the shower is based on charged particle energy loss. In this scheme, Geant tracks the showering process in which pair production and Bremsstrahlung take place successively, down to a set threshold energy. Below that energy, the energy is considered to be deposited in that Pb-glass cell. There is no consideration for the propagation of optical photons, such as the transparency of the glass or the photo-cathode quantum efficiency. The energy threshold was set to be above the typical cutoff for pair production ($\sim 2$ MeV), and well above the optical photon range. ($\sim 1$ eV) If the cell is long enough for the shower to fully develop, nearly 100 \% of the original photon energy would register as ``measured". As the photon energy increases, the shower develops further into the detector, opening up the possibility for energy loss due to ``punch throughs". For 200 GeV collisions, the 18 radiation lengths of the FPD cells are sufficient to contain most of the shower.  \\

While the shower shape is a two dimensional function, it only takes a one dimensional representation at a fixed x or y value to determine the free parameters. (We have found that using two to three copies of the function is enough to reproduce the full two dimensional structure in all cases we've looked at.) To do this, we take those events that lie along the vertical line that goes through the center of a given cell, $x \approx 0$. We choose the vertical line to minimize the effect of incident angle, as the FPD was placed on the horizontal plane that includes the beam line. From these events, we create a two dimensional distribution that correlates the fraction of the photon energy deposited in a cell with $y$ in equation \ref{eq:31}. The function $G(0,y)$, given by equation \ref{eq:Gxy}, is then fit against this distribution to constrain the free parameters. \\

\begin{figure} [t]
\centering
\includegraphics[width=.8\textwidth]{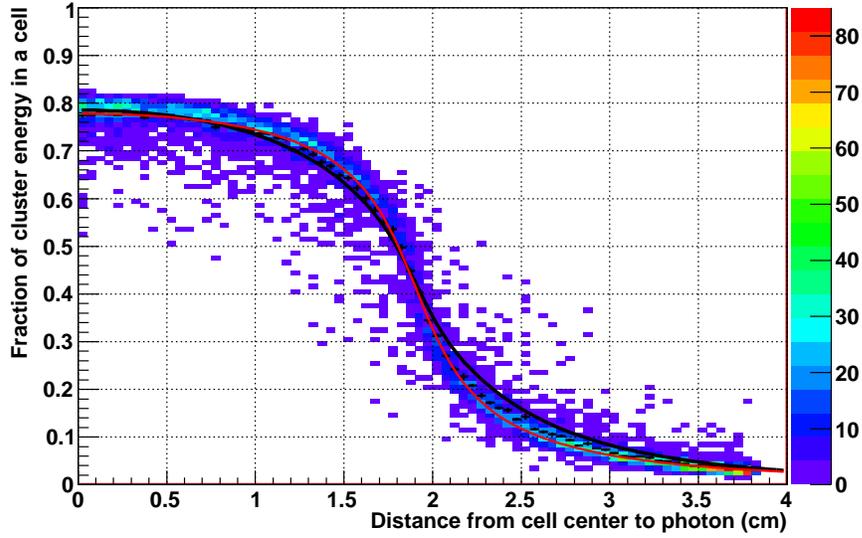}
\caption[Shower shape from Geant based on energy loss]{Geant photon only simulation: Fraction of the cluster energy deposited in a cell vs. the distance from the center of the cell to the photon location. BLACK: Shower function using default values in table \ref{Tdefault}. RED: Best fit to the distribution}
\label{hEYsim}
\end{figure}

Figure \ref{hEYsim} shows the shower shape distribution based on a 30 GeV single photon only Geant simulation. Also shown are the black curve for the ``default" shower shape described in table \ref{Tdefault}, and the red curve for the fit against this distribution. We see that the agreement is reasonable, with both functions getting the high tower fraction at $y=0$ correctly. The most significant deviation occurs between $y=2 \sim 3$ cm, which would affect the energies in the nearest neighbor cells of the high tower.\\
\begin{figure} [t]
\centering
\includegraphics[width=1.\textwidth]{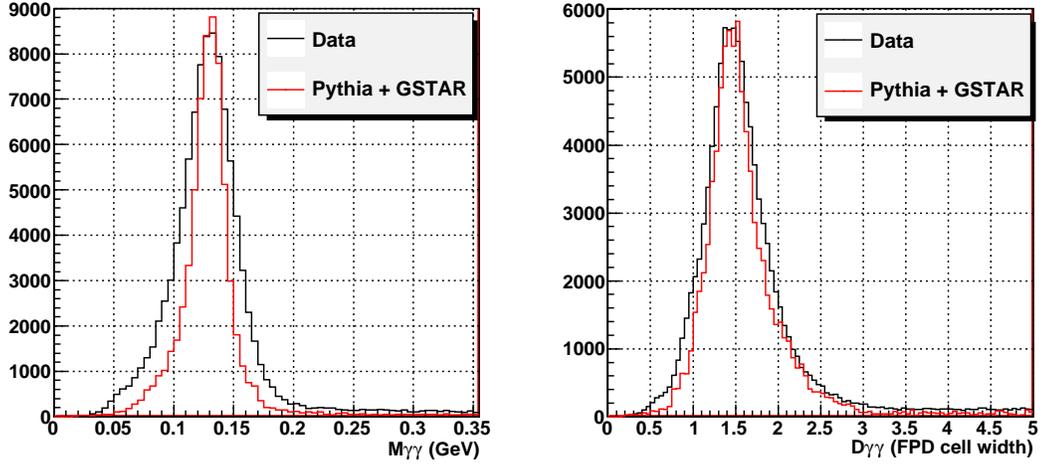}
\caption[Mass and separation, data - MC (E-loss) comparison]{LEFT: Di-photon invariant mass distribution. RIGHT: Di-photon separation distribution. Both are for events with 40 GeV $< E_{\gamma\gamma} <$ 45 GeV. BLACK: Data, RED: Pythia + GSTAR simulation, energy loss based shower}
\label{Mdist}
\end{figure} 

Given that the default shower shape largely agrees with the simulation, the question then is whether this is also the right shape for the data. Unfortunately, there are indirect evidences that suggest that the shower shape is not being simulated correctly. Here we briefly go over two examples. \\

The first example is the width of the $\pi^0$ mass peak. Figure \ref{Mdist} shows the comparisons of the di-photon invariant mass and separation distributions between the data and full simulation. As illustrated, the mass peak in the data is consistently found to be $60\sim70$ \% wider than what is seen in full Pythia + Geant simulation. (Pythia is a high energy physics event simulator. \cite{pythia}) Such a large difference in the mass resolution is very difficult to explain away by reasons other than shower simulation. The uncertainty in the interaction vertex can often broaden the mass peak, but the magnitude of the broadening makes it an unlikely cause. Explaining the 15$\sim$20 \% width seen in the data requires uncertainty of similar magnitude in the distance between the interaction vertex and the detector, which is nominally around 8 meters. This means the error in vertex position needs to be over 1 meter, which is extremely unlikely. (The nominal estimate for the uncertainty of STAR BBC based vertex is around 30 cm, and the apparent width of the vertex distribution is no wider than 60 cm.) The cell by cell calibration non-uniformity can also cause mass peak broadening, but the magnitude of random error in calibration would have to be on the order of 20 \%, which would manifest itself as an order of magnitude non-uniformity in counting rate. In addition, the width of the mass peak does not change significantly even for the $\pi^0$'s that are well confined in a particular pair of cells. On the other hand, the separation distributions are much better matched between simulation and data, suggesting that the culprit may be the resolution in total energy or error in energy sharing. \\

\begin{figure} [b]
\centering
\includegraphics[width=.8\textwidth]{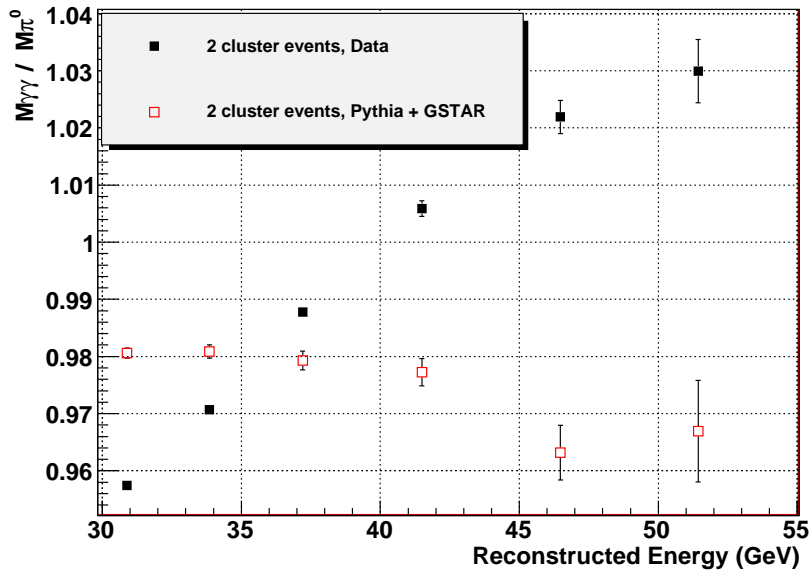}
\caption[$M_{\gamma\gamma}$ vs. energy, data - MC (E-loss) comparison ]{Di-photon invariant mass centroid / $M_{\pi^0}$ vs. reconstructed energy for two cluster events. BLACK: Data. RED: Pythia + GSTAR simulation based on energy loss}
\label{MvsE}
\end{figure} 

The second example is illustrated in figure \ref{MvsE}. In the data, there is a clear energy dependence of the location of the $\pi^0$ mass peak, where the mass increases at a rate of a few percent per 10 GeV. This is in fact a phenomenon that we have always seen in the previous FPD analyses, and it was dealt with under the assumption that the cause was a shift in gain as a function of energy. Through further simulation study, however, it was later found that the small energy dependence seen in the simulation was caused not by a gain shift, but by an energy dependent shift in two photon separation measurement caused by the discrepancy between the shower shape and the shower function. But whatever the case may be, if the shower shape is the same in the data and Geant, we would expect the effect to be well simulated given the choice of a shower function. This is clearly not what we see in figure \ref{MvsE}, where the simulation shows very little energy dependence. Both data and simulation were reconstructed using the same shower function, which is the red curve in figure \ref{hEYsim}. If we use the default shower function, we find that the simulation develops a mild energy dependence of mass (due to energy dependence of separation), but the energy dependence in data gets even more severe. \\

These and other evidences made it necessary to directly measure the shower shape in the data. In the following section, we will discuss how this was done, and how the result compares to the shower shape in Geant. 

\section{Measuring the Shower Shape in Data}
In order to determine the photon shower shape that appears in the data, the following three requirements need to be met. First, we need a sample of isolated photons that have sufficient energy to provide meaningful information out to the tail of the shower, given the limitation of $\sim 200$ MeV per count ADC granularity. Second, we need to know the true positions of these photons relative to the detector geometry, in order to produce the distribution of the deposited energy as a function of the distance from the photon center. And finally, the gain calibration in at least some part of the detector has to be uniform from cell to cell to within 2 \%, so that we can constrain the shower shape with a reasonable accuracy and avoid event selection bias.\\  

The photon sample was obtained by imposing a narrow mass cut to the identified $\pi^0$ events in the data with more than 30 GeV of energy and greater than 2.5 cell separation between the two photons. Figure \ref{hmd} shows the di-photon invariant mass and the separation distributions for the selected event sample. As can be seen from the left panel, we have a very clean $\pi^0$ mass peak even with less than ideal calibration and shower function. On the right, we see that a good fraction of the events have greater than 2.5 cell separation at this energy. \\

\begin{figure} [t]
\centering
\includegraphics[width=1.\textwidth]{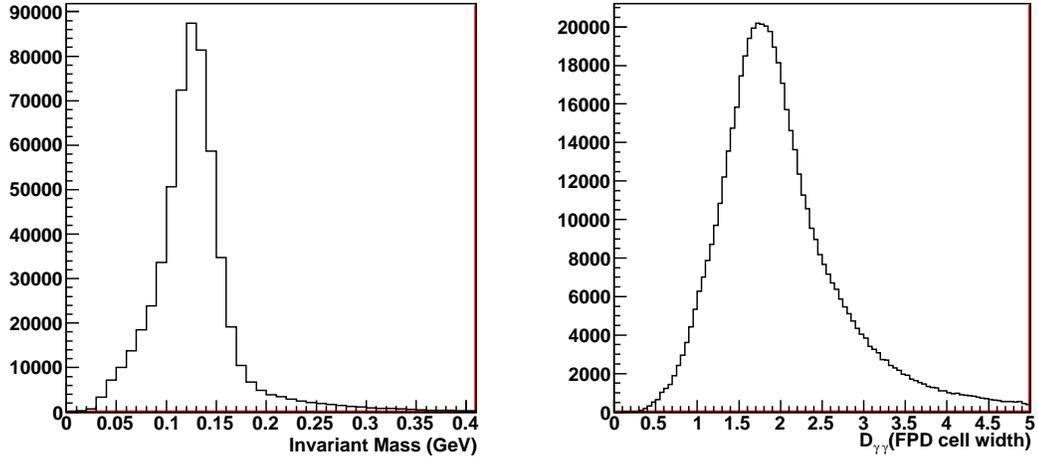}
\caption[Mass and separation for shower shape analysis sample]{For the events selected for the shower shape analysis, LEFT: Di-photon invariant mass ($M_{\gamma\gamma}$) distribution, RIGHT: Di-photon separation in units of FPD cell width ($D_{\gamma\gamma}$).}
\label{hmd}
\end{figure}

The selected events are highly asymmetric $\pi^0$'s ($Z_{\gamma\gamma} \approx $ 0.8) with the average energy for one of the photons reaching around 35 GeV, which is nominally equal to 175 ADC counts. Figure \ref{hze} shows the energy sharing and the larger of the two photon energy distributions for these events. \\

\begin{figure} [t]
\centering
\includegraphics[width=1.\textwidth]{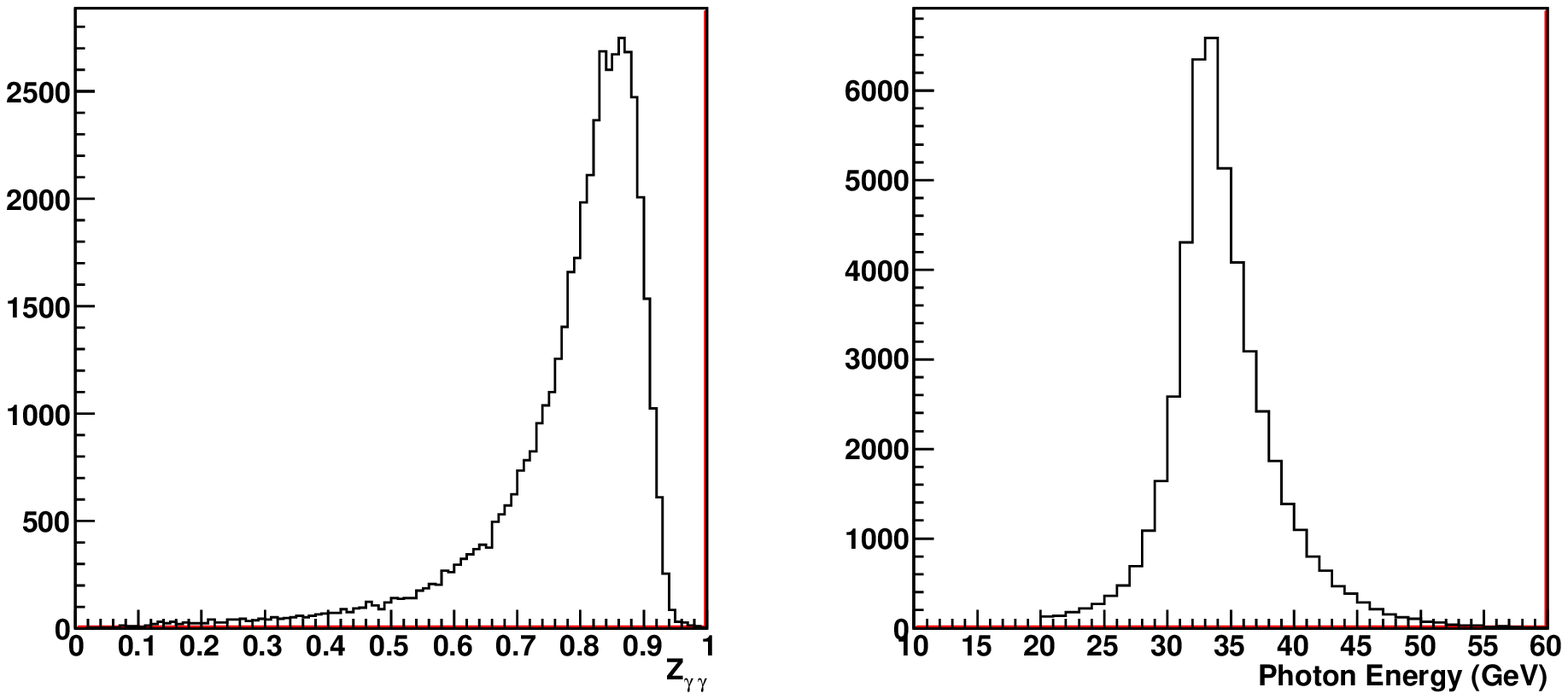}
\caption[Energy sharing and dominant photon energy for shower shape analysis sample]{For the events selected for the shower shape analysis, LEFT: Di-photon energy sharing($Z_{\gamma\gamma}$) distribution, RIGHT: The larger of the two photon energy distribution.}
\label{hze}
\end{figure}

The remaining two requirements are more difficult to satisfy. Both our knowledge of the true photon position and the calibration uniformity requires an accurate knowledge of the shower shape, the very quantity we are trying to measure. The reasons for this are the following.\\

\begin{figure} [t]
\centering
\includegraphics[width=.8\textwidth]{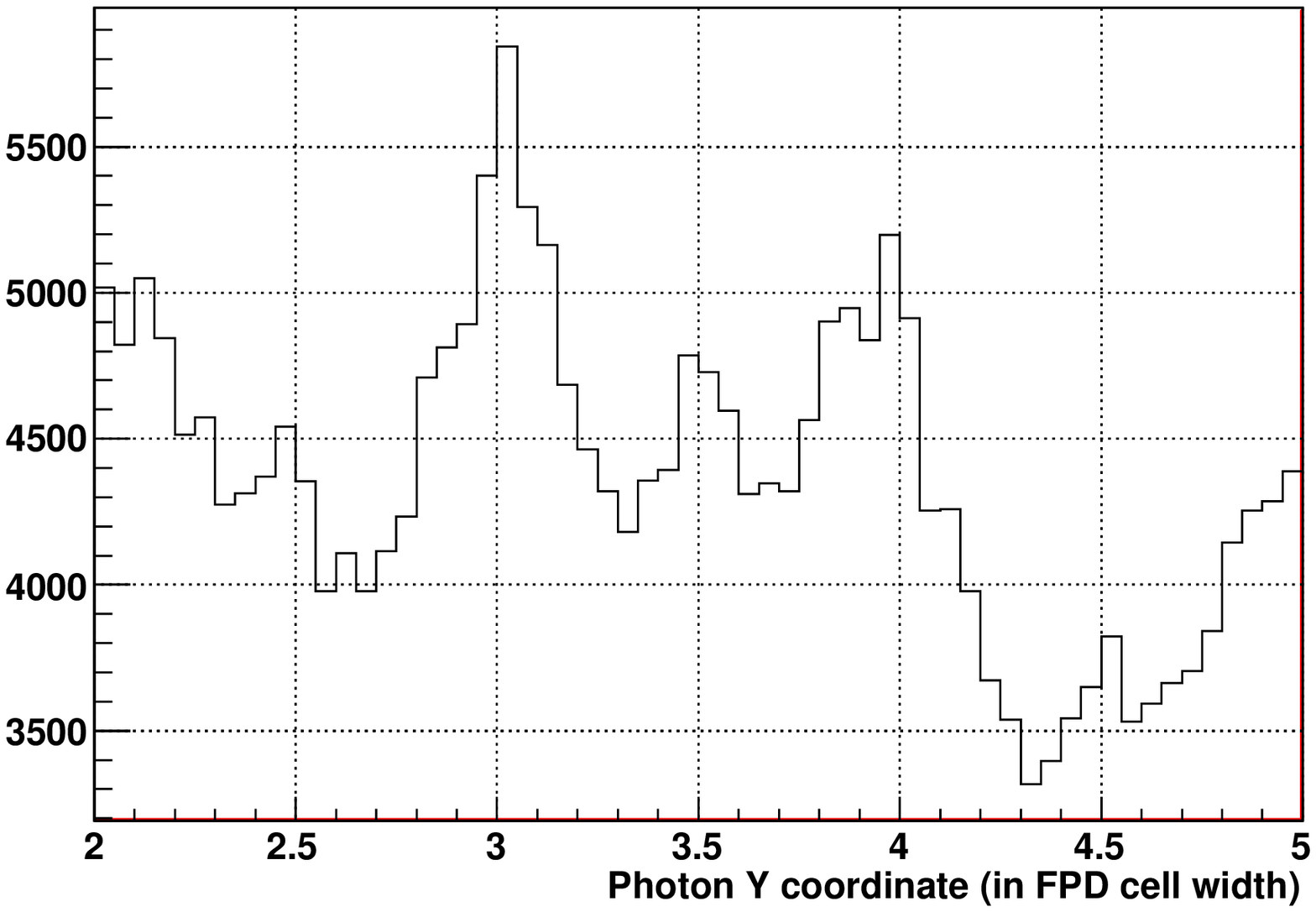}
\caption{Reconstructed photon Y-coordinate distribution in data, for the second column.}
\label{hy}
\end{figure}

First, the discrepancy between the shower function used in the reconstruction and the actual shower shape can systematically shift reconstructed photon positions. We have generally found that the distribution of the photon position within a cell is highly non-uniform, a convincing sign that the position accuracy was compromised. Figure \ref{hy} shows the distribution of photon Y-coordinates in the data, where we see a strong tendency for the reconstruction to pile up photons at the cell boundaries. This suggests that the slope of this version of the shower function near the boundary was too steep, forming a center of attraction. While it is easy to change this pattern by tweaking the parameters of the shower function, (for instance to make the photons accumulate at the cell center instead) we found that making it completely uniform through ad-hoc changes of the parameters was extremely difficult. In other words, while the intra-cell uniformity of the photon position distribution is a good diagnostic for the match between the shower function and the actual shower shape, it has too many degrees of freedom to be a useful handle to constrain the shower function. \\

Second, the calibration has a strong dependence on the shower function. The primary method of determining cell by cell calibration in the FPD, as is common with most electromagnetic calorimeters, is $\pi^0$ mass analysis. The difficulty with this method is that the invariant mass is a function of multiple variables, such as the two photon separation and energy sharing. In order to use the mass to calibrate the energy, all other variables need to be well understood. More details on the $\pi^0$ based calibration can be found in chapter \ref{ch:calib}, especially equation \ref{eq:mgg}. For the current discussion, however, it suffices to note that the two photon separation plays a crucial role in calibrating the FPD. \\

For a typical 35 GeV $\pi^0$ used for calibration, the average two photon separation is around 1.5 cell width in the FPD. Since the photon shower has meaningful contribution well beyond 0.5 cell width from the photon center, the two showers often overlap, making the separation measurement very sensitive to the detailed matching between the shower function and the real shower shape. (Calibrating at significantly lower energy was not an option as the hardware trigger threshold was set nominally to 30 GeV)\\

It is possible to solve both of these problems if we know, for some part of the detector, the true relative photon counting rates above some energy. For this, we confine our attention to the two groups of three cells in which the counting rate is expected to be nearly uniform within a group. They are from the central three rows, on the second and the third column. (The first column was dropped to avoid any edge issues.) Because of the proximity of these cells to the beam line, the average pseudo-rapidities for the three vertically neighboring cells do not vary much. Based on the full Pythia + Geant simulation, we indeed find the expected counting rate to be nearly flat within a group of three. While the pseudo-rapidity dependence of the cross-section in Pythia may not exactly match that in nature, the nearly constant average pseudo-rapidities for the three cells in a group ensure that we are not very sensitive to the possible discrepancy between the simulation and data. Figure \ref{count} shows the average pseudo-rapidities for one such group of three cells, along with the counting rate comparison between data and simulation. \\

\begin{figure} [t]
\centering
\includegraphics[width=.8\textwidth]{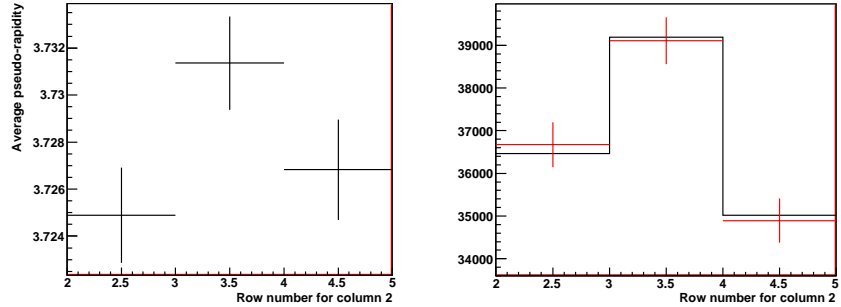}
\caption[Pseudo-rapidities and counting rates for cells included in shower shape analysis]{LEFT: Average pseudo-rapidity of three vertically neighboring cells. The central row has a small vertical offset relative to the beam line. RIGHT: Counting rate comparison between data (black) and Pythia + Geant (red) for the same three cells.}
\label{count}
\end{figure}

In reality, because of the steeply falling cross-section as a function of energy in the forward region, the counting rate is very sensitive to the calibration. (1 \% change in gain roughly corresponds to 10 \% change in counting rate.) This in turn can cause a bias in our photon sample that would make the shower look more peaked in the center than it really is, by accepting most of the events from hot cells. On the flip side, given that we know the true relative counting rate, this high sensitivity can be used to achieve the desired uniformity in relative calibration. Based on the data-MC comparison shown in the right-hand panel of figure \ref{count}, we expect the relative calibration for a group of three cells to be within well under 1 \%. \\

\begin{figure} [t]
\centering
\includegraphics[width=.8\textwidth]{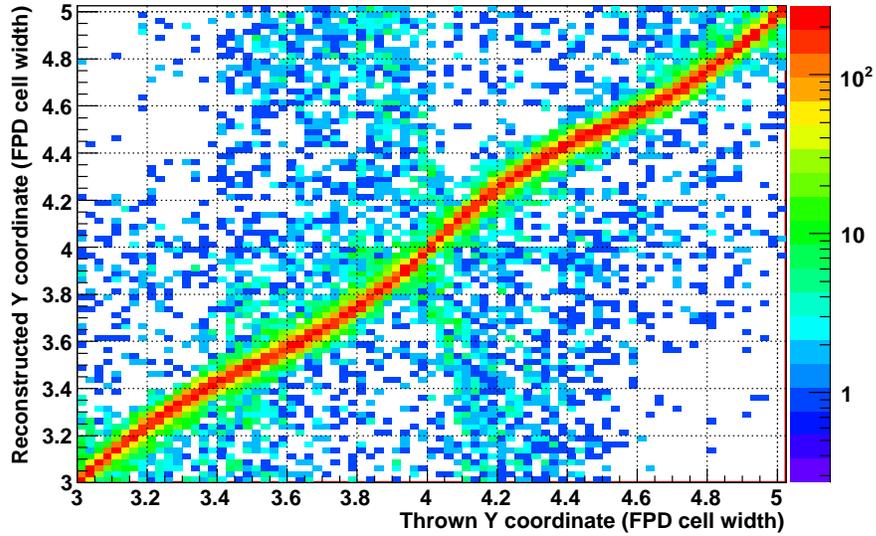}
\caption[Y-coordinate smearing, photon only Geant]{Reconstructed Y-coordinate vs. thrown Y-coordinate of photons in photon-only Geant simulation. The Y-coordinate is the vertical axis in the STAR coordinate system.}
\label{hyy}
\end{figure}

Based on our knowledge of the true relative counting rate, we can also find the functional form of the relationship between the true position and the reconstructed position of a photon. Furthermore, this is possible regardless of the details of the pathology introduced by the incorrect shower function. Figure \ref{hyy} is an example of the correlation between the thrown and the reconstructed Y-coordinates in Geant photon only simulation. (While we may not fully trust the simulation on the exact shower shape, we can still rely on it to gain qualitative understanding of the pathology caused by the incorrect shower function.) As it was generated with a version of the shower function that did not match the Geant shower well, we see a periodic modulation that indicates systematic miscalculations of photon positions. However, we also note that the structure is mostly confined in each cell, meaning that the problem is the redistribution of the photon coordinates within a cell, not migration among cells. Assuming that the relative calibration is in good shape, the highest tower in a cluster is always the one that contains the photon. The reconstruction, regardless of the details of the shower shape, does utilize this fact well. Therefore as long as we focus on the vertical direction, (in STAR coordinates the Y-direction) where there is very little incident angle to smear the shower and cause a spill over, we can safely assume that the photon thrown within the horizontal boundaries of a cell will reconstruct within those boundaries.  \\

Under these assumptions, we obtain the relationship between the true and reconstructed position within a cell as follows: \\
\begin{equation}
\begin{split}
N &= \text{Total number of photons above some energy in a chosen cell}\\
x &= \text{True photon coordinate}\\
x^{\prime} &= \text{Reconstructed photon coordinate}\\
f(x) &= \text{True counting rate as a function of true coordinate}\\
g(x^{\prime}) &= \text{Observed counting rate as a function of reconstructed coordinate}\\
\end{split}
\end{equation}\\
We define the coordinates within a cell in units of cell size, so that both $x$ and $x^{\prime}$ run from 0 to 1. Then we have,
\begin{align}
\int\limits_{0}^{1} f(x) dx = \int\limits_{0}^{1} g(x^{\prime}) dx^{\prime}& = N\\
\int\limits_{0}^{x_0} f(x) dx = \int\limits_{0}^{x_0^{\prime}} g(x^{\prime}) dx^{\prime}& = \text{some fraction of N}
\end{align}
Here, $x_0^{\prime}$ is an arbitrary value of reconstructed coordinate we chose. Since $f(x)$, $g(x^{\prime})$, and $x_0^{\prime}$ are all known, it is straightforward to calculate $x_0$, which is the true photon coordinate corresponding to $x_0^{\prime}$. In our case, the relationship is even more simplified as we picked a region where we expect $f(x)$ to be flat due to almost constant pseudo-rapidity within a cell. Then the true coordinate as a function of the reconstructed coordinate is given simply by
\begin{equation}
x_0 = \frac{1}{N} \int\limits_{0}^{x_0^{\prime}} g(x^{\prime}) dx^{\prime}
\end{equation}
For more details on this method, refer to \cite{NETshower}. \\

\begin{figure} [t]
\centering
\includegraphics[width=.8\textwidth]{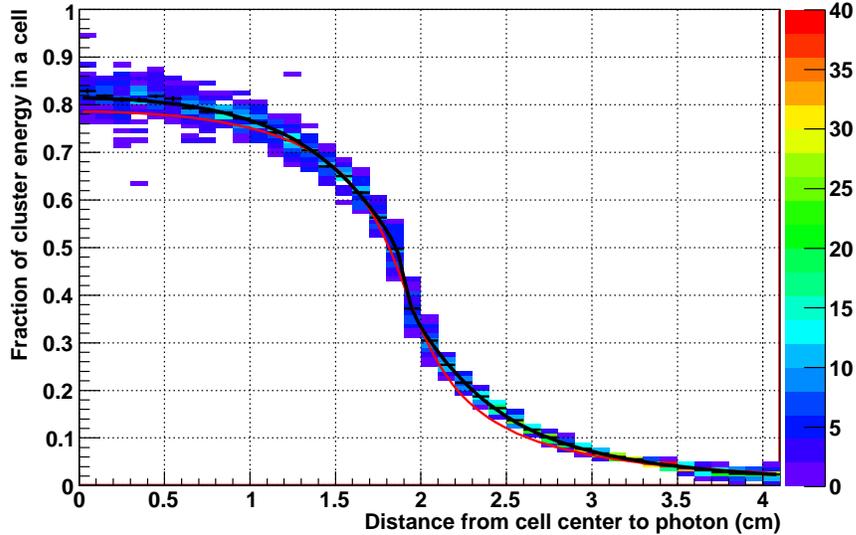}
\caption[Shower shape from data]{Data: Fraction of the cluster energy deposited in a cell vs. the distance from the center of the cell to the photon location. BLACK: Best fit to the distribution. RED: Fit to the Geant energy loss based shower shape (figure \ref{hEYsim})}
\label{hEYc}
\end{figure}

Now that we have all the pieces together, we can look at the photon shower shape in the data. Figure \ref{hEYc} shows the shower shape distribution analogous to the one shown in figure \ref{hEYsim} in the previous section. Overlaid in red is the best fit to the shower shape in Geant, which is again the same red curve as in figure \ref{hEYsim}. As we suspected in the previous section, the showers in the data and simulation do not agree perfectly. We see that in the data, a photon that hit very close to the cell center leaves about 82 \% of the energy in that cell, whereas in Geant it only leaves about 78 \%. The fact that the shower is ``narrower" (higher fraction at the origin) in the data is especially strange. It is easy to imagine effects that can broaden the shower in the simulation, for instance by adding hadronic background, but there is essentially nothing we can tune in the simulation to make it narrower. \\

It can be said that the difference we see in figure \ref{hEYc} is relatively small, and that such a difference may not be significant in some analyses. However, in the current analysis we are attempting to measure the cross-section and asymmetry at significantly higher energy than what was done previously. Many systematic errors that were insignificant at 40 GeV may affect us much more seriously at 60 GeV, one such example being the energy dependent gain shift. (Details on this subject can be found in section \ref{sec:massshift}.) Because the calibration has to be done at low energy due to statistics, the further we move up from that energy, the more important it is to treat this energy dependent shift correctly. Furthermore, even at low energy, a few percent difference in shower shape can become the source of a significant error in calibration by affecting both the energy and separation measurements. Such error in calibration again becomes amplified at higher energy, becoming a source of a dominant systematic error. \\

\section{\v{C}erenkov Photon Based Shower}\label{sec:chkv}

Given that Pb-glass is an optically transparent medium that isn't a very good scintillator, \v{C}erenkov photons are likely the main source of the optical photons seen by the phototube. It is then perhaps not surprising that the shower simulation that does not take into account any optical physics comes short in some areas. The reason that the full \v{C}erenkov simulation was originally omitted was simple economics, as simulating optical photons properly can take orders of magnitude more CPU time. \\

With the \v{C}erenkov effect based shower, Geant simulates the generation and propagation of the optical photons due to the \v{C}erenkov effect. The photons are generated within some energy range, (approximately between 1.8 eV and 3.8 eV) with abrupt cut off on both ends of the energy scale. Generally, the frequency dependence of the quantum efficiency effectively makes the cutoff smooth. The user has to provide the optical properties of the medium, such as index of refraction (1.67 for Pb-glass) and attenuation length for dielectric, and surface reflection coefficients for metal. The refraction at the dielectric-dielectric interface is handled by Snell's law. If the photon survives the propagation through Pb-glass and quantum efficiency of the photo-cathode, it is recorded as measured. The number of photons measured per unit energy varies widely depending on the overall strength of the attenuation, and the distance between the shower maximum and the photo-cathode.\\

\subsection{Shower Width}

First we look at the question of narrowness, which is the most direct evidence we have of the shower shape discrepancy between simulation and data. Figure \ref{hEYnotrk} shows the \v{C}erenkov based shower shape when we accept all generated optical photons, with 100 \% transparency of the glass and 100 \% reflection at the surface. Under these ideal assumptions, we see that the shower is indeed even narrower than what we see in the data with energy fraction reaching 85 \% at the origin. \\

\begin{figure} [t]
\centering
\includegraphics[width=.8\textwidth]{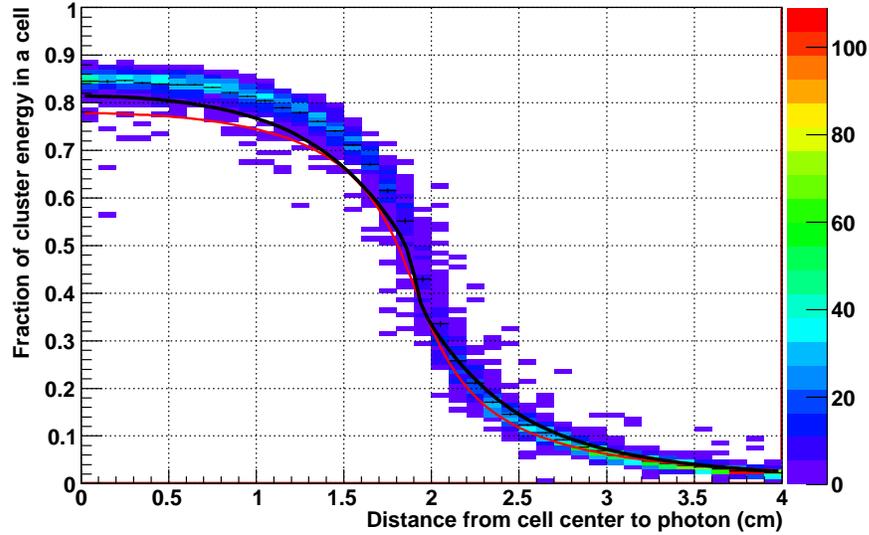}
\caption[Shower shape from \v{C}erenkov based Geant, no attenuation]{Shower shape based on \v{C}erenkov light collection, with no attenuation. BLACK: Best fit to the data shower shape. (figure \ref{hEYc}) RED: Fit to the Geant energy loss based shower shape (figure \ref{hEYsim})}
\label{hEYnotrk}
\end{figure}

From here, there are multiple parameters that can be tuned to make the profile wider, such as the absorption length in the Pb-glass, and the reflectivity of the Pb-glass and aluminized mylar interface. Generally, the effect of such tuning is that a higher level of attenuation (including reduced reflectivity) results in a wider shower profile. The reason is the following. First, the narrowness of the shower profile is largely determined by the ratio between the early and the late part of the shower. For the former, most of the photons are concentrated along the core. For the latter, the shower has already developed into a much wider shape (For more details on the longitudinal shower development, refer to the next section, especially figure \ref{4panel}). Second, the farther the photon has to travel to reach the photo-cathode, the more sensitive it is to factors that produce attenuation. Figure \ref{vz2} shows the detection probability of \v{C}erenkov photons as a function of the distance between the point of generation and the photo-cathode. We see that the effect of reduced attenuation is much more pronounced for the early part of the shower (large x-value in figure \ref{vz2}) than the later part (small x-value in figure \ref{vz2}).\\

\begin{figure} [t]
\centering
\includegraphics[width=.8\textwidth]{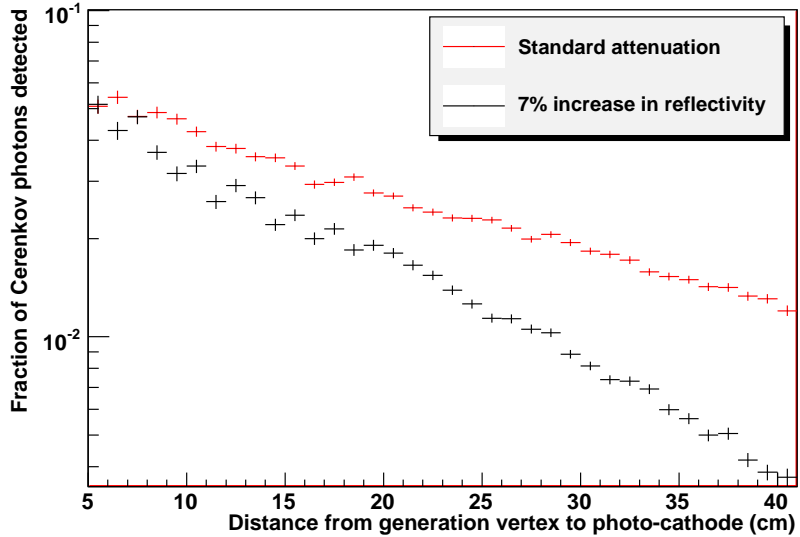}
\caption[\v{C}erenkov photon detection probability vs. distance from generation to photo-cathode]{\v{C}erenkov photon detection probability vs. distance from generation to photo-cathode. Standard values refer to the ones given in \cite{NETshower}. Surface reflectivity between glass and aluminum is a function of optical photon energy, and 7 \% reduction was applied across the energy range.}
\label{vz2}
\end{figure}

Unfortunately, our knowledge of the transparency of the Pb-glass and the reflectivity at the glass-aluminum interface is very limited. The optical properties of the FPD type Pb-glass and the aluminized mylar have been published \cite{NETshower}, but these values can only be used as starting points. The problem is that the attenuation length can vary due to radiation damage, and the surface reflectivity depends significantly on how much air gap exists between the glass and the wrapper. On the one hand, we have no way of knowing how much radiation damage was there during run 2006, (the same glasses have been in use since) and on the other hand, it is very difficult to measure the air gap in situ with micron level precision. (Glasses are stacked on top of each other, and their weight and relative flatness of the surface would all play into the thickness of the air gap). \\

\begin{figure} [b]
\centering
\includegraphics[width=1.\textwidth]{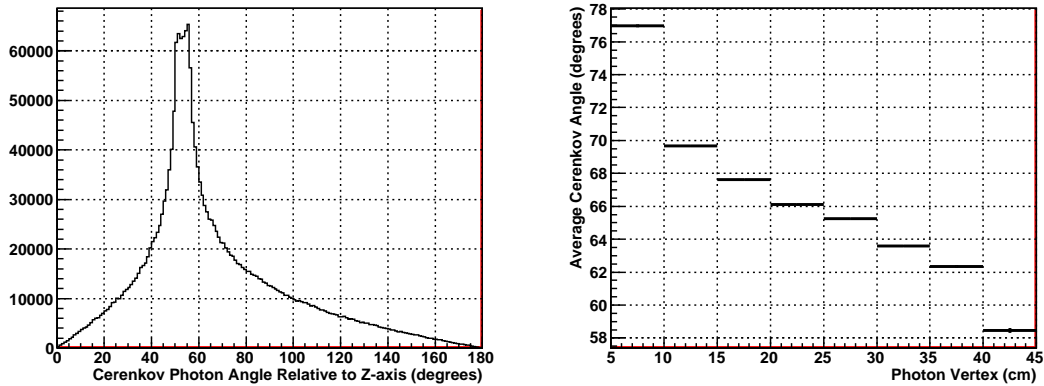}
\caption[\v{C}erenkov angle distribution, and average angle vs. distance from the cathode]{LEFT: Distribution of \v{C}erenkov angle relative to the longitudinal axis of the cell, normal incidence. RIGHT: Average \v{C}erenkov angle vs. distance between the photon generation point and the front face of the cell}
\label{ChAngle}
\end{figure} 

While we can perhaps make a reasonable guess on the transparency of the Pb-glass, as we don't expect the radiation damage to be severe enough to change it substantially from the previously measured values, the air gap poses a more difficult challenge. The existence of the air gap improves the reflection at the surface significantly, making its proper treatment important. The \v{C}erenkov angle is given by the following simple formula.
\begin{equation}
\theta_{Cherenkov} = cos^{-1}(\frac{1}{n\beta})
\end{equation}

For the Pb-glass with n=1.67, it is around 53 degrees relative the incident photon axis. Figure \ref{ChAngle} shows the distribution of \v{C}erenkov angle with respect to the detector Z-axis, and how its mean varies as the shower progresses into the detector. On the other hand, the total internal reflection angle for the Pb-glass and air interface is about 52 degrees relative to the surface. This means that if the shower initiating photon entered the glass column along its Z-axis, a large fraction of the \v{C}erenkov photons would be produced within the limit of total internal reflection. (If the photon trajectory was confined in a plane normal to the surface, the incident angle would be just above the limit of total internal reflection. Many more photons fall within the limit due to the extra dimension.) Somewhat counter-intuitively then, the air-glass interface can be more reflective than the aluminum-glass interface, which is expected to have around 90 \% reflectivity. \\

A few micron thick air gap should be sufficient to produce total internal reflection, as the wavelength of \v{C}erenkov photons is around 400 nm. It is very difficult to know the extent to which this level of air gap exists when the cells are stacked. Further complicating the matter, the FORTRAN based Geant3 has a floating point rounding error that affects very thin volumes. Consequently, we are not able to implement a few micron thick air gap properly, as the photons very often overshoot the air gap and ``reflect" off of the third volume beyond the aluminized mylar. \\

Given these limitations, we take the point of view that the primary factor that determines the physics is the total attenuation of the optical photons regardless of the source. This means that, for example, reduction in reflectivity can be offset by increase in transparency in the glass, and vice versa. With this assumption, the air gap was removed all together, and the thickness of the aluminum wrapper was widened from 10 to 100 microns to reduce the thin volume problem. In order to compensate for the loss of total internal reflection, the standard \cite{NETshower} surface reflectivity of aluminum was increased by 7.5 \% across the energy range. The unaltered standard \cite{NETshower} values are used for the attenuation length of the Pb-glass. Figure \ref{absref} shows the absorption length and the reflection coefficients used for the current analysis.\\

\begin{figure} [t]
\centering
\includegraphics[width=1.\textwidth]{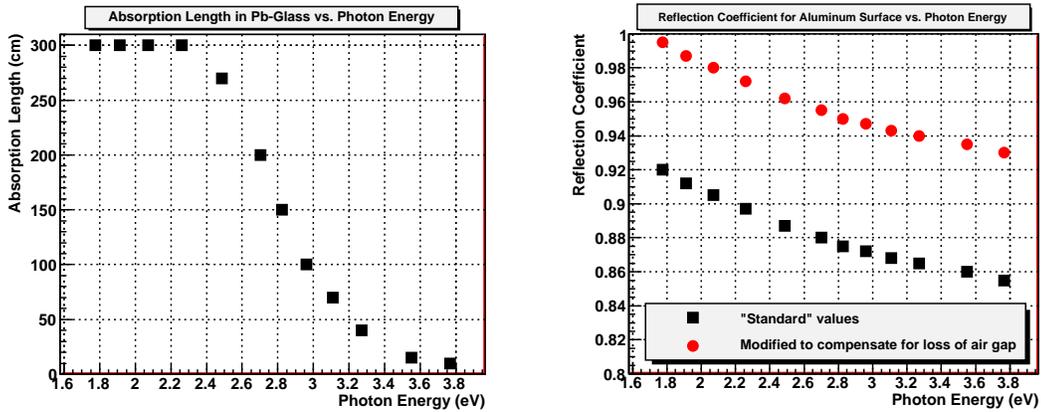}
\caption[Pb-glass absorption length and aluminized mylar reflection coefficient]{LEFT: Absorption length in the Pb-glass (cm) vs. photon energy (eV). RIGHT: Reflection coefficient for the aluminum surface vs. photon energy (eV). BLACK: ``Standard" values \cite{NETshower}. RED: Modified values to compensate for the loss of air gap.}
\label{absref}
\end{figure} 

There are obvious holes in this argument, such as the fact that the angle dependence is very different between total internal reflection and reflection off aluminum. This may have an effect in the shower shape, since the \v{C}erenkov angle changes with the progression of the shower as shown in the right-hand panel of figure \ref{ChAngle}. Nevertheless, we have found the above described setting to be acceptable for our purposes. Figure \ref{hEYchkv} shows the resulting shower shape, overlaid with the fit to data and previous energy loss based simulation. We find that the energy fraction at the origin now agrees very well, which was impossible to achieve with energy loss simulation. The remaining discrepancies are likely due to the less than satisfactory treatment of the surface, and the resulting ad-hoc correction in the Pb-glass attenuation length. Overall, while the agreement is far from perfect, it does improve on the issue of narrowness.   \\

\begin{figure} [t]
\centering
\includegraphics[width=.8\textwidth]{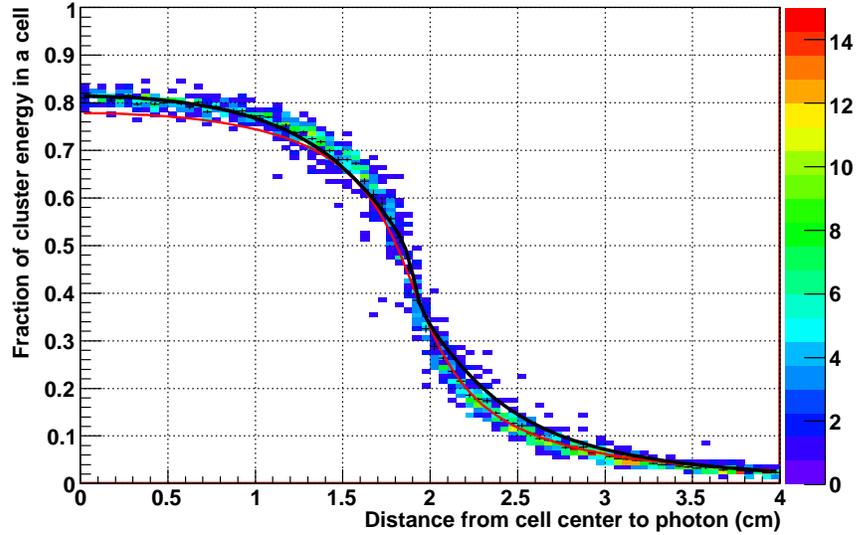}
\caption[Shower shape from \v{C}erenkov based Geant, with realistic attenuation]{Shower shape based on \v{C}erenkov light collection, with no air gap and 7.5 \% increase in surface reflectivity. BLACK: Best fit to the data shower shape. (figure \ref{hEYc}) RED: Fit to the Geant energy loss based shower shape (figure \ref{hEYsim})}
\label{hEYchkv}
\end{figure}

\subsection{Width of the $\pi^0$ Mass Peak}

Secondly, we look at the issue of the mass resolution, which was illustrated in figure \ref{Mdist}. With the energy loss based simulation, it was found that the width of the $\pi^0$ mass peak was much too narrow compared to the data. It was noted that only a small portion of that difference was attributable to the simulation of two photon separation. \\

With the \v{C}erenkov based shower simulation, we find that the mass resolution is much better simulated, as shown in the left-hand panel of figure \ref{Mdist2}. The simulation is still slightly narrower than the data, but improvement is dramatic. From the right-hand panel of the figure \ref{Mdist2}, we see that the separation distribution has also broadened, now matching the width in data almost exactly. However, comparing figure \ref{Mdist} and figure \ref{Mdist2}, it is clear that the additional smearing of the separation distribution is much smaller than what is seen in the mass resolution.\\

\begin{figure} [t]
\centering
\includegraphics[width=1.\textwidth]{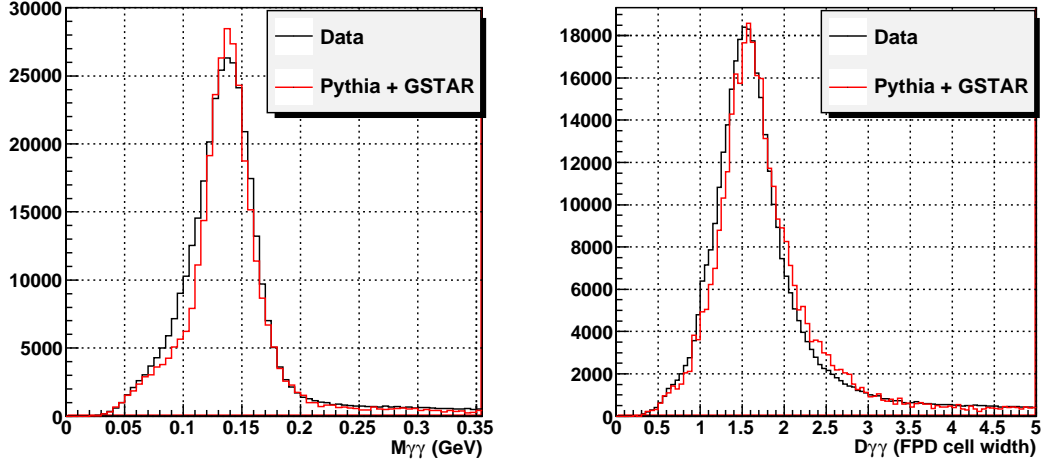}
\caption[Mass and separation, data - MC (\v{C}erenkov) comparison]{LEFT: Di-photon invariant mass distribution. RIGHT: Di-photon separation distribution. Both are for events with 40 GeV $< E_{\gamma\gamma} <$ 45 GeV. Simulation was normalized to make the comparison easier. BLACK: data, RED: Pythia + GSTAR simulation, \v{C}erenkov based shower}
\label{Mdist2}
\end{figure} 

The main cause of the mass peak broadening is the energy. There are two main ways in which the energy measurement based on the \v{C}erenkov simulation differs from the one based on the energy loss. First, the energy resolution is degraded by more than a factor of 7. Second, the ratio of deposited energy to thrown energy now has a dependence on photon energy. Both of these effects come from the attenuation of optical photons, meaning that they are found in the number of detected optical photons after attenuation, but not in the number of generated photons. Figure \ref{MeanSigma} illustrates these differences, where a comparison is made for three types of simulation. The first is the energy loss based simulation. The second is a version of the \v{C}erenkov simulation where the attenuation length was set to 38 meters for all optical photons, and reflection was set to almost 100 \%. Around 4800 optical photons were detected per GeV of thrown energy. The third is the \v{C}erenkov simulation using the finalized attenuation parameters explained in the previous section. Around 1400 optical photons were detected per GeV.\\ 

\begin{figure} [t]
\centering
\includegraphics[width=1.\textwidth]{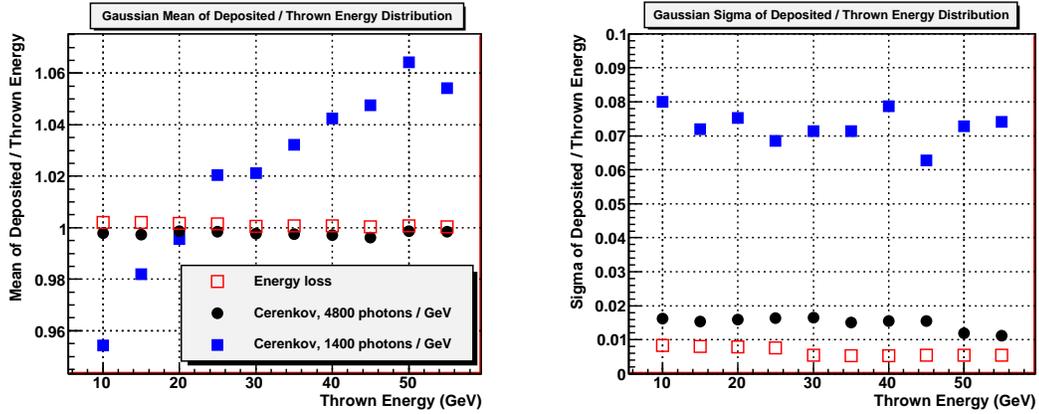}
\caption[Energy dependence of the gain and energy resolution]{LEFT: Gaussian mean of the deposited / thrown energy distribution vs. thrown photon energy. RIGHT: Gaussian $\sigma$ of the deposited / thrown energy distribution vs. thrown photon energy. RED: Energy loss simulation. BLACK: \v{C}erenkov simulation with extremely low attenuation. BLUE: \v{C}erenkov simulation with realistic attenuation.}
\label{MeanSigma}
\end{figure} 

It is clear that if the optical attenuation is extremely low, the \v{C}erenkov simulation behaves rather similarly to the energy loss simulation in this regard. (But it will be much narrower in profile.) But once we put in more realistic estimates of the attenuation, we see the emergence of the energy dependent gain, and the severe broadening of the energy resolution. It was also found that these effects are largely independent of how such attenuation is achieved, whether through the opacity of the Pb-glass, or the absorption at the glass-aluminum interface. In other words, given the narrowness of the shower shape, one can predict the energy dependence or the energy resolution reasonably well without regard to the details of the optical physics.\\

\subsection{Energy Dependent Mass Shift}\label{sec:massshift}

Finally, we look at the issue of the energy dependent mass shift illustrated in figure \ref{MvsE}. The energy dependent gain shift seen in figure \ref{MeanSigma} is clearly a mechanism by which the \v{C}erenkov simulation can generate the mass shift. Figure \ref{MvsE2} shows the data and simulation comparison of the energy dependent mass shift, where the simulation is based on \v{C}erenkov, and both the data and simulation were reconstructed using the same shower function. When comparing the figures \ref{MvsE} and \ref{MvsE2}, the improvement is immediately clear. While the data still has a slightly steeper slope, overall the trend is very well matched.\\

\begin{figure} [t]
\centering
\includegraphics[width=.8\textwidth]{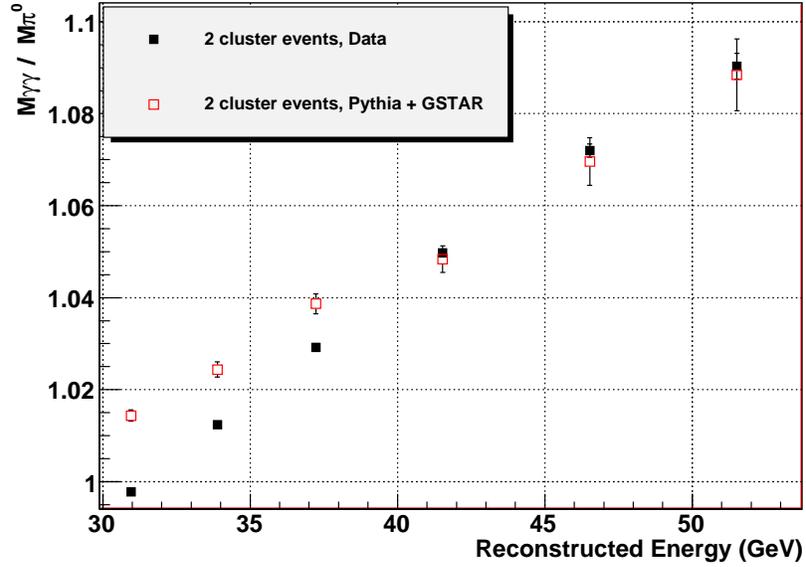}
\caption[$M_{\gamma\gamma}$ vs. energy, data - MC (\v{C}erenkov) comparison]{Di-photon invariant mass centroid / $M_{\pi^0}$ vs. reconstructed energy for two cluster events. BLACK: Data. RED: Pythia + GSTAR simulation based on \v{C}erenkov}
\label{MvsE2}
\end{figure} 

As already mentioned, one obvious cause of this effect in the simulation is the energy dependent gain shift. However, we should note that the slope seen in figure \ref{MvsE2} is roughly 4 \% per 10 GeV, whereas the slope of the energy mean shift in figure \ref{MeanSigma} was closer to 2 \% per 10 GeV. Clearly, there is something else that is adding to the effect. \\

\begin{figure} [b]
\centering
\includegraphics[width=1.\textwidth]{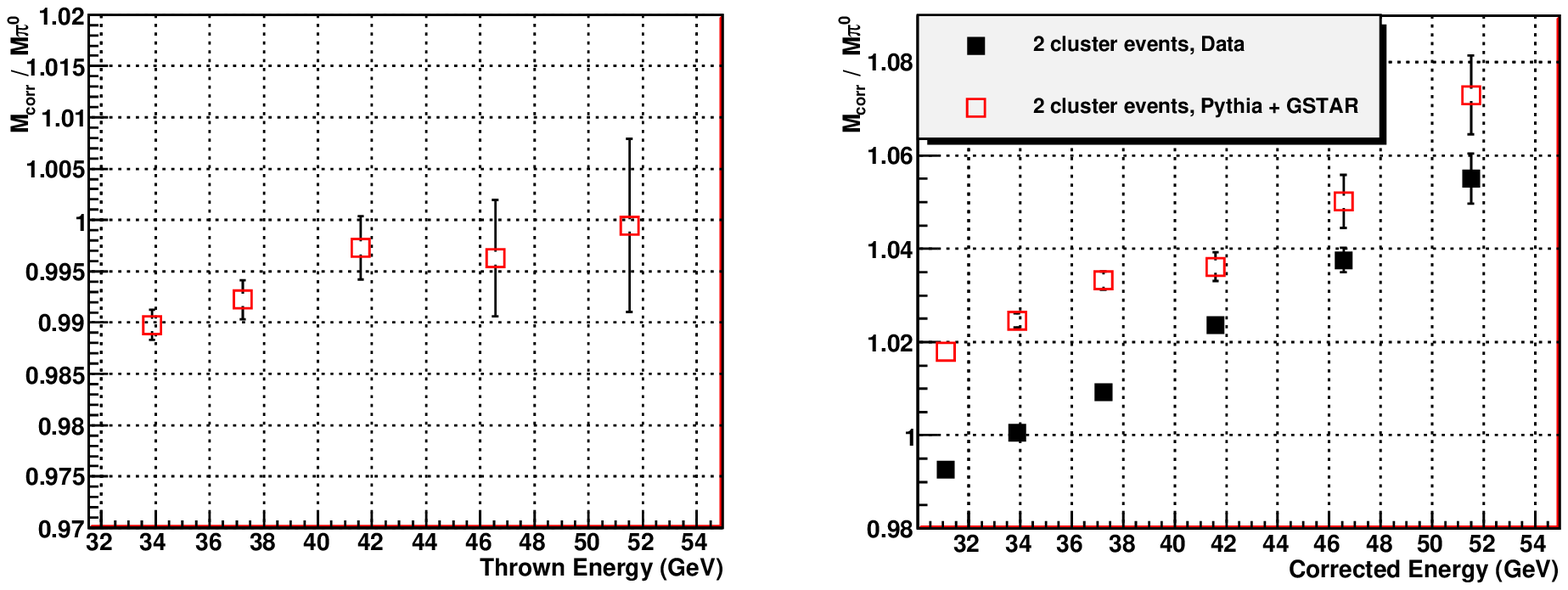}
\caption[$M_{\gamma\gamma}$ vs. thrown and measured energies, after correction]{LEFT: Corrected di-photon invariant mass centroid / $M_{\pi^0}$ vs. thrown energy for two cluster events. RIGHT: Corrected di-photon invariant mass centroid / $M_{\pi^0}$ vs. corrected energy for two cluster events. BLACK: Data. RED: Pythia + GSTAR simulation based on \v{C}erenkov}
\label{MvsEc}
\end{figure} 

The culprit turns out to be the energy resolution, which causes a significant bin migration when coupled with the rapidly falling cross-section. Whenever an observable is binned in deposited/reconstructed energy, there will be events that migrate to and from the neighboring true energy bins. Because of the rapidly falling cross-section, however, the majority of such migrant events are the ones that move up in observed energy relative to true energy. Since these events by definition have their energy overestimated, the average energy in any given bin will also be overestimated. If the observable depends on the energy, as is the case with the invariant mass, the average will be shifted according to its energy dependence. Figure \ref{MvsEc} illustrates the difference between binning in thrown energy and observed energy, where the observed energy has been corrected to take out the energy dependent gain shift. With corrections, the energy dependent shift in mass is largely gone on the left-hand side panel, where the observed mass is binned in thrown energy. On the other hand, the same events show significant energy dependence on the right-hand side panel, where binning is in observed energy.\\

The degree to which the average energy in a particular observed energy bin is overestimated depends primarily on two factors. The first is the energy resolution. Clearly, the poorer the resolution is, the more migrant events there will be. The second is the first derivative of the cross-section with respect to energy. Generally, a steeper local slope of the cross-section will result in a greater overestimation of the average energy by increasing the ratio of the upward migrant events (events that move up in observed energy relative to true energy) from lower energy bins to the events that are native to the bin. Depending on how these two factors combine, the degree of overestimation can in principle take any functional form versus energy. \\

It is useful to present a set of conditions in which the degree of overestimation becomes constant as a function of energy. While this is certainly not the only set of conditions that has such an effect, it will help explain the observed trend in energy dependent mass shift. First, we start with the assumption that the cross-section is purely exponential in energy,
\begin{equation}
\frac{d\sigma}{dE}=e^{a-bE}
\end{equation}
where both $a$ and $b$ are positive parameters. We define the energy smearing function $S$, which is a probability density function that returns the likelihood of measuring observed energy $E_{observed}$ for an event with true energy $E_{true}$. The ``width" of $S$ can be interpreted as the energy resolution. The second assumption that we make is that the functional form of $S$ only depends on the difference between true energy and observed energy, independent of the energy scale. (This means that fractionally, the energy resolution gets better as the true energy increases.)
\begin{equation}
S(E_{observed},E_{true}) = S(E_{observed}-E_{true})
\end{equation}

As a simple example, we take an arbitrary energy bin, and calculate the ratio between the number of events that came from lower energy bins, and the number of events whose true energy belonged to this bin. We assume that the binning is uniform, with the fixed bin width of $E_{bin}$. Then the number of events whose true energy is between $E_0$ and $E_0+E_{bin}$ is given by,
\begin{equation}
N(E_0)=\int\limits_{E_0}^{E_0+E_{bin}} e^{a-bx} dx
\end{equation}
The ratio of upward migrant events to the native events is then given by,
\begin{equation}
f(E_0)=\frac{1}{N(E_0)}\int\limits_{E_{thsh}}^{E_0} e^{a-bx} \int\limits_{E_0-x}^{E_0-x+E_{bin}} S(y) dxdy 
\end{equation}
where $E_{thsh}$ is the threshold energy, such as the trigger threshold. We make one final assumption, which is that $E_0$ is sufficiently away from $E_{thsh}$ in the scale of $S$. That is, $S$ dies out well before it reaches $E_0-E_{thsh}$, and the contributions from the lower energy bins do not come from the threshold region.\\

We then calculate the same ratio for some other point in energy, at $E=E_0+\delta E$. 
\begin{equation}
f(E)=\frac{1}{N(E_0+\delta E)}\int\limits_{E_{thsh}}^{E_0+\delta E} e^{a-bx} \int\limits_{E_0+\delta E-x}^{E_0+\delta E-x+E_{bin}} S(y) dxdy\\
\end{equation}
By substituting $x^{\prime}=x-\delta E$,  
\begin{equation}
f(E)=\frac{1}{N(E_0+\delta E)}\int\limits_{E_{thsh}-\delta E}^{E_0} e^{a-b(x^{\prime}+\delta E)} \int\limits_{E_0-x^{\prime}}^{E_0-x+E_{bin}} S(y) dx^{\prime}dy
\end{equation}
But the integral from $E_{thsh}-\delta E$ to $E_{thsh}$ is zero because $E_0-E_{thsh}$ is large in the scale of $S$, which makes $S(\text{vicinity of }E_0-E_{thsh})=0$. Therefore,
\begin{equation}
\begin{split}
f(E)&\approx \frac{1}{N(E_0+\delta E)}\int\limits_{E_{thsh}}^{E_0} e^{a-b(x^{\prime}+\delta E)} \int\limits_{E_0-x^{\prime}}^{E_0-x+E_{bin}} S(y) dx^{\prime}dy\\
&=\frac{e^{-b \delta E}}{e^{-b \delta E} \cdot N(E_0)}\int\limits_{E_{thsh}}^{E_0} e^{a-bx^{\prime}} \int\limits_{E_0-x^{\prime}}^{E_0-x+E_{bin}} S(y) dx^{\prime}dy\\
&=f(E_0)
\end{split}
\end{equation}

For all bins, the ratio of the number of events that move in from the low energy bin to the number of events that truly belong to that bin is constant. While the calculation of the ratio of the average observed energy to true energy would be somewhat more complicated, it is easy to see that the same argument will apply. We conclude that for an exponential cross-section coupled with a smearing function that only depends on the difference in energy, the degree of energy overestimation is constant for all bins sufficiently away from the threshold.\\

\begin{figure} [t]
\centering
\includegraphics[width=1.\textwidth]{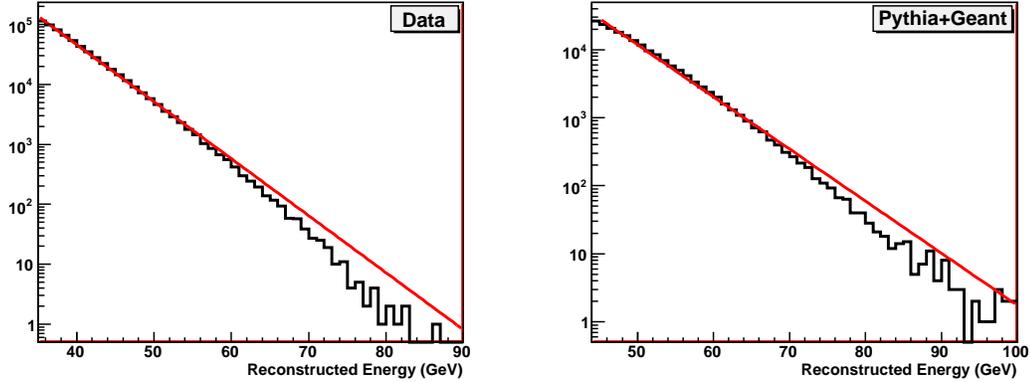}
\caption[Energy distribution in data and simulation with exponential fits]{Energy distribution in the $\pi^0$ mass region. LEFT: data, RIGHT: Pythia + Geant simulation. Also shown in red are exponential fits.}
\label{EDsim}
\end{figure} 

In figure \ref{MeanSigma}, we have found that the energy resolution is fractionally energy independent. Therefore, relative to the above described condition on the smearing function, the real smearing becomes more severe with increasing energy. This will results in shifting the observed mass up for higher energy bins. Furthermore, while the functional form of the cross-section is locally very close to the exponential, it tends to get softer at higher energy. Figure \ref{EDsim} show the energy distributions in the $\pi^0$ mass region for the data and full Pythia + Geant simulation, overlaid with exponential fits. Such deviations from exponential will also create an energy dependent mass shift in the observed direction. Therefore, the two effects add constructively, producing a positive slope for observed mass when binned in observed energy. \\

\section{Incident Angle Effect}

During RHIC run 6, the FPD was located at the ``middle" position, which is about 30 cm from the beam line to the edge of the detector. The FPD itself has a width of about 25 cm. Since the distance from the interaction point to the FPD is about 800 cm, the average incident angle was roughly 0.05 radians, or 3 degrees. Even though the angle is very small, we have found that it causes a meaningful shift in shower shape when measured in the horizontal direction along which the incident angle is the largest. For the current analysis, we have folded in the effect of the incident angle into the shower function. \\

\begin{figure} [t]
\centering
\includegraphics[width=1.\textwidth]{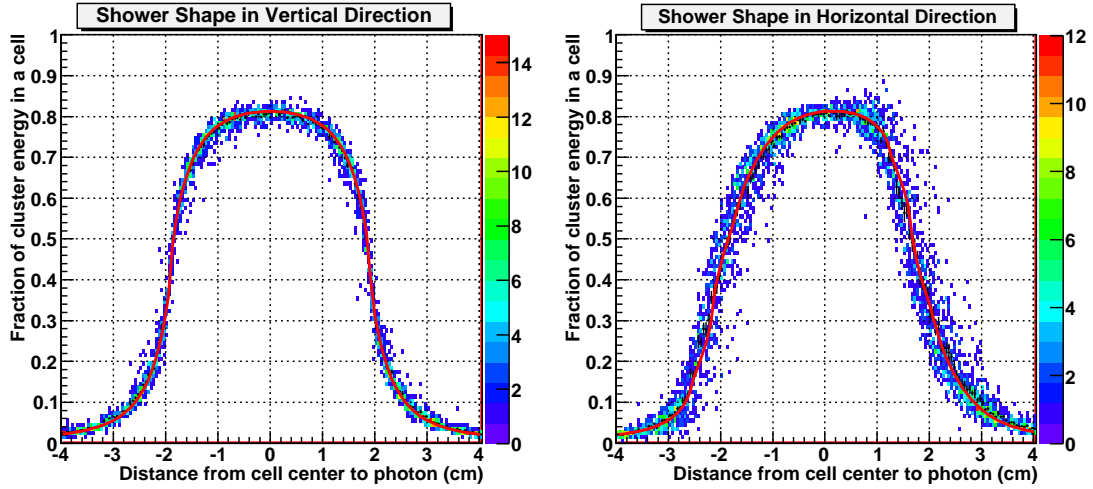}
\caption[Shower shape in horizontal and vertical direction, \v{C}erenkov based Geant]{Shower shape based on \v{C}erenkov light collection. LEFT: Vertical direction. RIGHT: Horizontal direction. RED: Shower function with incident angle effect}
\label{hEXYchkv}
\end{figure} 

Figure \ref{hEXYchkv} shows the shower shape based on \v{C}erenkov simulation, in both vertical and horizontal direction. The vertical distribution is in fact identical to the one shown in figure \ref{hEYchkv}. The effect of the incident angle is clear when comparing the two distributions. It is certainly true that the difference is in many ways much greater than the difference between the energy loss and \v{C}erenkov simulations, at least in regard to the shower shape. The overlaid red curve is the final version of the shower function used in the analysis, which incorporates the incident angle effect. The function includes two angle parameters that can be set for individual photons based on the location of the cluster. For the remainder of this brief section, we will discuss how this shower function was constructed.\\

\begin{figure} [t]
\centering
\includegraphics[width=1.\textwidth]{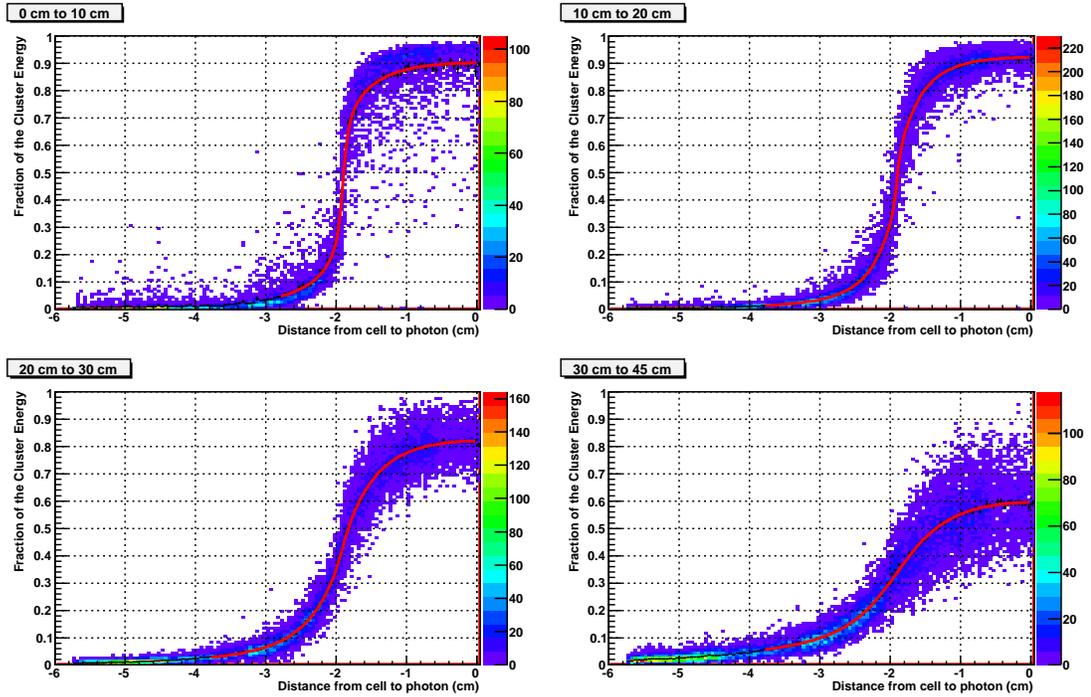}
\caption[Shower shape in four longitudinal segments, \v{C}erenkov based Geant]{Shower shape in four longitudinal segments of the 45 cm long Pb-glass of the FPD. The segment boundaries are 10 cm, 20 cm, and 30 cm from the front face of the glass. RED: Fit using the basic shower function described by equation \ref{eq:ShowerF}}
\label{4panel}
\end{figure}  																																			

While the most rigorous way to fold in the incident angle would be a full three dimensional shower simulation, it is too cumbersome and likely unnecessary for small angles in the FPD. We took a much simpler approach, by dividing the shower into four longitudinal segments, and creating a shower function for each ``slice". Each segment was made by collecting only those photons that were generated in a certain longitudinal interval in the Pb-glass cell. Each of the four shower distributions is fit to the standard shower function. The four incomplete shower functions can then be added with appropriate weights and offsets to form the full shower function, where the offsets are given by the incident angle, and the relative average positions from the shower maximum.\\
																										
Figure \ref{4panel} shows the transverse shower profile in each of the four longitudinal segments, with segment boundaries at 10, 20, and 30 cm from the front face of the glass. The distribution is normalized per segment, meaning that the shower shape is measured relative to the energy deposited by only those photons that originated from that segment. Notice that the shower is much more spread out towards the end of the glass, closer to the photo-cathode. Before we can combine them to produce the final shower shape, weighting factors need to be applied to the four functions based on the ratio of the average energy per segment to the total photon energy. Figure \ref{Eseg} shows the distribution of the ratio of the segment energy to total energy for the four segments. The weighting factors were taken from the means of these four distribution. The full list of parameters and weighting factors appears in table \ref{Tshower}. \\

\begin{figure} [t]
\centering
\includegraphics[width=0.6\textwidth]{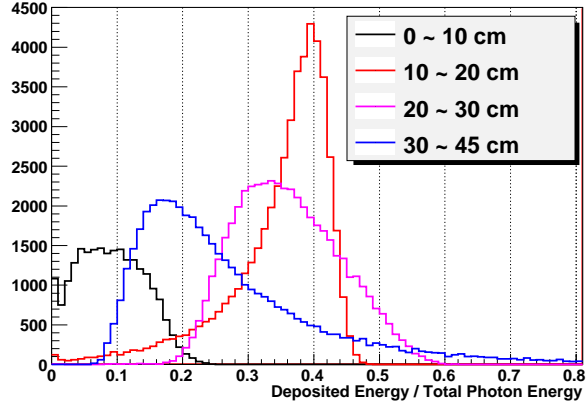}
\caption[Distribution of the ratio of the segment energy to total energy for the four segments]{Distribution of the ratio of the segment energy to total energy for the four segments. The segment boundaries are 10 cm, 20 cm, and 30 cm from the front face of the glass. RED: Fit using the basic shower function described by equation \ref{eq:ShowerF}}
\label{Eseg}
\end{figure}  																																	
The main limitation of this analysis, apart from the coarseness of the four segment approximation, is that no concern is given to the event by event correlations. In figure \ref{Eseg}, there are small numbers of events that deposited 70 \% of the total photon energy within the very last segment. Clearly, such an anomalous shower development during the last 15 cm of the glass has to be paired with equally anomalous developments elsewhere in the glass, all of which are being averaged out in the current analysis. But because of the incident angle being so small, we have found that such omission is justified on a practical ground. For the future analysis that may deal with much larger incident angles, however, the issue will likely have to be revisited.\\																																					\begin{table}
\centering
\begin{tabular}{ | c || c | c | c | c | c | c | c | c |}
    \hline
      & weight & z-offset & $a_1$ & $a_2$ & $a_3$ & $b_1$ & $b_2$ & $b_3$ \\ \hline  
    \hline
    $0 \sim 10$ cm & 0.09 & 6.73 & 0.999 & 0.432 & -0.416 & 0.069 & 0.493 & 0.085\\ \hline 
    $10 \sim 20$ cm & 0.39 & 15.1 & 0.248 & 1.000 & -0.229 & 0.041 & 0.399 & 0.946\\ \hline 
    $20 \sim 30$ cm & 0.33 & 24.5 & 0.141 & 0.896 & 0.0000 & 0.117 & 0.508 & 0.199\\ \hline 
    $30 \sim 45$ cm & 0.18 & 35.6 & 0.790 & 0.992 & -1.074 & 0.662 & 6.685 & 19.23\\ 
    \hline
\end{tabular}
\caption[Parameters for the incident angle dependent shower function.]{Parameters for the incident angle dependent shower function. The z-offset and $b_i$'s are in units of cm.}
\label{Tshower}
\end{table}

\section{Photon Position Resolution}

Despite various improvements in the shower shape function, such as the \v{C}erenkov based simulation and the inclusion of an incident angle effect, we still find that the distribution of the reconstructed photon coordinates is less than uniform. Figure \ref{hxynew} shows the X and Y coordinate distribution from data and simulation, where the data is based on the most up to date calibration constants. Clearly, the intra-cell structure remains, although it is much reduced compared to the original reconstruction algorithm used for the previous FPD analysis. While the simulation exhibits a somewhat different structure, the magnitudes of non-uniformity in data and simulation are comparable. \\

\begin{figure} [t]
\centering
\includegraphics[width=1.\textwidth]{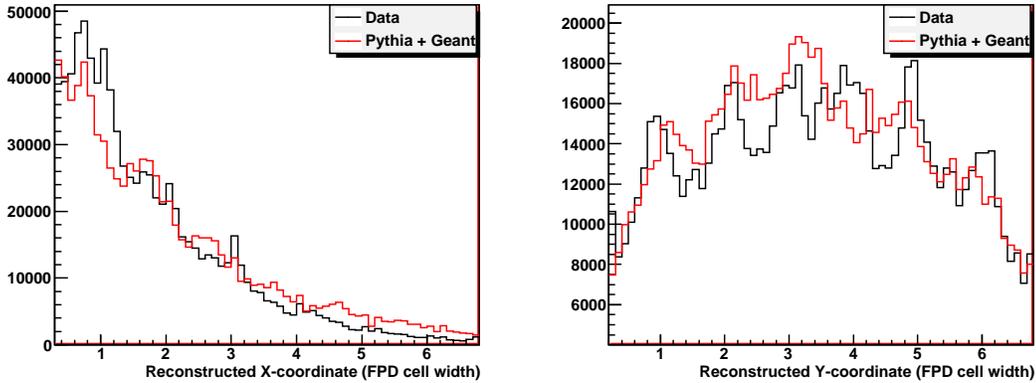}
\caption[Photon X-Y coordinate uniformity]{Reconstructed photon coordinate distributions for data and Pythia + Geant simulation, LEFT: X-coordinate, RIGHT: Y-coordinate.}
\label{hxynew}
\end{figure}  			

Finally, we note that this is an area that needs to be improved in the future, and that the shower shape is still not fully understood. There still remain measurable differences in shape between the shower in simulation and the shower in data, and we need to understand the precise optical properties of the Pb glass and the glass-mylar interface in order to set up the Geant simulation more accurately. Furthermore, the upgraded STAR forward calorimetry, the Forward Meson Spectrometer (FMS), covers a much wider range of pseudo-rapidity, which likely necessitates a more sophisticated approach to the incident angle dependence. These remaining issues are left for the future FMS analyses.\\

\end{doublespace}

\chapter{Off-Line Calibration}\label{ch:calib}
\setlength{\parindent}{1cm}
\begin{doublespace}

\section{Introduction}
The off-line, software calibration of the FPD relies primarily on $\pi^0$ reconstruction. Since $\pi^0$'s are copiously produced, about 98.8 \% of which then decay into two photons, the $\pi^0$ peak in the di-photon invariant mass spectrum provides useful information on the absolute as well as relative (within the detector) energy scale. However, as briefly discussed in chapter \ref{ch:shower}, the energy measurement is only one part of the equation that describes the invariant mass, which also depends on various other variables like the two photon separation ($D_{\gamma\gamma}$), two photon energy sharing ($Z_{\gamma\gamma}$), and the interaction vertex ($Z_{Vertex}$). Therefore, for the invariant mass to be an accurate indicator of the energy scale, all other quantities that it depends on have to be brought under control, at least for the part of the phase space used for calibration. In this chapter, we will study the dependence of invariant mass on a number of variables based on data and simulation.

\section{$\pi^0$ Mass Based Calibration Scheme}	
First, it is useful to describe the general scheme of the calibration. For each two photon event selected for calibration, the invariant mass is associated with two FPD channels that contain the reconstructed coordinates of the photons. (For $\pi^0$'s used for calibration, the separation is large enough that the two photons do not reconstruct within a single cell.) For each channel, a mass histogram is accumulated based on this association, which is used to determine the centroid of the $\pi^0$ mass peak. Figure \ref{m7x7} shows 49 mass histograms, which correspond to the 49 channels of the well calibrated FPD North detector. A Gaussian fit is visible for each channel. Once the centroids are determined, channel by channel gain corrections are made based on the ratio between the observed centroid and the target mass. This constitutes one iteration, and if our method is successful, the mass centroids will converge towards the target mass after enough iterations. \\

\begin{figure} [t]
\centering
\includegraphics[width=1.\textwidth]{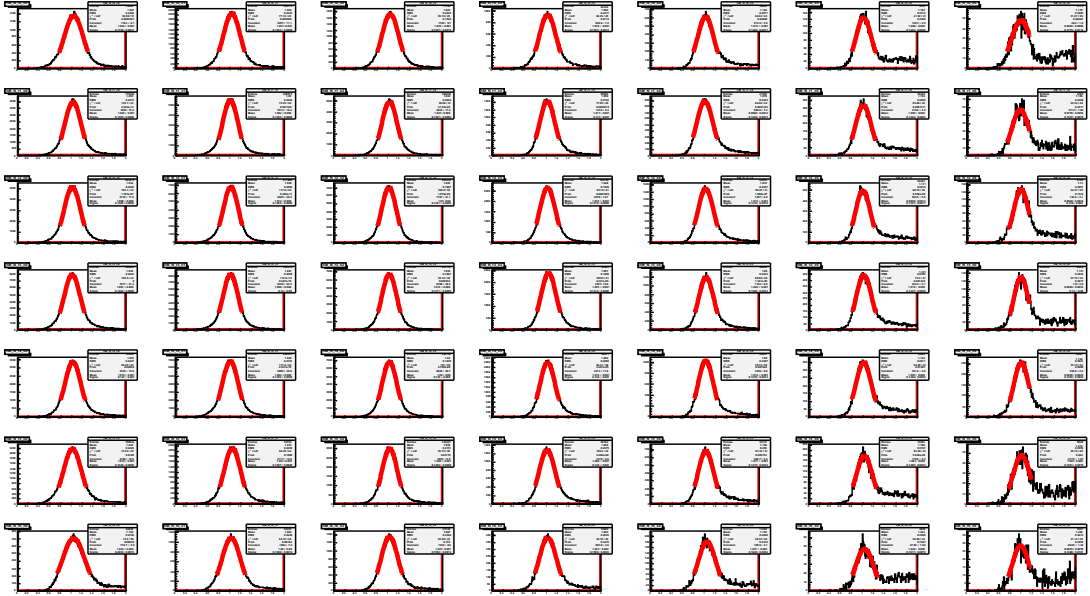}
\caption[$M_{\gamma\gamma}$ distribution for the 49 channels of north FPD]{Distributions of di-photon invariant mass divided by $M_{\pi^0}$ for the 49 channels of north FPD. Pseudo-rapidity of the cells decreases from left to right. Also shown are Gaussian fits overlaid with the distributions, used for finding the mass centroid.}
\label{m7x7}
\end{figure}  	

The di-photon invariant mass can be written as the following.
\begin{equation}\label{eq:mgg}
\begin{split}
M_{\gamma\gamma}&=\sqrt{2 \cdot E_1 E_2(1-cos(\theta_{open}))}\\
&= 2\sqrt{E_1 E_2} \cdot sin(\frac{\theta_{open}}{2}) \\ 
&= E_{total} \cdot \sqrt{1-{Z_{\gamma\gamma}}^2} \cdot sin(\frac{\theta_{open}}{2}) \\
&\approx E_{total} \cdot \sqrt{1-{Z_{\gamma\gamma}}^2} \cdot \frac{D_{\gamma\gamma}}{2 \cdot Z_{Vertex}} 
\end{split}
\end{equation}
\begin{equation}
\begin{split}
E_{total} &= E_1+E_2\\
Z_{\gamma\gamma} &= \frac{E_1-E_2}{E_{total}}\\
D_{\gamma\gamma}&=\text{Di-photon separation}\\
\theta_{open}&=\text{Di-photon opening angle}\\
Z_{Vertex}&=\text{Interaction vertex}\\
\end{split}
\end{equation}

The last line in equation \ref{eq:mgg} uses a small angle approximation, as $D_{\gamma\gamma} \sim$ 5 cm, and $Z_{Vertex} \sim$ 8 meters. Since the gain correction factors are equal to the ratio of the target mass to the mass centroid, which in turn is proportional to the square root of the photon energy, the corrections tend to be conservative, and the convergence generally requires many iterations.  \\

The calibration process can in principle bring the energy scale to the correct value, but only insofar as the target mass is chosen correctly. Ideally, the target mass would be the known value of the $\pi^0$ mass, at 135 MeV. However, that is only true if all of the quantities that make up the observed mass are accurately measured. In reality, this is often not the case. The most problematic is the separation measurement ($D_{\gamma\gamma}$), which at the level of precision that we require, (1 \%$\sim$2 \%) has a strong dependence on variety of factors. The energy sharing ($Z_{\gamma\gamma}$) is also a difficult measurement, but owing to the functional shape of the energy sharing term ($\sqrt{1-{Z_{\gamma\gamma}}^2}$) that largely suppresses the actual variation in $Z_{\gamma\gamma}$, it is much less problematic.\\

\section{1 Cluster vs. 2 Cluster Events}
\begin{figure} [t]
\centering
\includegraphics[width=.8\textwidth]{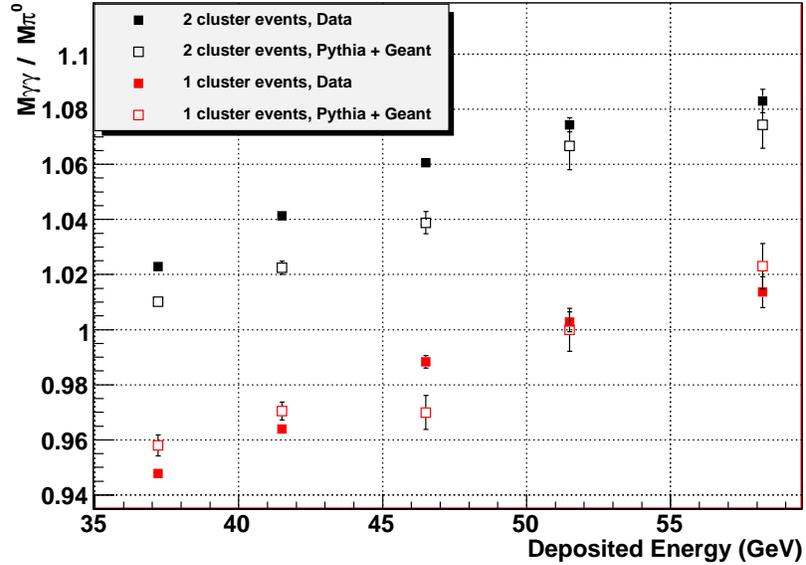}
\caption[Di-photon invariant mass / $M_{\pi^0}$ vs. deposited energy for 1 and 2 cluster events, for data and Pythia + Geant simulation]{Di-photon invariant mass / $M_{\pi^0}$ vs. deposited energy for 1 and 2 cluster events, for data and Pythia + Geant simulation. Deposited energy refers to the sum of energy observed in one FPD module. The invariant mass in this figure was calculated by using the deposited energy instead of the reconstructed two photon energy.}
\label{m12}
\end{figure}  	

Depending on the separation and topology, a two photon event can yield either two separate clusters of energy, or a single cluster. Because many aspects of the reconstruction are affected by the result of clustering, it is useful to think of them as two distinct event classes. The most obvious manifestation of their differences is shown in figure \ref{m12}, where we see a clear difference in the location of reconstructed mass centroid between one and two cluster events. One cluster events return masses that are roughly 5 \% lower than that of the two cluster events, consistently throughout the energy range. Since the effect is reasonably well simulated by Pythia + Geant, we turn to simulation to study its origin.\\

For figure \ref{m12}, we used the total deposited energy in the FPD instead of the usual reconstructed energy to calculate the mass. By doing so, we remove the reconstruction dependence of the total energy term in equation \ref{eq:mgg}, reducing the effect of clustering in mass calculation. Nevertheless, the total deposited energy is still affected by the clustering when binned in observed energy. Figure \ref{EandE} shows the dependence of the deposited energy measurement on the number of clusters. On the left-hand panel, we see that when binned in thrown energy, indeed the two types of events produce identical energy measurements. It is only when binned in deposited energy, as shown in the right-hand panel, that we see that the energy in two cluster events is overestimated by about 5 \% relative to the energy in one cluster events. \\

\begin{figure} [t]
\centering
\includegraphics[width=.9\textwidth]{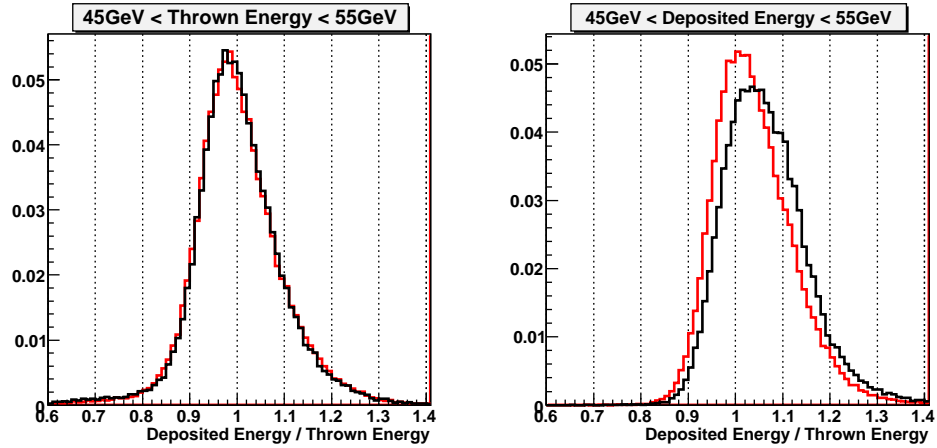}
\caption[Deposited energy / thrown energy distribution, for 1 cluster (red) and 2 cluster (black) events.]{Deposited energy / thrown energy distribution, for 1 cluster (red) and 2 cluster (black) events. LEFT: Binned in thrown energy. RIGHT: Binned in deposited energy. Pythia + Geant simulation.}
\label{EandE}
\end{figure}  	

The difference is due to the bin migration effect described in chapter \ref{sec:chkv}. Here, the faster falling event rate as a function of energy for the two cluster events results in a greater degree of overestimation. The energy slope difference between the two types of events is simply due to the fact that as the energy increases, the separation decreases, and the $\pi^0$ events are more likely to yield only one cluster. (For more details on number of clusters vs. energy, see figure \ref{DvsE})\\

When using deposited energy to calculate the mass, there are two terms in equation \ref{eq:mgg} that depend on the reconstruction: $D_{\gamma\gamma}$ and $Z_{\gamma\gamma}$. Figure \ref{DandZ} shows the reconstructed over generated distributions for $D_{\gamma\gamma}$, and the term in equation \ref{eq:mgg} involving $Z_{\gamma\gamma}$. Both are binned in observed energy. As mentioned earlier, the $Z_{\gamma\gamma}$ term has a very narrow resolution, and does not behave differently for the two classes of events. On the other hand, the peak of the separation measurement for two cluster events is higher by about 2 \% compared to one cluster events. Furthermore, the width and shape of the distributions are significantly different. \\ 

\begin{figure} [t]
\centering
\includegraphics[width=.9\textwidth]{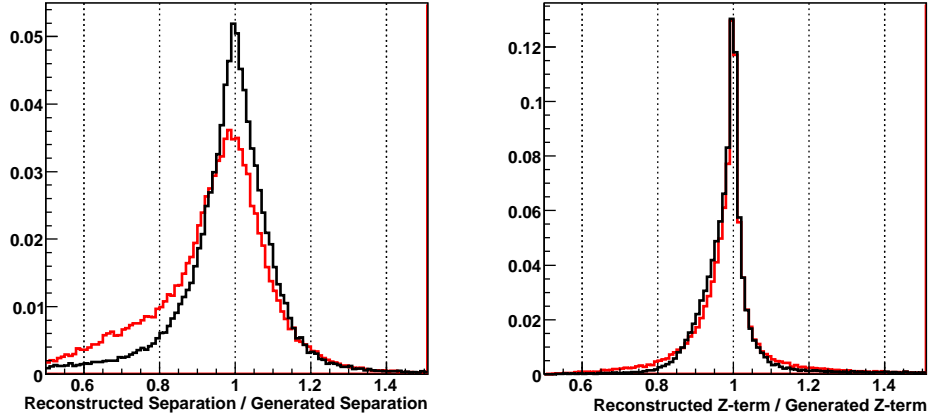}
\caption[Reconstructed / generated distributions for di-photon separation and $\sqrt{1-Z_{\gamma\gamma}^2}$]{Reconstructed / generated distributions for di-photon separation (LEFT) and $\sqrt{1-Z_{\gamma\gamma}^2}$ (RIGHT), $45\text{ GeV} < E_{\gamma\gamma} < 55\text{ GeV}$, for 1 cluster (red) and 2 cluster (black) events. Pythia + Geant simulation.}
\label{DandZ}
\end{figure}  	

Generally, events that produce two clusters tend to have cleaner topologies, in that it is easier to separate out the two photons. The effect can be seen in the left-hand panel of figure \ref{DandZ}, where the separation resolution is significantly better for the two cluster events. For this reason, we choose two cluster events for calibration purposes. After all, the energy scale may be anchored for a subset of the data, and the rest can be bootstrapped from there. As long as we understand the differences in non-energy variables like separation and energy sharing between the two classes of events, it is possible to translate the difference in mass into difference in energy, and apply energy corrections accordingly to the events that are not included in the calibration. \\

\section{Separation Dependence of Mass}
\begin{figure} [b]
\centering
\includegraphics[width=1.\textwidth]{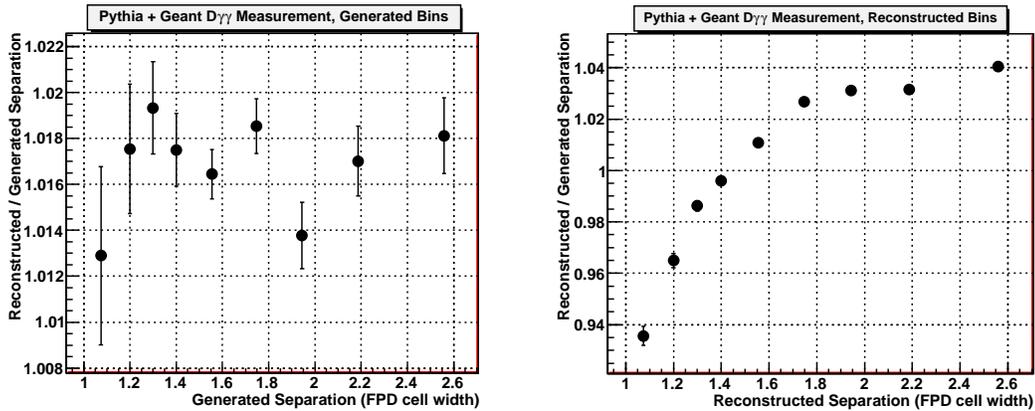}
\caption[Reconstructed / generated di-photon separation distributions]{Reconstructed / generated di-photon separation distributions. LEFT: Binned in generated separation. RIGHT: Binned in reconstructed separation. Pythia + Geant simulation.}
\label{DD}
\end{figure}  	
In comparing figure \ref{EandE} and figure \ref{DandZ}, we note that the width of the separation resolution is comparable to that of the energy resolution. In addition, the $\pi^0$ separation for a small range of energy is a highly peaked distribution, as shown in figure \ref{Mdist2}. For the energy range used for calibration (35 GeV $\sim$ 40 GeV), the peak is just under 1.5 FPD cell width, with a steep slope on either side of the peak. It is then not surprising that the separation measurement, when binned in measured separation, will suffer from the bin migration problem similarly to the energy measurement when binned in observed energy. Figure \ref{DD} illustrates this point, where the observed separation binning produces a systematic underestimation to the left side of the peak, and an overestimation to the right side of the peak. \\

\begin{figure} [b]
\centering
\includegraphics[width=.8\textwidth]{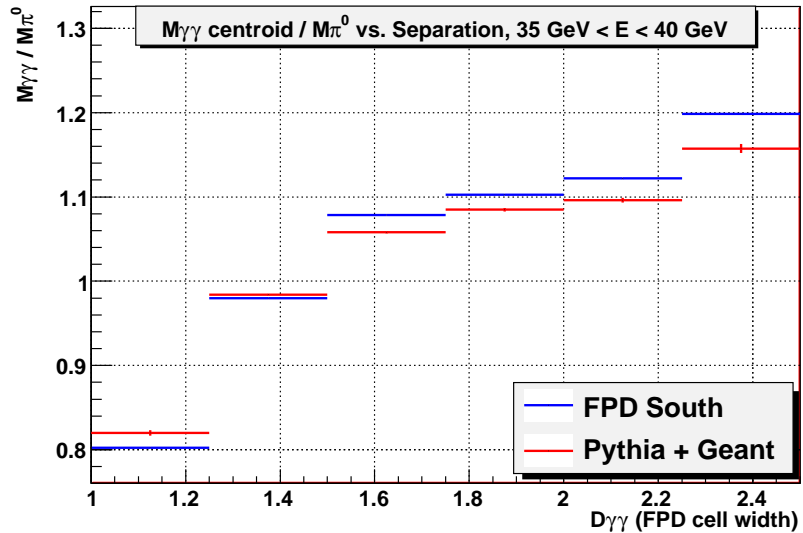}
\caption[$M_{\gamma\gamma} / M_{\pi^0}$ vs. reconstructed separation]{$M_{\gamma\gamma} / M_{\pi^0}$ vs. reconstructed separation, $35\text{ GeV} < E_{\gamma\gamma} < 40\text{ GeV}$, for data and Pythia + Geant simulation.}
\label{MvsD}
\end{figure}  	

For two cluster events only, now we look at the separation dependence of mass. Figure \ref{MvsD} shows the location of mass centroids as a function of reconstructed separation, for two FPD detectors and Pythia + Geant simulation. The absolute scale difference between data and simulation is somewhat arbitrary, as it depends on how we choose to calibrate. Overall, while the general trends of over- and under-estimation around the peak is reproduced in the simulation, the mass in data tends to exhibit a greater degree of variation as we move away from the peak on either sides. The origin of this discrepancy is not exactly understood, since both the shape and the width of the separation distribution seem to be well simulated in the \v{C}erenkov based scheme, as seen in figure \ref{Mdist2}. We can only guess that the remaining discrepancy in the shower simulation has to be the culprit, although the exact mechanism remains unknown. \\

Identifying the range of separation in which the measurement is accurate relative to the true separation is not crucial for our purpose. Rather, our focus is to find a class of events for which the simulation matches the data well. As long as the simulation is accurate, whatever systematic errors that are present in the separation measurement will be folded into the definition of the target mass, which need not be 135 MeV. In principle, we have no direct way of evaluating the quality of the simulation in this regard. On the one hand, due to the limitations of the shower simulation, small separation events in which much of the showers overlap may bring in systematics that are not visible in figure \ref{MvsD}. (For instance, the total reconstructed energy measurement, or the energy sharing). On the other hand, the effect of bin migration will average out if we integrate over all values of separation. Added to that is a need to preserve enough statistics to calibrate on the far edges of the detector, which sets a limit on how stringent we can be on event cuts. \\

We have tried to impose a minimum separation requirement in order to minimize the effects of the shower shape discrepancy. However, it turns out that if the minimum separation is too high, we lose too much statistics, and if the minimum separation is around the peak of the separation distribution, the result becomes too sensitive to the precise location of the cut. At the end, we chose to include all two cluster events in the calibration process, noting that the very small separation events that may be less well simulated are only a very small part of the event sample. \\

\section{Internal Angle Dependence of Mass}
Even with the event cuts discussed so far, the invariant mass shows a significant dependence on the internal azimuthal angle of the two photons, which we will call $\theta$. The angle $\theta$ is defined as following:
\begin{equation}
\theta = \frac{y_1-y_2}{|y_1-y_2|} \cdot cos^{-1}(\frac{x_1-x_2}{D_{\gamma\gamma}})
\end{equation}
where $x_1$ and $y_1$ are the coordinates of the photon with larger energy, and $x_2$ and $y_2$ are those of the photon with smaller energy. $\theta$ of 0 and $\pi$ correspond to a horizontal orientation of the two photons, and $\theta = \pm 0.5\pi$ to a vertical orientation. \\ 

\begin{figure} [t]
\centering
\includegraphics[width=.8\textwidth]{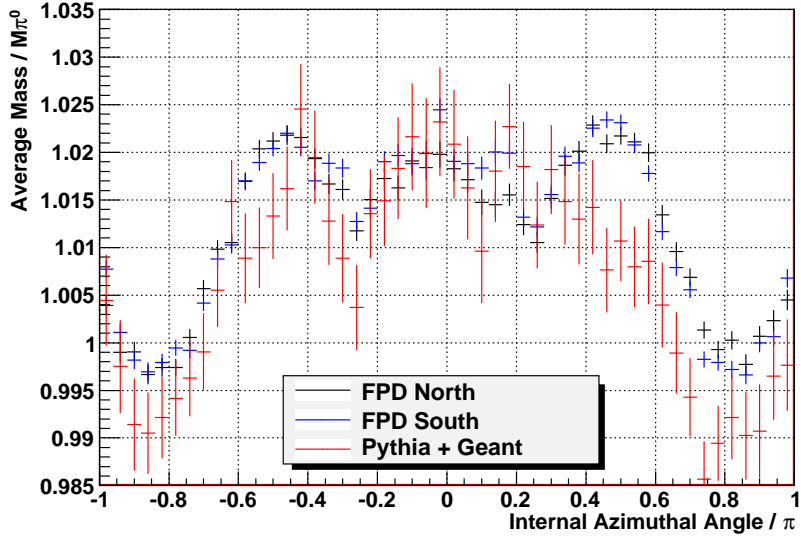}
\caption{$M_{\gamma\gamma} / M_{\pi^0}$ vs. internal azimuthal angle $\theta / \pi$, for data and Pythia + Geant simulation. }
\label{MvsAng}
\end{figure}  

Figure \ref{MvsAng} shows the $\theta$ dependence of the mass for events satisfying the ``calibration" cut, for data and simulation. There is over 2 \% variation in mass depending on the orientation of the two photons, with a pronounced multi-peak structure. The largest mass values are associated with the vertically oriented pairs ($\theta = \pm 0.5\pi$), and the horizontally oriented pairs where the larger energy photon is farther away from the beam ($\theta = 0$). The smallest mass values occur for the horizontal pairs with the opposite energy configuration. Overall, the effect is reasonably well simulated, which is one of the improvements of using the \v{C}erenkov photon based shower. \\

\begin{figure} [t]
\centering
\includegraphics[width=.8\textwidth]{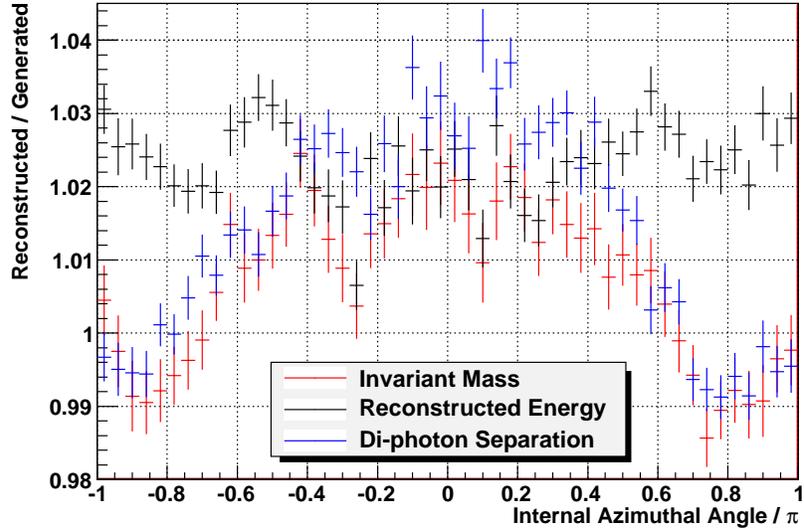}
\caption[Reconstructed over generated $M_{\gamma\gamma}$, $E_{\gamma\gamma}$, and $D_{\gamma\gamma}$ vs. internal azimuthal angle $\theta / \pi$, Pythia + Geant simulation]{Reconstructed over generated $M_{\gamma\gamma}$, $E_{\gamma\gamma}$, and $D_{\gamma\gamma}$ vs. internal azimuthal angle $\theta / \pi$, Pythia + Geant simulation}
\label{SvsAng}
\end{figure}  	
	
Figure \ref{SvsAng} illustrates how the mass, energy, and separation measurements depend on the angle $\theta$. It is clear that the separation measurement is largely responsible for the observed structure. The reconstructed energy exhibits $\theta$ dependence as well, but it is limited to about 1 \%, and for the most part washed out by the much larger variation in separation. (The average values shown here may not necessarily agree with the centroids of the distribution. They are obtained by taking a profile of a two dimensional distribution in a chosen range of mass.) \\

The precise form of this structure is not fully understood. However, the strong angle dependence itself should not come as a surprise, given the complexity of the reconstruction algorithms, and the fact that not all directions are equal on a square grid. Photon pairs that are diagonally oriented have a significantly different topology in energy distribution compared to vertical or horizontal decays, and often times yield a different number of clusters. Furthermore, our version of the incident angle effect does not take into account the energy dependence of the shower depth. In general, a shower that develops deeper into the detector will reconstruct further away from the beam. The relatively small size of the FPD, and the fast falling cross-section may also combine to produce edge effects that are non-uniform in directions. We conclude that while the exact cause of this structure has not been identified, there are a number of plausible mechanisms that can generate such an effect, and that it is after all well simulated. The potential systematics it may introduce to the calibration should be well contained in the target mass definition, which can in principle be chosen differently for each cell.

\section{Ambiguity Due to Energy Bin Migration}
Due to the large bin migration effect in energy measurement, there is an ambiguity as to how the absolute calibration should be normalized. One strategy is to demand that the energy measurement be accurate in observed energy bins. Practically, this involves calculating the target mass in the simulation in an observed energy bin, for instance for $35\text{ GeV} < E_{recon} < 40\text{ GeV}$. The problem with this strategy is that while the average energy measurement in a given observed energy bin is normalized correctly, for a beam of photons at a particular energy, the calibration will be underestimated. Figure \ref{gErE} illustrates how the normalization of energy measurement changes depending on the type of binning used, much like what we saw with the invariant mass measurement in figure \ref{MvsEc}, and energy measurement in figure \ref{EandE}. \\

\begin{figure} [t]
\centering
\includegraphics[width=.9\textwidth]{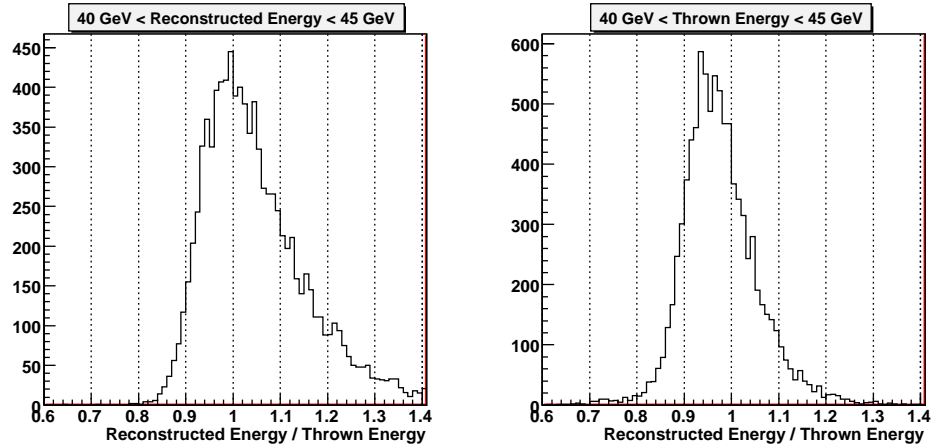}
\caption[Energy measurement when calibrated in reconstructed energy bins]{Reconstructed energy / thrown energy distributions when calibrated in reconstructed energy bins. LEFT: Binned in reconstructed energy. RIGHT: Binned in thrown energy. Pythia + Geant simulation.}
\label{gErE}
\end{figure}  	

In this scheme, if a $\pi^0$ with 30 GeV of energy hits the detector, the reconstructed energy is likely to be lower than 30 GeV. It is only because there are disproportionately larger numbers of lower energy $\pi^0$'s, some of which reconstruct to over 30 GeV of energy, that the normalization becomes correct when looking at an ensemble of events within an observed energy bin. In other words, the calibration scheme compensates for the overestimation of energy due to bin migration, effectively folding in the shape of the cross-section and the width of the energy resolution. \\

The advantage of this strategy is clear. If there is little need to unfold the effects of bin migration (for instance when measuring a quantity that does not vary rapidly as a function of energy, such as the spin asymmetry), it allows us to simply report the measurement at average energy in each observed energy bin. It also puts the mass peaks close to their natural values in the binning that we actually use for data analysis. \\

However, if the quantity that we are interested in has a strong dependence in energy (such as cross-section), then given the energy resolution, some level of unfolding is unavoidable. In this case, the above discussed strategy produces a largely off-diagonal and asymmetrical mixing matrix (one that maps true energy to observed energy), since a large portion of the events in a given true energy bin is likely to reconstruct in a lower reconstructed energy bin. Furthermore, the validity of the calibration is intimately tied to the shape of the cross-section, which may be different for different types of events. For instance, if we want to look at three photon events in the FPD, which may have a different energy dependence of the cross-section compared to two photon events, then we would need a different set of correction factors to take into account the different degree of bin migration.\\

An alternative strategy is to pursue a photon-by-photon normalization. Practically, this involves calculating the target mass in the simulation in a generated energy bin. The resulting target mass, when looked at in an observed energy bin, is higher than standard PDG value of the $\pi^0$, as the effect of bin migration, which has not been folded into the calibration, causes energy overestimation. Consequently, the average energy in any observed energy bin will require corrections. However, the mixing matrix needed for this correction will be more diagonal and symmetric in this scheme compared to the previous one. For a different measurement, the same calibration can be used in conjunction with a mixing matrix calculated specifically for the shape of the cross-section and energy resolution for that type of events.\\

The difference between the two methods is largely a matter of convenience. Either way, the bin migration effect has to be resolved, the only difference being whether it is included in the calibration, or in the unfolding process. For the current analysis, we chose the second approach, in which we calculate the gain correction factors independently of the cross-section or energy resolution. We found that a more diagonal unfolding matrix is easier to deal with for the cross-section analysis. The target mass in this scheme is set at 138.5 MeV, about 2.5 \% higher than the nominal value. \\

\section{Energy Dependent Gain Shift}

\begin{figure} [t]
\centering
\includegraphics[width=.8\textwidth]{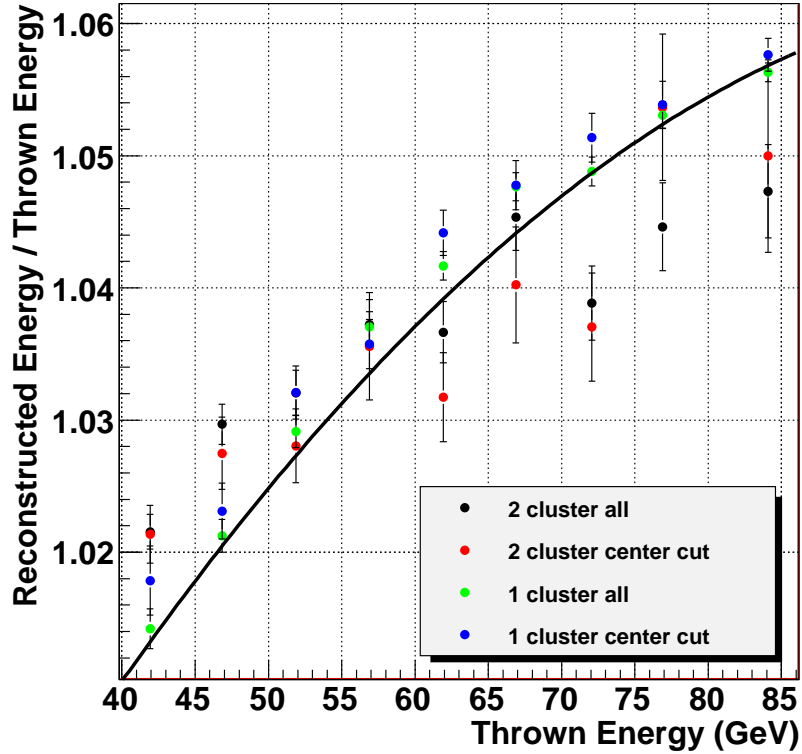}
\caption[Reconstructed energy / thrown energy vs. thrown energy in $\pi^0$ only Geant simulation]{Reconstructed energy / thrown energy vs. thrown energy in $\pi^0$ only Geant simulation. An equal number of $\pi^0$'s were generated for each bin. Also shown in black is the function used to correct the energy in data and simulation.}
\label{calibEdep}
\end{figure}  	

In chapter \ref{ch:shower}, we discussed the energy dependent mass shift, where we pointed out that a part of the shift comes from the actual increase in gain as a function of energy, and the rest comes from bin migration that gets more severe at higher energy. Since we chose to normalize the calibration on a photon-by-photon basis without folding in the bin migration effect, we take the same approach in correcting for the energy dependent gain shift. Figure \ref{calibEdep} shows the actual shift in gain as a function of thrown energy, based on $\pi^0$ only Geant simulation. The simulation sample has a flat energy distribution, but it is not a concern since the binning is in thrown energy. Also shown is a quadratic function used to correct the energy dependent gain shift in simulation and data. This only corrects that part of the shift that is caused by the actual shift in gain, and not the part caused by the bin migration. The latter is dealt with when we unfold the energy smearing effect, discussed in chapter \ref{ch:anal}.\\

\section{Calibration Uncertainty}

It is difficult to quantify the uncertainty in calibration precisely. The fundamental issue is, as is often the case, the remaining discrepancy between data and simulation. As a last part of the discussion on calibration, here we list the most likely causes of calibration error. \\

Firstly, it is possible that due to the shower shape discrepancy discussed in chapter \ref{ch:shower}, the target mass definition obtained in simulation is incorrect. The most likely cause would be the simulation of di-photon separation, which depends critically on the shower shape. However, there are reasons to believe that this issue is under control within $1 \sim 2$ \%, by studying the width of mass peaks, and the overall shape of the mass spectrum as a function of energy. Above all, various ``tweaking" of the shower function in the course of calibration yielded shift in invariant mass by no more than 2 \%, but sometimes accompanied by recognizable patterns within the detector. (For instance, the central row would change more than the others.)\\

\begin{figure} [t]
\centering
\includegraphics[width=.8\textwidth]{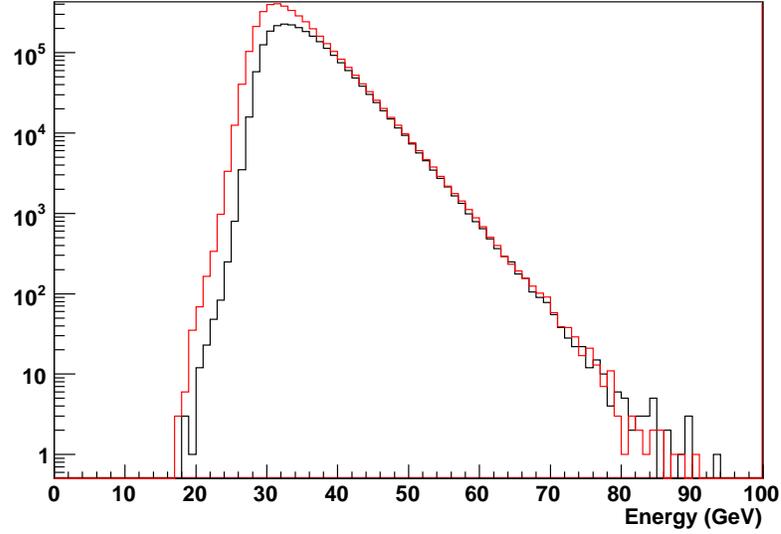}
\caption[Counting rates for north and south FPD]{FPD counting rate after rate equalization is enforced. BLACK: north, RED: south.}
\label{NvsS}
\end{figure}  	

Secondly, there is a persistent discrepancy in counting rate between the north and the south FPD. Calibrating solely based on invariant mass results in the north counting rate being about 30 to 40 \% lower than that of the south. If we enforce some level of counting rate equalization, the target masses of the two modules end up differing by $1 \sim 2$ \%. It is true that the on-line gain calibration for the north was overestimated relative to the south, resulting in an effectively higher trigger threshold for the north. However, the effect of trigger threshold should be limited to the threshold region, (nominally 35 GeV) whereas we see a constant difference in counting rate across the measured energy range. This issue is not fully understood, but based on the study of mass spectrum as a function of energy, it was concluded that forcing equal counting rate at the expense of invariant mass was a more desirable approach. As can be seen in figure \ref{NvsS}, the rate equalization actually results in an almost identical exponential slope for the north and south in the higher energy region. If there is a relative difference in calibration between the two modules, one that is artificially boosted will exhibit a harder slope (less steep) than the other. \\ 

In the end, we conservatively estimate the absolute calibration uncertainty at 3 \%, and the relative calibration at 1.5 \%. While these uncertainties matter little for the spin asymmetry measurement, they are one of the main components of the systematic error for the cross-section measurement.\\

\end{doublespace}

\chapter{Data Analysis}\label{ch:anal}
\setlength{\parindent}{1cm}
\begin{doublespace}

\section{Introduction}
In this chapter, we will cover various correction factors needed for the physics measurements. They include the background estimations for $\pi^0$ and $\eta$ signals, the detection efficiencies, and the unfolding of the energy smearing. The latter two are used only for the cross-section measurement.

\section{Backgrounds}
As we saw in invariant mass distributions in previous chapters, the $\pi^0$ signal has a relatively low level of background. Little more is needed than to simply apply a cut in the invariant mass spectrum to achieve a reasonable level of signal purity. The only potential issue with the $\pi^0$ signal is the misidentification of single photons into $\pi^0$'s. The $\eta$ signal, on the other hand, has a background spectrum that has a strong dependence on energy. Overall, the mass spectrum is judged to be reasonably well simulated, and we rely on the Pythia + Geant simulation to study the detailed composition of the mass spectrum.\\

\subsection{Di-Photon Invariant Mass Spectrum}	

With the \v{C}erenkov-based shower simulation, the Pythia + Geant simulation does reasonably well in reproducing the observed di-photon invariant mass spectrum in the data. Figure \ref{Mspect} shows data-MC comparison of the mass spectrum in three energy bins. In chapter 2, it was shown that the simulated $\pi^0$ mass peak was still slightly narrower than that in the data. The same can be said about the $\eta$ mass peak, while the continuum in between is well represented in the simulation. However, the difference is small enough to be understood in terms of some minor discrepancy in smearing, rather than the composition of the mass spectrum being significantly different between simulation and data. Assuming that the former is indeed the case, we proceed to study the background based on the simulation.\\	

\begin{figure} [t]
\centering
\includegraphics[width=1.\textwidth]{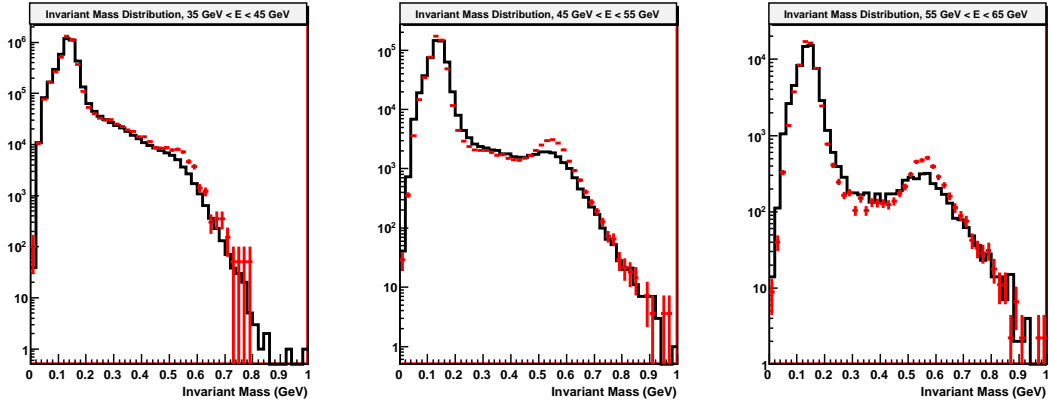}
\caption[Di-photon invariant mass spectrum in three energy bins, data and simulation comparison]{Di-photon invariant mass spectrum in three energy bins, BLACK: data, RED: simulation. From left, $35\text{ GeV} < E_{\gamma\gamma} < 45\text{ GeV}$, $45\text{ GeV} < E_{\gamma\gamma} < 55\text{ GeV}$, and $55\text{ GeV} < E_{\gamma\gamma} < 65\text{ GeV}$. The simulation was normalized to data in each energy bin separately. Two and only two photon events are included.}
\label{Mspect}
\end{figure}  	

First, we look at the types of tracks that make up the mass spectrum. Since two reconstructed photons have to be in the FPD for the current analysis, one would hope that most of the events do come from two real photons. However, the number of photons can be overestimated due to mis-reconstructed hadrons, or split single photons. It is also possible to have two or more photons merge during reconstruction and appear to be one. \\

\begin{figure} [t]
\centering
\includegraphics[width=0.8\textwidth]{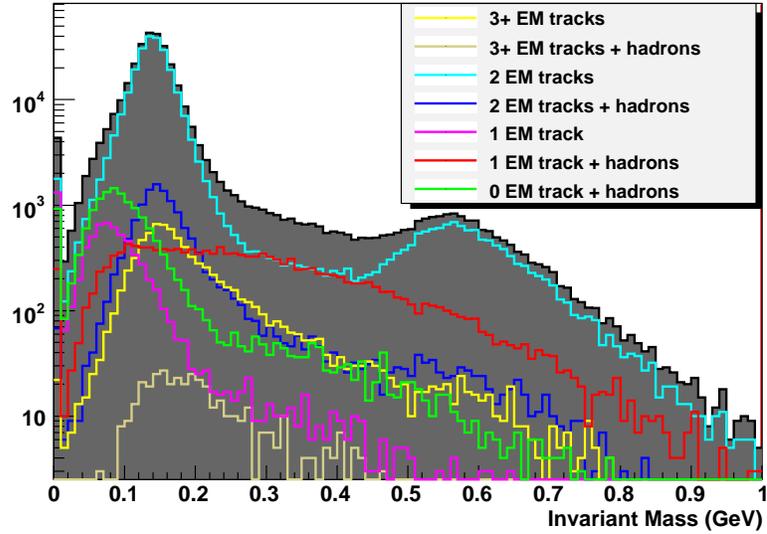}
\caption[Simulated di-photon invariant mass spectra 1]{Simulated di-photon invariant mass spectra for various event types based on the track composition. We select events with $E_{\gamma\gamma} > 45\text{ GeV}$ yielding two and only two reconstructed photons from Pythia + Geant simulation.}
\label{Mbg}
\end{figure}  	

Figure \ref{Mbg} superimposes mass spectra from various track compositions for events with a total energy greater than 45 GeV. It is clear that the vast majority of events in the $\pi^0$ and $\eta$ mass region come from two photons with no hadronic tracks in the vicinity. It is also one of the two major sources of the continuum between $\pi^0$ and $\eta$, the other being one photon events that have hadronic background. The events in which there were three photons, or two photons with hadrons, tend to be relatively more pronounced in the high mass tail of the $\pi^0$ peak. The events in which there was only one photon, or no photon and only hadrons, tend to reconstruct in the lower mass tail of the $\pi^0$ peak. Both of these cases have a very small contribution to the total mass spectrum. Overall, we conclude that two photon only events dominate the signal region, while the continuum has about equal contributions from real two photon events and one photon plus hadron events. \\

\begin{figure} [t]
\centering
\includegraphics[width=0.8\textwidth]{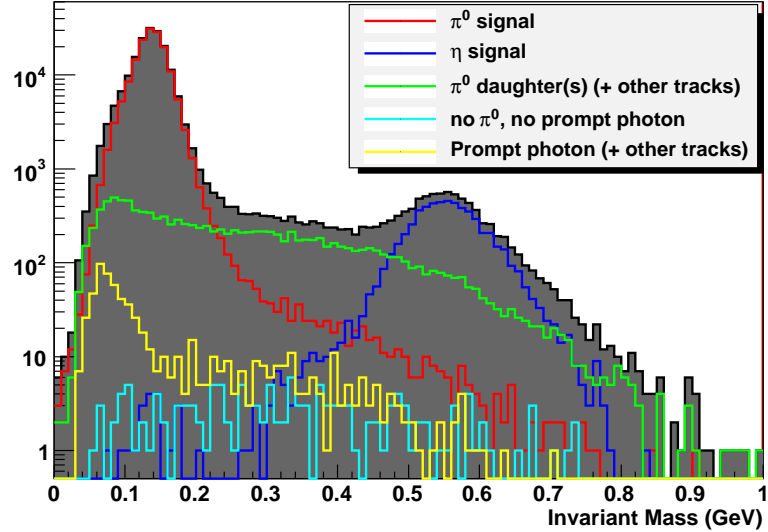}
\caption[Simulated di-photon invariant mass spectra 2]{Simulated di-photon invariant mass spectra, sorted by the parentage of the tracks. We select events with $E_{\gamma\gamma} > 45\text{ GeV}$ yielding two and only two reconstructed photons from Pythia + Geant simulation.}
\label{Mbg2}
\end{figure}  	

Secondly, we identify the parentage of the tracks and estimate the signal fractions in the mass regions for $\pi^0$ and $\eta$. Figure \ref{Mbg2} shows the breakdown of the mass spectrum based on parentage for events with total energy greater than 45 GeV. The two mass peak regions are clearly dominated by signal, while most of the continuum comes from combinations of $\pi^0$ decay photons, $\eta$ decay photons, and hadrons. More specifically, it largely consists of $\pi^0$ combinatorics background, and a $\pi^0$ decay photon plus hadron events. The mass spectrum from events involving a prompt photon is shown separately, which may be an important source of single photons at high energies where $\pi^0$ decays become highly clustered. While the accuracy of the Pythia prompt photon cross-section in this kinematic region is unknown, the fact that it is roughly three orders of magnitude suppressed in the usual $\pi^0$ mass cut window ($0.08\text{ GeV} < M_{\gamma\gamma} < 0.19\text{ GeV}$) suggests that it is extremely unlikely to be a major source of background for $\pi^0$.\\

\begin{figure} [t]
\centering
\includegraphics[width=1.\textwidth]{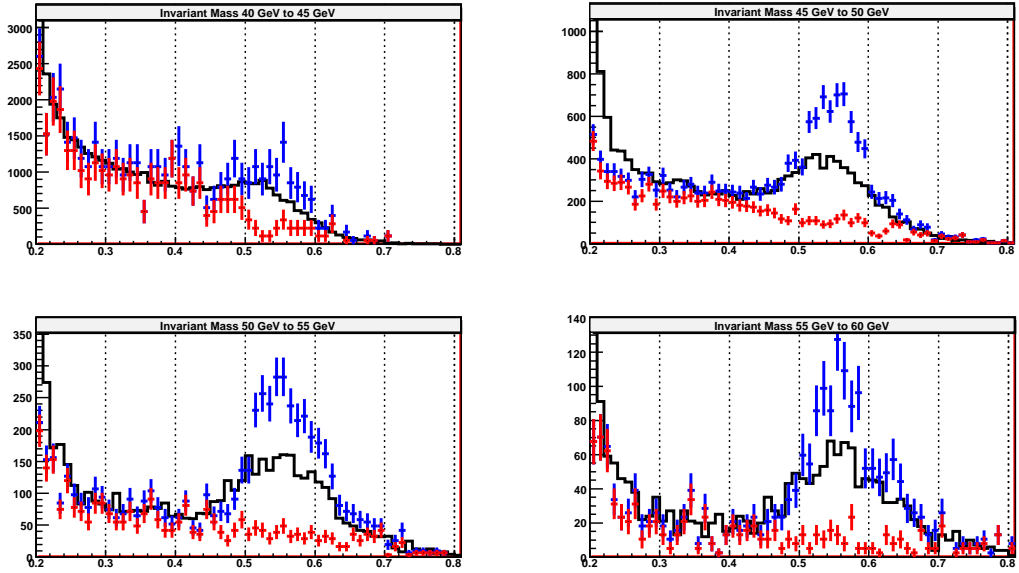}
\caption[Background estimation for the $\eta$ signal, data and simulation comparison]{Background estimation for the $\eta$ signal, data and simulation comparison with 5 GeV bins from 40 GeV to 60 GeV. BLACK: data, BLUE: simulation total, RED: simulation non-$\eta$ events. The simulation was normalized to data in each energy bin separately. Two and only two photon events are included.}
\label{Mspect2}
\end{figure}  	

Based on the analysis that produced figure \ref{Mbg2}, but over a wider energy range, we can calculate the signal fractions for $\pi^0$ and $\eta$ mass cut regions as a function of energy, as seen in the simulation. While this is perfectly adequate for the $\pi^0$, the same is not true for the $\eta$. As shown in figure \ref{Mspect}, there is a significant difference in the simulated strength of the $\eta$ signal and what's seen in the data. Figure \ref{Mspect2} shows the invariant mass spectrum from 40 GeV to 60 GeV in four bins, emphasizing the $\eta$ mass region. We see that the continuum is reasonably well simulated, and the main discrepancy is in the size of the $\eta$ signal. As the simulation is normalized to match the data in the total number of events for each panel, and the number of events is dominated by $\pi^0$'s, this essentially means that the $\pi^0 - \eta$ ratio in the simulation is different from that in the data. In fact, we have no reason to expect them to be the same. Based on the assumption that the background is well simulated but the signal isn't, we can estimate the signal fraction in data. \\

\begin{figure} [t]
\centering
\includegraphics[width=0.7\textwidth]{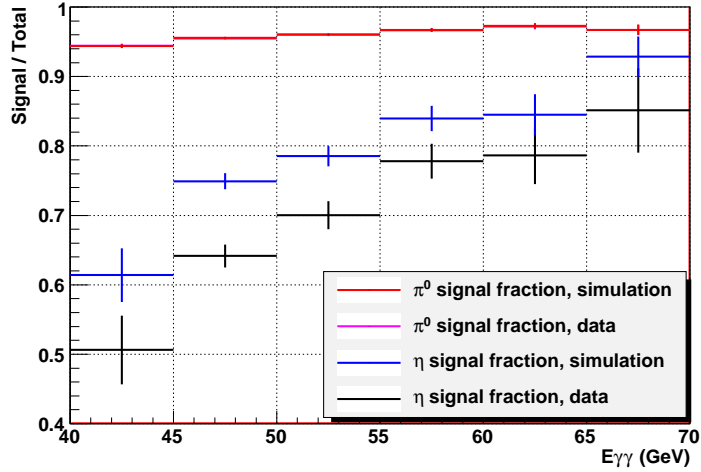}
\caption[Signal fraction estimates for data and simulation]{Signal over total number of events fraction for $\pi^0$ and $\eta$ mass regions for $0.08\text{ GeV} < M_{\gamma\gamma} < 0.19\text{ GeV}$ and $0.45\text{ GeV} < M_{\gamma\gamma} < 0.65\text{ GeV}$, respectively. Pythia + Geant simulation. The red and magenta almost entirely overlaps for all bins.}
\label{sig}
\end{figure} 

The results are shown in figure \ref{sig}. The values used for the mass cuts are $0.08\text{ GeV} < M_{\gamma\gamma} < 0.19\text{ GeV}$ for the $\pi^0$, and $0.45\text{ GeV} < M_{\gamma\gamma} < 0.65\text{ GeV}$ for the $\eta$. As expected, there are significant differences in the $\eta$ background estimate between data and simulation, especially at lower energy. As the background estimate is nominally a part of the efficiency correction, this discrepancy is applied separately as an adjustment factor to the efficiency correction. Finally, we note that the $\pi^0$ mass region has signal purity over 90 \% across the energy range, while the $\eta$ mass region reaches around 80 \% purity by 60 GeV. \\

\subsection{$\pi^0$-$\gamma$ Separation}\label{sec:pigamma}

As we are attempting to measure $\pi^0$'s at higher energy than what was done in the previous measurements with the FPD, one of the main concerns for the background is the single photon splitting. The left-hand panel of figure \ref{DvsE} shows the distribution of di-photon separation as a function of energy for $\pi^0$'s in the FPD. The y-axis is measured in units of FPD cell size, which is equal to 3.81 cm. We see that the peak separation in the energy range where we make most of the measurements is less than two cells wide, which becomes smaller than one cell wide above 60 GeV of $\pi^0$ energy. The right-hand panel of figure \ref{DvsE} shows the fraction of $\pi^0$ events that yield one and two clusters. The clustering algorithm has the ability to break off a cluster into two based on topology, but even with such functionality, the one cluster case dominates at high energy. At this point, our ability to distinguish a single photon cluster from a double photon one becomes crucial. \\

\begin{figure} [b]
\centering
\includegraphics[width=1.\textwidth]{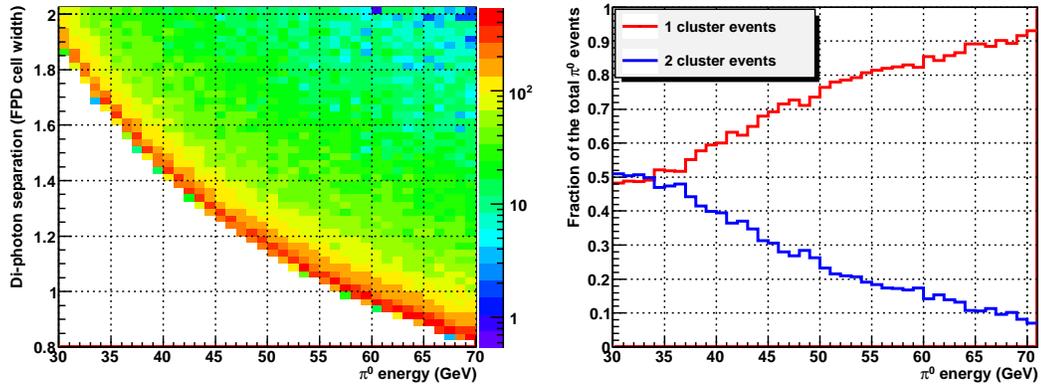}
\caption[Generated separation vs. thrown energy for the FPD, and 1 and 2 cluster events fraction vs. thrown energy]{LEFT: Generated separation vs. thrown energy for the FPD. RIGHT: 1 and 2 cluster events fraction vs. thrown energy. Results are based on $\pi^0$ only Geant simulation.}
\label{DvsE}
\end{figure} 

For the previous FPD analyses, we have relied on energy weighted cluster moments to identify clusters that contain two photons versus one. The details of this method, along with the details of the clustering algorithm, can be found in Yiqun Wang's thesis \cite{yiqun}. Here, we will only briefly discuss the overall scheme. First, for each cluster, we calculate the first and second moments in energy. 
\begin{equation}\label{linw1}
\bar{x}=\frac{\sum_{i}E_i \cdot x_i}{\sum_{i}E_i} 
\end{equation}
\begin{equation}
\bar{y}=\frac{\sum_{i}E_i \cdot y_i}{\sum_{i}E_i}
\end{equation}
\begin{equation}
\sigma_x = \sqrt{ \frac{\sum_{i}E_i \cdot (\bar{x}-x_i)^2}{\sum_{i}E_i} }
\end{equation}
\begin{equation}
\sigma_y = \sqrt{ \frac{\sum_{i}E_i \cdot (\bar{y}-y_i)^2}{\sum_{i}E_i} }
\end{equation}
\begin{equation}\label{linw2}
\sigma_{xy} = \sigma_{yx} = \sqrt{ \frac{\sum_{i}E_i \cdot (\bar{x}-x_i)(\bar{y}-y_i)}{\sum_{i}E_i} }
\end{equation}
The index $i$ runs up to the total number of cells included in the cluster. With the three second moments, we form a $2 \times 2$ matrix. 
\begin{equation}
M=
\begin{pmatrix}
\sigma_x^2 & \sigma_{xy}^2 \\
\sigma_{xy}^2 & \sigma_y^2
\end{pmatrix}
\end{equation}
This symmetric matrix has two eigenvectors, of which the one with the larger eigenvalue maximizes the second moment in energy, while the one with the smaller eigenvalue minimizes it. We call the second moments in these two directions $\sigma_{max}$ and $\sigma_{min}$, respectively. \\

\begin{figure} [t]
\centering
\includegraphics[width=1.\textwidth]{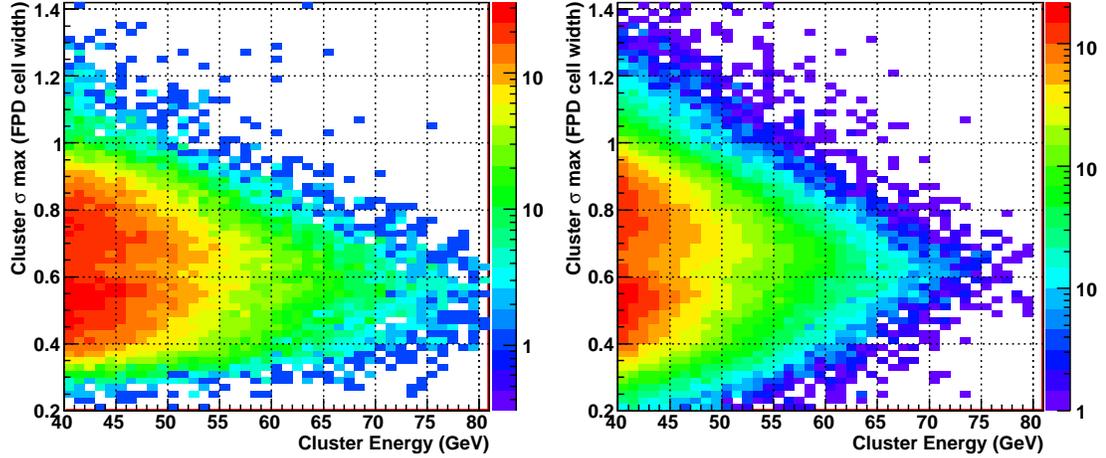}
\caption[Energy weighted $\sigma_{max}$ vs. cluster energy]{Energy weighted $\sigma_{max}$ vs. cluster energy. LEFT: Pythia + Geant, RIGHT: data.}
\label{oldcl2}
\end{figure} 

Qualitatively, the second moment in energy is loosely related to the invariant mass in that it is the product of the energy and the distance. It is then not too surprising that the $\sigma_{max}$ for a given energy range can be a useful indicator of the photon content of a cluster. Figure \ref{oldcl2} shows the distribution of $\sigma_{max}$ as a function of cluster energy, for single photon and $\pi^0$ only Geant simulations. The $\sigma_{max}$ is nearly constant for single photon clusters, with its mean at around 0.55. The $\sigma_{max}$ for $\pi^0$ clusters, on the other hand, show strong dependence on energy, as expected from the decreasing two photon separation. \\

The goal is to make a cut in the $\sigma_{max}$ vs. cluster energy space, in such a way as to maximize the separation between the single photon and the $\pi^0$ signal. If a cluster has a $\sigma_{max}$ that is large for its energy, it is considered a two photon cluster. If the $\sigma_{max}$ is small, then it is considered a single photon cluster. Ambiguous cases are handed over to the shower fitting stage, where the photon content is determined by performing both single and double photon fits, and by looking at the quality of each fit. Figure \ref{oldcl} shows the $\sigma_{max}$ distributions, presented in three energy bins between 40 GeV and 70 GeV. Up to the energy of about 50 GeV, the $\sigma_{max}$ for the two types of clusters are reasonably well separated. Beyond that, however, the two distributions merge significantly, making the distinction between single photon and $\pi^0$ much harder.\\

\begin{figure} [t]
\centering
\includegraphics[width=1.\textwidth]{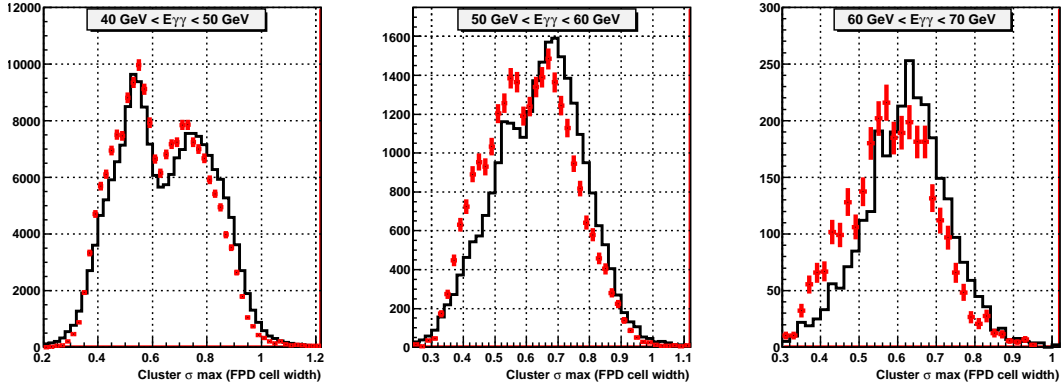}
\caption[Energy weighted $\sigma_{max}$ distribution in energy bins]{Energy weighted $\sigma_{max}$ distribution in energy bins. BLACK: Data, RED: Pythia + Geant simulation. From left, $40\text{ GeV} < E_{cluster} < 50\text{ GeV}$, $50\text{ GeV} < E_{cluster} < 60\text{ GeV}$, and $60\text{ GeV} < E_{cluster} < 70\text{ GeV}$. The simulation was normalized to data in each energy bin separately.}
\label{oldcl}
\end{figure} 

An obvious question to ask is then why we do not try both single and double photon fits on all clusters, and determine the photon content that way. The difficulty with this method is that the $\chi^2$ evaluation in shower fitting needs to be very well understood in order to compare the qualities of the two different types of fits fairly. Because our understanding of the shower shape and its fluctuation were limited, we had found the tried and true moment analysis to be more reliable. However, for future analysis, the reconstruction algorithm is being updated to include more sophisticated means of separating $\pi^0$'s from single photons.\\ 

For the current analysis, we take a more gradual approach, and make a fairly minor change to the moment analysis that nonetheless improves the accuracy of the method significantly. Namely, instead of using the energy, we now the use logarithm of the energy as the weight to calculate the moments. As the logarithm tends to emphasize the very low energy fluctuations too much, there needs to be a sensible low energy cutoff, which comes naturally via the required offset to keep the logarithmic weights positive. So instead of the equations \ref{linw1} to \ref{linw2}, we have,
\begin{equation}
\bar{x}=\frac{\sum_{i} log(E_i+E_{offset}) \cdot x_i}{\sum_{i} log(E_i+E_{offset})} 
\end{equation}
\begin{equation}
\bar{y}=\frac{\sum_{i} log(E_i+E_{offset}) \cdot y_i}{\sum_{i} log(E_i+E_{offset})}
\end{equation}
\begin{equation}
\sigma_x = \sqrt{ \frac{\sum_{i} log(E_i+E_{offset}) \cdot (\bar{x}-x_i)^2}{\sum_{i} log(E_i+E_{offset})} }
\end{equation}
\begin{equation}
\sigma_y = \sqrt{ \frac{\sum_{i} log(E_i+E_{offset}) \cdot (\bar{y}-y_i)^2}{\sum_{i} log(E_i+E_{offset})} }
\end{equation}
\begin{equation}
\sigma_{xy} = \sigma_{yx} = \sqrt{ \frac{\sum_{i} log(E_i+E_{offset}) \cdot (\bar{x}-x_i)(\bar{y}-y_i)}{\sum_{i} log(E_i+E_{offset})} }
\end{equation}
where the $i^{th}$ term in the sum is skipped if $log(E_i+E_{offset}) < 0$. Therefore $1-E_{offset}$ can be thought of as the low energy cutoff. For the current analysis, $E_{offset}$ is somewhat arbitrarily chosen to be 0.5 GeV. This choice was based on the fact that the nominal resolution of the ADC was 200 MeV per count, and we wanted to make sure that one or two count fluctuations were suppressed. Later analysis showed that 0.5 GeV turned out to be a reasonable choice, even though the rationale behind it was perhaps not the most relevant. \\

\begin{figure} [t]
\centering
\includegraphics[width=1.\textwidth]{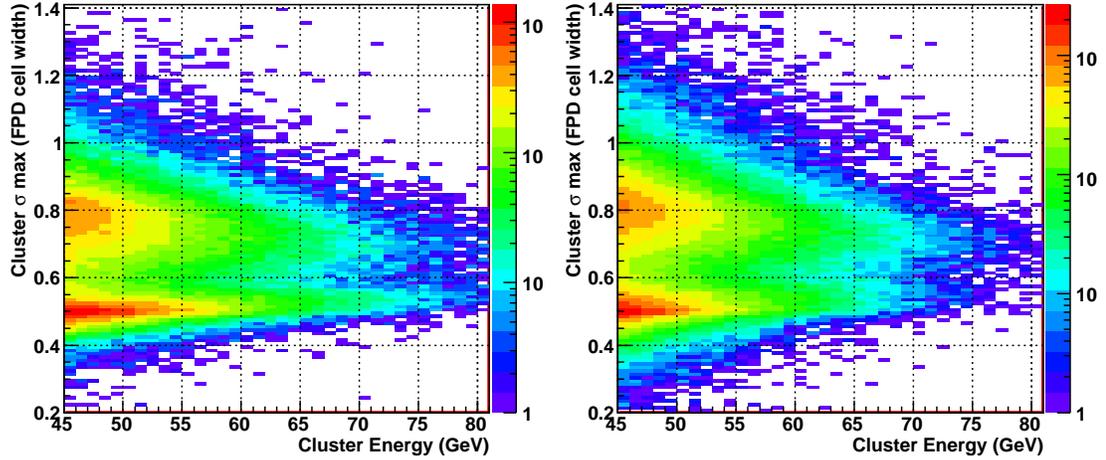}
\caption[Logarithm of energy weighted $\sigma_{max}$ vs. cluster energy]{Logarithm of energy weighted $\sigma_{max}$ vs. cluster energy. LEFT: Pythia + Geant simulation. RIGHT: Data. }
\label{newcl2}
\end{figure} 

Given these changes, the rest of the analysis follows the same pattern as before. Figure \ref{newcl2} shows the $\sigma_{max}$ vs. cluster energy for simulation and data, using log-weighted moments. There is a visibly well defined valley up until the statistics run out. Figure \ref{newcl} shows the $\sigma_{max}$ distribution in three energy bins from 45 GeV to 75 GeV.  Compared to figures \ref{oldcl2} and \ref{oldcl}, the improvement in $\pi^0$ and single photon separation is very significant. With this change, we can apply a simple cut in the $\sigma_{max}$ vs. cluster energy space and classify all clusters at the moment analysis level, without relying on the fit results. \\

\begin{figure} [t]
\centering
\includegraphics[width=1.\textwidth]{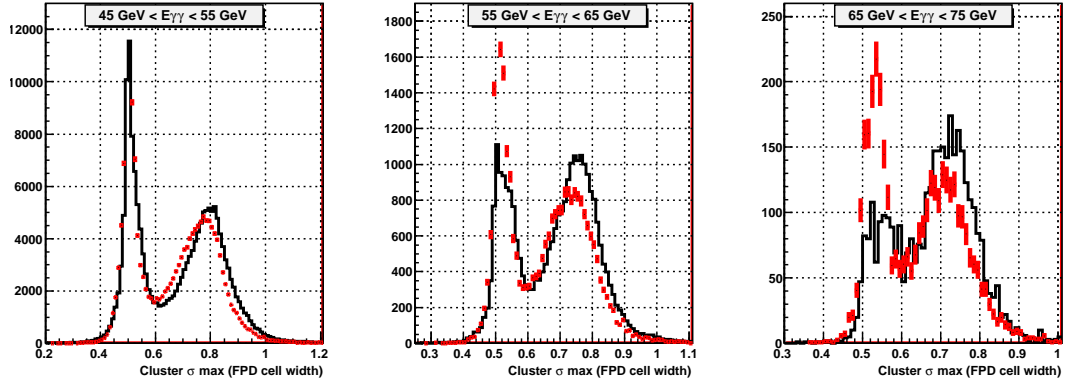}
\caption[Logarithm of energy weighted $\sigma_{max}$ distribution in energy bins]{Logarithm of energy weighted $\sigma_{max}$ distribution in energy bins. BLACK: Data, RED: Pythia + Geant simulation. From left, $45\text{ GeV} < E_{cluster} < 55\text{ GeV}$, $55\text{ GeV} < E_{cluster} < 65\text{ GeV}$, and $65\text{ GeV} < E_{cluster} < 75\text{ GeV}$. The simulation was normalized to data in each energy bin separately.}
\label{newcl}
\end{figure} 

Furthermore, the match between the data and simulation is better than it was with energy weighted $\sigma_{max}$, suggesting that the remaining discrepancy in shower shape likely resides in soft towers that are excluded in the new moment analysis scheme. Granted, there is a noticeable difference between the relative size of the single and double photon peaks, but that is not too surprising since we do not expect Pythia to be able to predict the relative cross-section between single photon and $\pi^0$ precisely in this very forward region. \\

To summarize, the logarithm of energy weighted moment analysis provides a markedly improved separation between single photons and $\pi^0$'s over the previous linear energy weighted scheme. The signal to background analysis discussed in the previous section (i.e. figure \ref{sig}) already incorporates these new improvements. Due to the general expectation that prompt photons, which do not go through fragmentation, may become dominant at high energy relative to $\pi^0$'s, it is essential that we have a good handle on the single photon background to $\pi^0$ signal. This applies not only to the cross-section analysis, but also to the spin measurements where we do not a priori know the type of spin asymmetry that prompt photons may carry. Conversely, these improvements, along with the continuing work on the reconstruction algorithm, bode well for the possible future forward prompt photon analysis.

\section{Efficiency Correction}\label{sec:eff}
In order to translate the counting rates into measurements of the cross-sections for $\pi^0$ and $\eta$ mesons, we need to know their detection efficiency. The total efficiency is a combination of geometrical acceptance, reconstruction efficiency, and the branching ratio into the $2\gamma$ channel. In chapters \ref{ch:shower} and \ref{ch:calib}, we have found that the Pythia + Geant full simulation (based on the \v{C}erenkov effect) describes the details of the shower shape and reconstruction reasonably well. Consequently, we expect both the geometrical acceptance and the reconstruction efficiency to be well simulated.  \\ 

\subsection{Geometrical Acceptance}
\begin{figure} [b]
\centering
\includegraphics[width=0.9\textwidth]{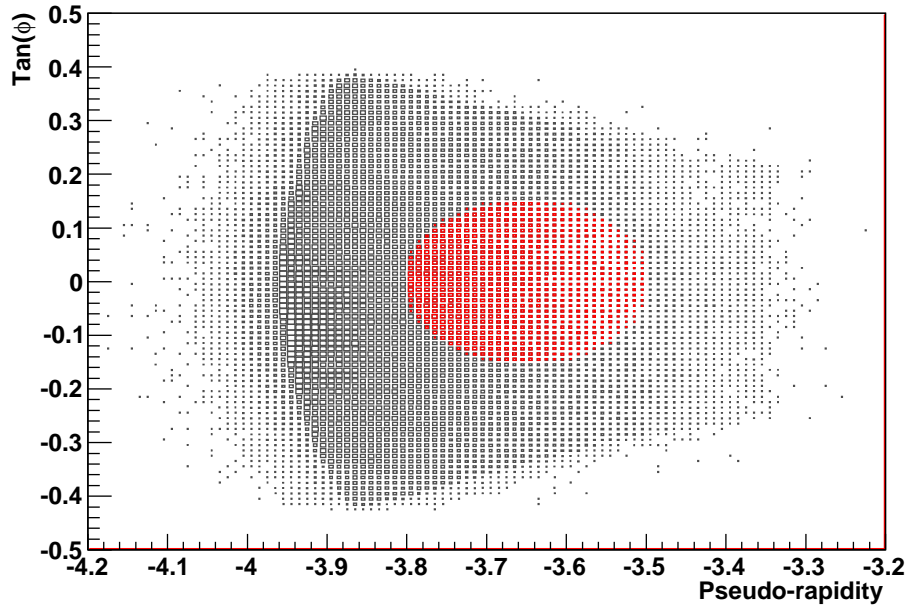}
\caption[Distribution of di-photon center of mass in pseudo-rapidity ($\mathsf{y}$) and tangent of azimuthal angle ($\phi$) space in data]{Distribution of di-photon center of mass in pseudo-rapidity ($\mathsf{y}$) and tangent of azimuthal angle ($\phi$) space in data. Indicated in red are the events that pass the ``center cut" (equation \ref{eq:center})}
\label{center}
\end{figure}  	

For the calculation of the geometrical acceptance of the FPD, we consider $\pi^0$ and $\eta$ mesons whose centers of mass satisfy the following condition.
\begin{equation}\label{eq:center}
(\mathsf{y}+3.65)^2 + tan^2(\phi) < 0.15
\end{equation}
where $\mathsf{y}$ is the pseudo-rapidity and $\phi$ is the azimuthal angle of the meson. (We reserve $\eta$ for the name of the meson to avoid duplicate notations.) This cut is informally called the ``center cut", due to the fact that it covers roughly the central region of the FPD acceptance. The purpose of this cut is to enhance the $\eta$ signal relative to its background. Because $\eta$ mesons barely fit inside the FPD, the signals likely come from those events in which the center of mass of the $\eta$ is in the central part of the FPD. Figure \ref{center} illustrates the overall coverage of the FPD in pseudo-rapidity and $tan(\phi)$ space, along with the location of the center cut. As described in chapter \ref{ch:result}, both the cross-section and the spin asymmetry are measured using the center cut.\\

\begin{figure} [t]
\centering
\includegraphics[width=.75\textwidth]{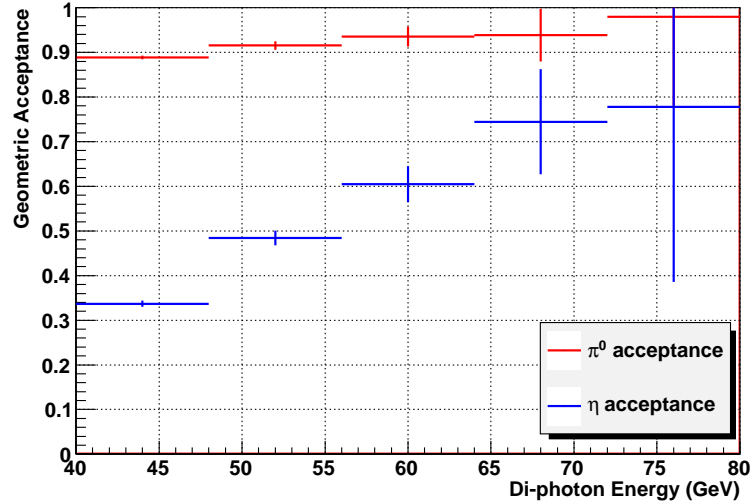}
\caption[Geometrical acceptance of the FPD vs. thrown energy for $\pi^0$ and $\eta$]{Geometrical acceptance of the FPD vs. thrown energy for $\pi^0$ and $\eta$ mesons whose thrown coordinates satisfy the center cut.}
\label{accp}
\end{figure}  	

For the $\pi^0$ and $\eta$ mesons that are thrown into the center cut region, the acceptance of the FPD is shown in figure \ref{accp}. Only those $\pi^0$'s and $\eta$'s that decay into two photons are counted. We consider an event accepted if both of the decay photons fall within the fiducial volume, which is defined to be 0.25 cell width away from the edge of the detector in this case. The acceptance is defined to be purely geometrical, and no aspects of the reconstruction plays a role in its calculation. We see that the $\pi^0$ acceptance is very good, as the typical two photon separation is on the order of one FPD cell width at around 55 GeV of energy. As expected, the $\eta$ acceptance suffers at low energy due to its large opening angle, but it improves significantly as the energy increases. Above 65 GeV, the difference in acceptance between $\pi^0$ and $\eta$ is only about 20 \%. \\

\subsection{Reconstruction Efficiency}
	
Figure \ref{receff} shows the reconstruction efficiency, which is defined to be the fraction of events that are reconstructed ``correctly" out of the events that were ``accepted" in the FPD as described above. An event is considered correctly reconstructed if the algorithm finds two and only two photons whose invariant mass falls within the mass cut used for the actual measurement ($0.08\text{ GeV} \sim 0.19\text{ GeV}$ for $\pi^0$ and $0.45\text{ GeV} \sim 0.65\text{ GeV}$ for $\eta$). We do not consider the effects of energy smearing and resulting bin migration here, which is covered separately later in this chapter. Also shown in the figure are the major modes of reconstruction failure.\\

\begin{figure} [t]
\centering
\includegraphics[width=1.\textwidth]{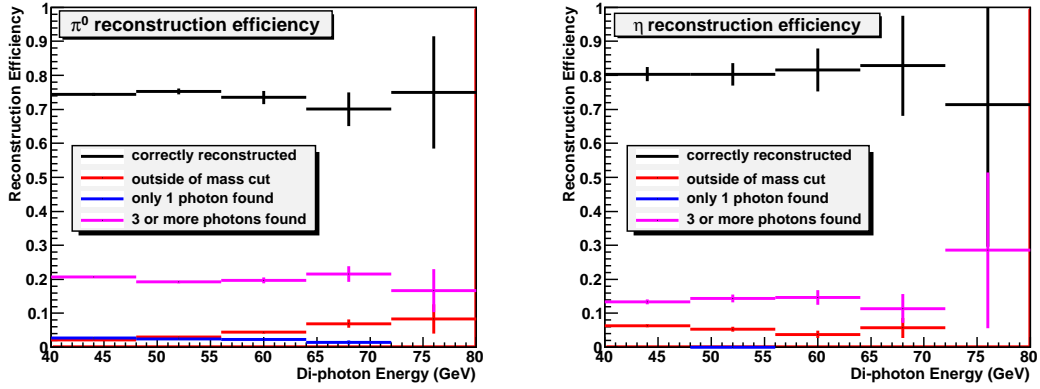}
\caption[Reconstruction efficiency of the FPD vs. thrown energy]{Reconstruction efficiency of the FPD vs. thrown energy for $\pi^0$ (LEFT) and $\eta$ (RIGHT) mesons that were accepted in the FPD. Correctly reconstructed means that two photons were found whose invariant mass satisfies the mass cut. ($0.08\text{ GeV} \sim 0.19\text{ GeV}$ for $\pi^0$ and $0.45\text{ GeV} \sim 0.65\text{ GeV}$ for $\eta$.) Also shown are the three main modes of reconstruction failure. }
\label{receff}
\end{figure}  

For both $\pi^0$ and $\eta$, the reconstruction efficiency is around 80 \% throughout the energy range, which is reasonably good. The most significant mode of failure is the one in which more than two photons were reconstructed, which accounts for about 20 \% of the total events for $\pi^0$'s and 15 \% for $\eta$'s. Since the result is based on the full simulation, the extra photon could be coming not only from the splitting of one real photon into two, but also from hadronic and electromagnetic backgrounds. We do not see significant merging of $\pi^0$ photons at high energy, nor do we see an alarming rate of single photon splitting. The latter would manifest itself as three photon events, especially for the $\eta$'s. The only potential exception is the very last energy bin for the $\eta$, where the significance is severely limited by poor statistics. (There are only four $\eta$ events in that bin) \\

\subsection{Total Efficiency Correction}
The total efficiency is defined as the ratio between the number of events that pass all our measurement cuts, and the number of $\pi^0$'s and $\eta$'s that were thrown in the kinematic region where we aim to make the measurement. In addition to the geometric acceptance and the reconstruction efficiency discussed so far, it also includes the correction from branching ratios, as well as the event cuts not included in the reconstruction efficiency. (For instance, the reconstructed position of the center of mass has to satisfy the center cut (equation \ref{eq:center})). The branching ratios into the $2\gamma$ channel are 98.8 \% and 39.3 \% for $\pi^0$ and $\eta$, respectively. Along with the limited geometrical acceptance, it is one of the main reasons for the reduction of $\eta$ efficiency relative to that of the $\pi^0$. The thrown particles are counted within a rectangle in pseudo-rapidity ($\mathsf{y}$) and azimuthal angle ($\phi$) space, defined by $-3.85 < \mathsf{y} < -3.45$ and $-0.2 < \phi < 0.2$.\\

\begin{figure} [t]
\centering
\includegraphics[width=1.\textwidth]{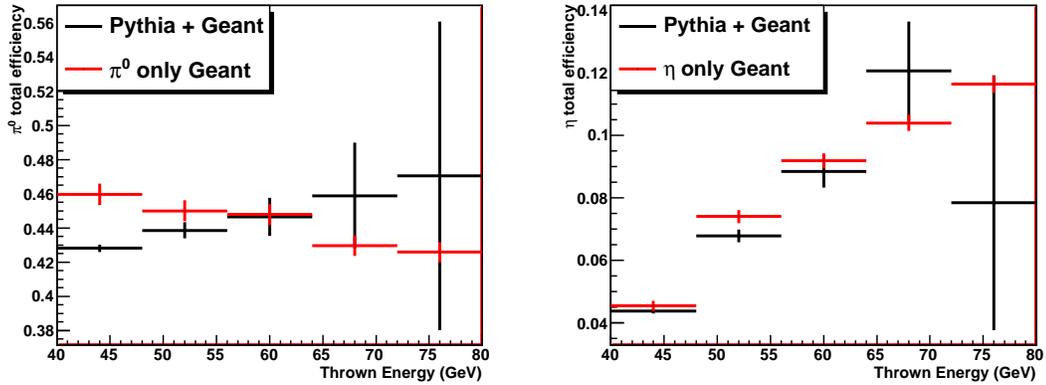}
\caption[Total efficiency correction, raw]{Total efficiency correction based on the full Pythia + Geant simulation, and single particle, flat distribution Geant simulation. LEFT: $\pi^0$, RIGHT: $\eta$.}
\label{flat}
\end{figure}  

Figure \ref{flat} shows the total efficiency correction for $\pi^0$ and $\eta$ from 40 GeV to 96 GeV. It is given as a function of thrown energy, and can be applied to the measured distribution once the effect of energy smearing is factored out and the ``true" energy distribution is obtained. Two different functions are shown, one from full Pythia + Geant simulation, and the other from single particle Geant simulations with flat energy and pseudo-rapidity distributions. The validity of the latter is clearly compromised by the lack of background, and the completely unrealistic shape of the cross-section. Normally, there would be no reason to look beyond the full simulation. However, as can be seen on the left-hand panel of figure \ref{flat}, the uncertainty in the $\eta$ efficiency is extremely large beyond 72 GeV. This region includes the last bin where we make a measurement, between 72 GeV and 80 GeV, which has only four $\eta$ events that pass all the cuts. Doubling the full simulation statistics, which would cost significant CPU time, would still only produce $\sim 30$ \% measurement. \\

On the other hand, if we compare the results from full simulation to that from single particle Geant for both $\pi^0$ and $\eta$, we see a noticeable agreement in the overall trend, with what seems to be a near constant offset across the energy range. The offset is likely caused by two main shortcomings of the single particle Geant simulation. Firstly, the lack of background would affect the reconstruction at some level. Secondly, the flat energy and pseudo-rapidity distributions mean that even with the same kinematic cut, the events are coming from different parts of the detector compared to the full simulation. On the other hand, the effect of energy bin migration is not a concern, as efficiency is calculated as a function of thrown energy. Ultimately, we make the assumption that the single particle Geant simulation contains information that is closely related to that in the full simulation. \\

\begin{figure} [t]
\centering
\includegraphics[width=.7\textwidth]{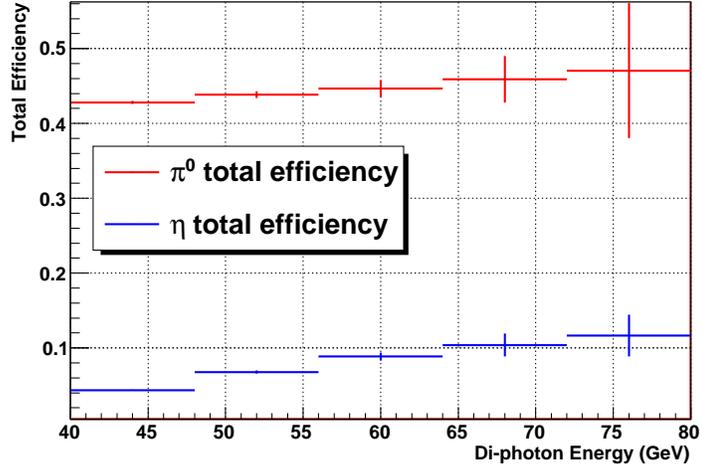}
\caption[Total efficiency correction, partially adjusted]{Total efficiency correction for $\pi^0$ and $\eta$ mesons based on the full Pythia + Geant simulation. The last bin for the $\eta$ has been corrected based on the $\eta$ only Geant simulation.}
\label{toteff}
\end{figure}  

Under this assumption, instead of using the nominal value calculated based on four events, we can estimate the $\eta$ efficiency for 72 $\sim$ 80 GeV bin based on the apparent trend found in the single particle Geant simulation. As the geometrical acceptance for $\eta$ improves with energy, it is reasonable to assume that the efficiency for the last bin should be no less than that for the previous bin. We use one standard deviation below the previous bin value as our minimum, which is equal to 0.068. At the same time, the single particle Geant simulation suggests that the rate of increase tapers off above 70 GeV. This leads to the assumption that the efficiency for the last bin should be no more than what we get by a linear extrapolation from the four previous points above 40 GeV. Accordingly, we set the maximum at 0.096. This gives us the estimated efficiency for the last bin as $0.082 \pm 0.014$. The final result for the total efficiency is shown in figure \ref{toteff}.\\

\section{Unfolding the Energy Smearing}
In order to measure the cross-section accurately, it is crucial to unfold the effect of energy bin migration. Due to the rapid fall of the cross-section as a function of energy, smearing can change the shape of the cross-section significantly. The details of the energy bin migration were discussed in chapter \ref{ch:shower} in the context of the energy dependent gain shift. Here, we focus on correcting the counting rates in energy bins to derive the true energy distribution from the observed one. \\

\subsection{Energy Smearing Matrix}

As described in chapter \ref{ch:calib}, we chose to generate the off-line gain correction factors so as to obtain the correct measurement of energy in true energy bins, independent of the shape of the energy distribution. This choice results in the apparent gain of the detector being on average about 2.5 \% too high in measured energy bins, due to bin migration. Simply shifting the average energy in each bin back down by this amount may not be sufficient, especially for the cross-section measurement. When the true shape of the smeared distribution varies rapidly, as is the case with our analysis, the effect of bin migration can change the shape of the distribution substantially. Consequently, it is necessary to unfold the effects of energy smearing to recover the true energy distribution.\\

In order to do so, we generate the energy smearing matrix, $\mathcal{S}$. $\mathcal{S}$ is an $N_{bin} \times N_{bin}$ square matrix, where $N_{bin}$ is the number of bins in which the cross-section is measured. Each column of $\mathcal{S}$ tells us the smearing probability of an event in a particular true energy bin into all of the apparent energy bins. We define vectors $\mid \tau \rangle$ and $\mid \alpha \rangle$, corresponding to the true and apparent energy distributions, respectively. The dimensions of these vectors are equal to $N_{bin}$. Then we have,
\begin{equation}
\mid \alpha \rangle = \mathcal{S} \mid \tau \rangle
\end{equation}

\begin{figure} [t]
\centering
\includegraphics[width=1.\textwidth]{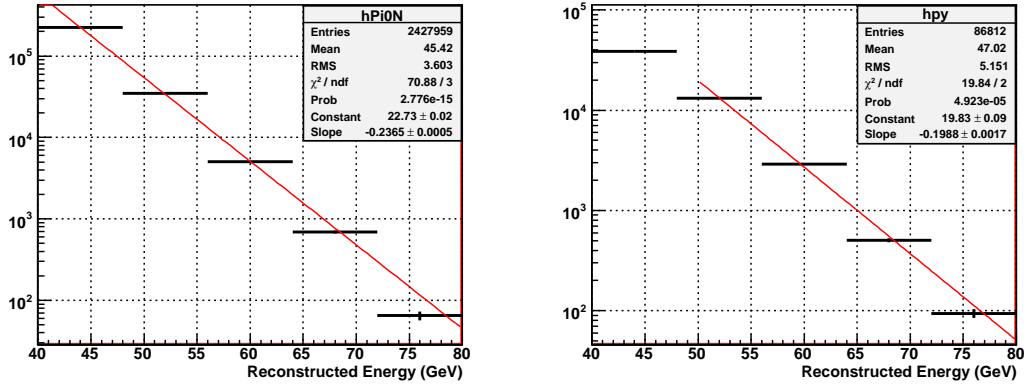}
\caption[$\pi^0$ reconstructed energy distribution]{$\pi^0$ reconstructed energy distribution. LEFT: data, RIGHT: Pythia + Geant simulation. The simulation has been weighted within each bin, but not overall. Due to the Pythia filter at 40 GeV, the first bin in the simulation does not get much contribution from previous bins, and therefore it is lower than it should be.}
\label{effEdist}
\end{figure}  

The idea is to obtain the matrix $\mathcal{S}$ in the full simulation, and apply it to the data to produce the true energy distribution. For the simulation-based matrix to be applicable to the data, there are two requirements. Firstly, the energy resolution must be well simulated throughout the energy range. Given the simulation and data comparisons discussed in previous chapters, this condition is likely satisfied. Secondly, the shape of the cross-section in the simulation within each true energy bin must match that of the data. We are not sensitive to the overall shape of the cross-section in the simulation, as each true energy bin does not affect the counting rate in any other true energy bin. However, the shape of the apparent energy distribution within a true energy bin should resemble that in the data, so as to correctly simulate the fraction of the events that would migrate up/down to the neighboring bins. The cross-section from our full simulation is generally harder in energy than that in the data, so weights were applied per bin to make it softer. (Applying an overall weight across the energy range is, as explained, unnecessary, and reduces statistics.) Figure \ref{effEdist} shows the comparison of the apparent energy distribution between data and full simulation. Also shown are exponential fits, which describe the slopes reasonably well but overshoot the highest energy bins. As discussed earlier, the observed distribution generally falls faster than an exponential at high energy. \\

\subsection{Calculation of the Smearing Matrix}

We obtain the matrix $\mathcal{S}$ in the following way. We first create a two dimensional energy distribution, where the X-axis corresponds to the true energy bins, and the Y-axis to the apparent energy bins, for the events that will be used to measure the cross-section. The bin size is chosen based on the magnitude of the smearing. While we use 5 GeV binning for the asymmetry measurement to capture the fine structure, this binning tends to be too small for the cross-section measurement. Too small a binning makes the smearing matrix overly off-diagonal, and we have chosen 8 GeV binning instead. The left-hand side of figure \ref{effmat} shows the 2-D energy distribution for $\pi^0$'s passing all the event cuts. We then divide each entry by the total number of events within the same column, and throw away the edge columns and rows to produce the matrix shown on the right-hand side of figure \ref{effmat}. This matrix tells us how events in each thrown energy bin are distributed in apparent energy, so it is our smearing matrix $\mathcal{S}$.\\

\begin{figure} [t]
\centering
\includegraphics[width=1.\textwidth]{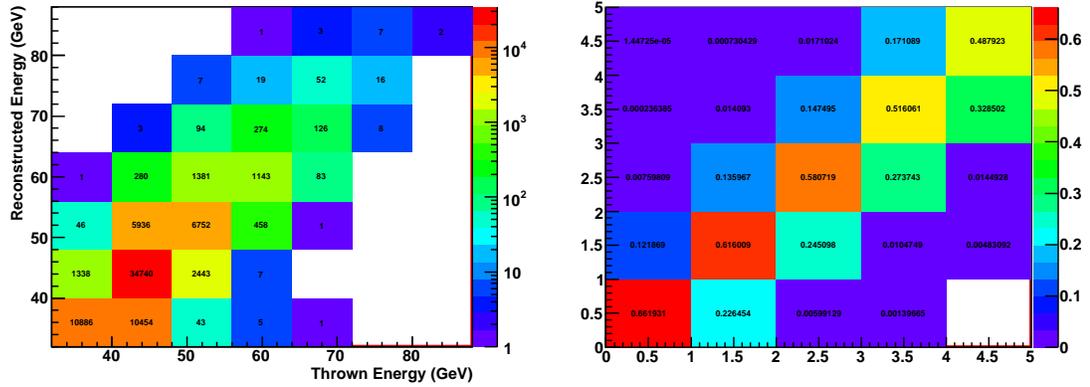}
\caption[Reconstructed vs. thrown energy distribution and smearing matrix $\mathcal{S}$ for $\pi^0$'s ]{LEFT: Reconstructed vs. thrown energy distribution for $\pi^0$'s. The Pythia level filter was set at 40 GeV, including hadronic energy. RIGHT: Smearing matrix $\mathcal{S}$ for $\pi^0$'s obtained from the energy distribution on the left. The binning is from 40 GeV to 80 GeV in 8 GeV interval, with a total of 5 bins.}
\label{effmat}
\end{figure}  

In chapter \ref{ch:calib}, we mentioned that the true energy bin based calibration produces the most diagonal and symmetric smearing matrix. This is not strictly true, however, as the matrix shown in figure \ref{effmat} has more strength below the diagonal than above. This is due to the shape of the cross-section within each bin, which falls rapidly with energy and has many more events in the lower bin boundary than in the upper bin boundary. \\

However, if we chose the alternative calibration scheme based on the energy measurements in apparent energy bins (which folds in the effect of bin migration), the matrix $\mathcal{S}$ would become even more off-diagonal and asymmetrical. If we normalized $\mathcal{S}$ by rows instead of columns, which would make $\mathcal{S}$ something like a reverse-smearing matrix, then such a calibration would make $\mathcal{S}$ more diagonal and symmetric. \\

\begin{figure} [t]
\centering
\includegraphics[width=1.\textwidth]{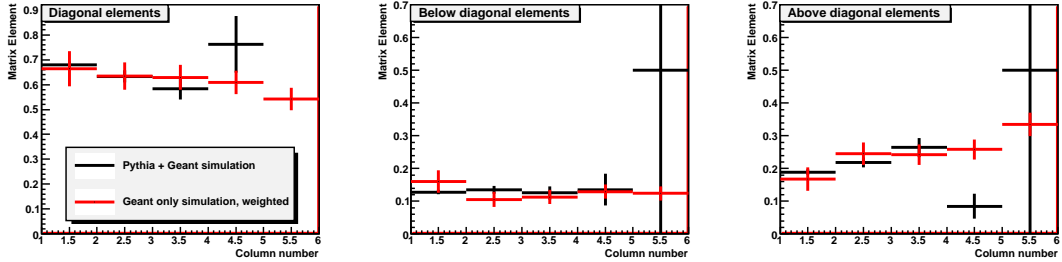}
\caption[Diagonal and off-diagonal elements for the smearing matrix]{Diagonal and off-diagonal elements for the smearing matrix, BLACK: full Pythia + Geant simulation, RED: weighted Geant-only simulation. From left, diagonal elements, below diagonal elements ($E_{recon} > E_{true}$), and above diagonal elements ($E_{recon} < E_{true}$). The X-axis is the column number for the $7 \times 7$ smearing matrix, counting from 0. Only the central 5 rows and columns are used for the analysis.}
\label{matcompare}
\end{figure}  

The main problem in calculating the smearing matrix is the limited simulation statistics at high energy, especially for the $\eta$. This is the same problem that was discussed in section \ref{sec:eff} about the efficiency correction, and here we again take a similar approach, and utilize the Geant-only $\eta$ simulation. The Geant only simulation has been weighted in true energy and pseudo-rapidity, based on the true distributions in the full simulation. The advantage of this method is that the energy weighting needs to be done only within a true energy bin, as there is no cross-talking between true energy bins. The resulting energy distribution resembles a saw tooth, but it is perfectly usable for our purposes. \\

Figure \ref{matcompare} shows the comparison between the full and the Geant only simulation for the diagonal and the two off-diagonal matrix elements. For the last two bins, there are too few events in the full simulation to really determine what the matrix should look like. On the other hand, we note that for the first three bins, which go up to 64 GeV, the agreement between the two simulations is very good. Therefore, we take the point of view that the Geant only simulation, when weighted properly, is a good surrogate for the full simulation. For the $\eta$, then, we construct the smearing matrix by combining the first three columns from the full simulation with the last two columns from the Geant only simulation. The additional systematics that could come from this assumption are handled in a later section with all other systematics. For $\pi^0$'s, we only use the full simulation.\\ 

\subsection{Unfolding Scheme}

Now that we have $\mathcal{S}$, the obvious approach is to invert $\mathcal{S}$ and apply it to the measured energy distribution in the data to get the true energy distribution.  
\begin{equation}
\mid \tau \rangle = \mathcal{S}^{-1} \mid \alpha \rangle
\end{equation}
The problem with this method is that the matrix itself has uncertainties associated with the limited simulation statistics, and any variation within these uncertainties leads to a variation in the true energy distribution. In other words, there are a range of true energy distributions that are consistent with the measured energy distribution, given our limited understanding of the smearing. While most of these distributions are qualitatively similar to each other, they are often not as smoothly varying as the apparent distribution. \\

\begin{figure} [b]
\centering
\includegraphics[width=.8\textwidth]{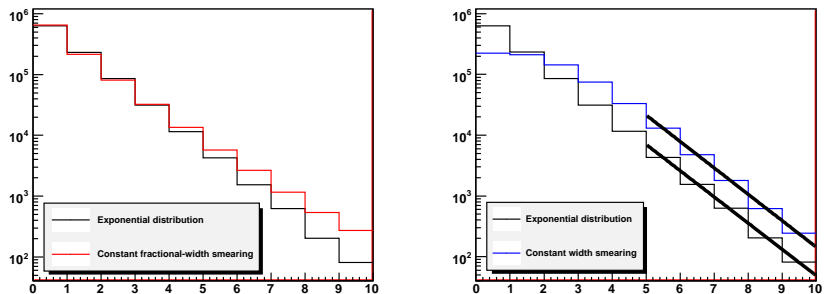}
\caption[Toy smearing of exponential distribution]{LEFT: Randomly generated exponential distributions with a constant 30 \% Gaussian smearing. RIGHT: Exponential distributions with a constant 1.5 units in X Gaussian smearing. Also shown are exponential fits to the original and smeared distributions.}
\label{effsmear}
\end{figure} 

Instead of directly applying $\mathcal{S}^{-1}$, we choose the opposite approach. Given the form of the apparent distribution shown in figure \ref{effEdist}, we assume that the form of the true energy distribution is a smoothly varying function that mostly behaves like an exponential, but falls off slightly faster at high energy. The reason for this choice is the following. If the underlying true distribution was a pure exponential, then a smearing effect that is fractionally constant across the energy range results in an apparent distribution that falls slower than an exponential as the energy increases. The smearing we have is fractionally constant, as seen in figure \ref{MeanSigma}, and as evidenced by the decrease of strength in diagonal elements at high energy in the smearing matrix shown in figure \ref{effmat}. It is when the width of the smearing is constant in energy that we find the apparent distribution to be also an exponential with the same exponential coefficient. This rather obvious point, which is related to the discussion in chapter \ref{ch:shower} regarding the energy dependent correction, is illustrated in figure \ref{effsmear}. Since the apparent energy distribution in data falls faster than an exponential at high energy (figure \ref{effEdist}), given the fractionally constant smearing, the underlying distribution has to fall even faster as the energy increases. The following purely empirical function was found to be sufficient for the job. 
\begin{equation}
f(E)=e^{a + bE + cE^2 + dE^3}
\end{equation}

The goal is not to constrain the true distribution according to a specific form of the function, but to merely impose a smoothness requirement. Parameters $c$ and $d$ are expected to be very small, providing corrections for the highest energy bins. We obtain an estimated true energy vector $\mid \tau_i \rangle$, corresponding to a particular parametrization $P_i$ of this function, by integrating the function for each bin. $\mid \tau_i \rangle$ can be transformed to an estimated apparent energy vector $\mid \alpha_i \rangle$ by multiplying by the smearing matrix $\mathcal{S}$. Therefore, by letting the parameters vary (scanning through the index $i$), we can fit this function via $\mid \alpha_i \rangle$ against the measured energy vector in data, $\mid \alpha \rangle$. By doing so, we select the $\mid \tau_i \rangle$ that has the assumed functional form, and when smeared to $\mid \alpha_i \rangle$, is the most consistent with $\mid \alpha \rangle$. This particular $\mid \tau_i \rangle$ is our best estimate of the true underlying energy distribution. \\ 

\begin{figure} [t]
\centering
\includegraphics[width=1.\textwidth]{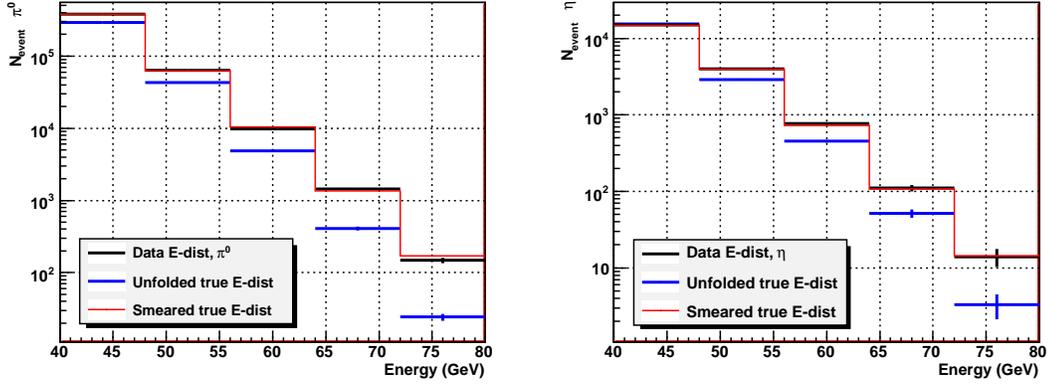}
\caption[Unfolded true energy distribution]{Measured energy distribution $\mid \alpha \rangle$, unfolded true energy distribution $\mid \tau_i \rangle$, and the apparent energy distribution $\mid \alpha_i \rangle$ derived from $\mid \tau_i \rangle$. LEFT: For $\pi^0$, RIGHT: for $\eta$.}
\label{effEdist2}
\end{figure}  

Figure \ref{effEdist2} shows the result of the fitting, for both $\pi^0$ and $\eta$. The blue points are the estimated true energy distribution, $\mid \tau_i \rangle$. The error bars on the blue points are statistical only, under the assumption that the functional form that we chose, and the smearing matrix $\mathcal{S}$, are both correct. The red points are $\mid \alpha_i \rangle \sim \mathcal{S} \mid \tau_i \rangle$. The reason that the equality does not hold is because the very first bin in $\mid \alpha_i \rangle$ requires a correction that is outside the scope of the matrix $\mathcal{S}$. $\mathcal{S}$ does not tell us how many events from the true energy bins that are outside of the matrix roll into the bins that are included. Given that the binning is 8 GeV wide, and that the width of the smearing is smaller in GeV at lower energy, most of the effect should be confined to the very first bin only. For the $\pi^0$'s, we can estimate the fraction of events in bin 0 (32 GeV to 40 GeV, which is below Pythia filter threshold) migrating into bin 1 based on the extrapolation of the true distribution function, and from the off diagonal elements in $\mathcal{S}$. For the $\eta$'s, however, there is an additional uncertainty. Around 40 GeV, the $\eta$ acceptance varies rapidly as a function of energy, so the actual number of events that can feed into the higher energy bins is significantly reduced from the expected cross-section at low energy. The resulting uncertainty is folded into the systematic error.\\

\subsection{Statistical and Systematic Uncertainties}

The statistical uncertainty is well defined in the observed energy bins, but it needs to be translated into the true energy space. Instead of using explicit transformation, we calculated it by putting random Gaussian variations into the measured distribution $\mid \alpha \rangle$ according to its statistical uncertainty. For each of those trial, we repeated the fitting process to obtain an ensemble of $\mid \tau_i \rangle$'s. The uncertainty for each bin was determined based on the degree of the variation within this ensemble.\\

The systematic uncertainty in the unfolding process comes largely from the errors in the matrix $\mathcal{S}$. Similar to the above described method, we estimate the effects of the matrix uncertainty by randomly varying each elements, and for each altered version of the matrix, repeating the fitting process. For the $\eta$ matrix, the uncertainty for the last two bins that come from the Geant only simulation are manually increased to allow for a more conservative estimation. The uncertainty due to the end treatment is handled in the same way, by introducing random variation to the expected number of feed-downs from bin 0 within a reasonable limit.  \\

\end{doublespace}
\chapter{Physics Results}\label{ch:result}
\setlength{\parindent}{1cm}
\begin{doublespace}

\section{Introduction}
In this chapter, we present four physics results. The first is the absolute cross-section, based on the previously determined integrated luminosity measurement. \cite{STARAN} The second is the cross-section ratio between $\pi^0$ and $\eta$. The third is the analyzing power as a function of di-photon invariant mass. Lastly, we present the analyzing power as a function of $x_F$ for $\pi^0$ and $\eta$, including background corrections.\\

\section{Cross-Sections for Forward $\pi^0$ and $\eta$ Production at High $x_F$}

\begin{figure} [t]
\centering
\includegraphics[width=1.\textwidth]{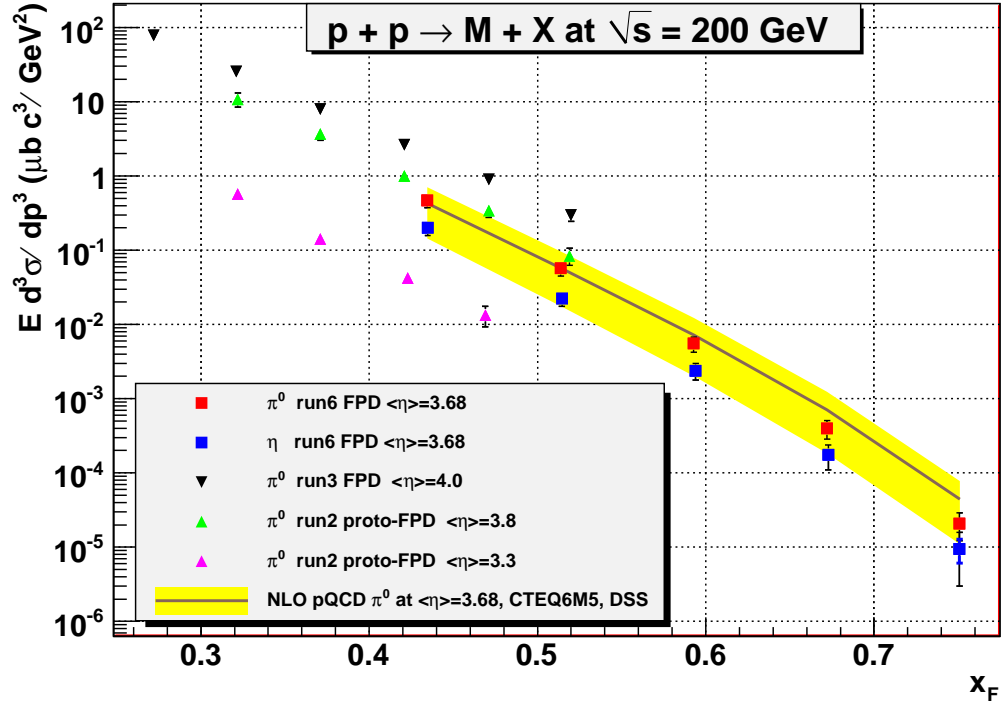}
\caption[Absolute cross-section for $\pi^0$ and $\eta$]{Absolute cross-section for $\pi^0$ and $\eta$ at average pseudo-rapidity of 3.7, with center cut. $p + p \rightarrow M + X$  at $\sqrt{s}=200$ GeV. Also shown are previously published results in similar kinematic regions. The theory prediction was obtained by interpolation. Black error bars indicate the combined statistical and systematic uncertainties.}
\label{Xsec}
\end{figure}  	

Figure \ref{Xsec} shows the absolute cross-section measurement for $\pi^0$ and $\eta$. The center cut (equation \ref{eq:center}) was imposed for both mesons. Also shown are the previously published results from RHIC runs 2 and 3, for three pseudo-rapidity points. Of the three, only the result at pseudo-rapidity of 4.0 was measured by the FPD in run 3. The other two points were based on the data taken by the prototype FPD, which utilized a substantially different detector technology. Finally, the next-to-leading-order (NLO) pQCD theory prediction is shown in yellow band, \cite{vogel} based on CTEQ6M5 parton distribution function \cite{CTEQ} and DSS fragmentation function \cite{DSS}. The uncertainty for the theory prediction indicates how much the result changes when the factorization/renormalization scale is increased from $\mu=p_T$ to $\mu=2p_T$. We note that the DSS fragmentation function includes in its calculation the published STAR data points at pseudo-rapidity of 3.3 and 3.8, along with other RHIC results. \cite{DSS}\\

\begin{table} [t]
\centering
\begin{tabular}{ | c || c | c | c | c | c |}
    \hline
     $\pi^0$ & 40-48 GeV & 48-56 GeV & 56-64 GeV & 64-72 GeV & 72-80 GeV \\ \hline  
    \hline
    $N_{raw}$ & 378739 & 63443 & 9813 & 1435 & 148 \\ \hline
    $N_{unfolded}$ & 290431 & 42852 & 4862 & 408 & 24 \\ \hline
    Efficiency & 0.4281 & 0.4386 & 0.4466 & 0.4589 & 0.4706 \\ \hline 
    Eff. Error & 0.48 \% & 1.06 \% & 2.49 \% & 6.77 \% & 19.2 \% \\ \hline 
    Unfld. Error & 1.04 \% & 1.13 \% & 3.11 \% & 7.66 \% & 13.9 \% \\ \hline 
    Calib. Error & 1.8 \% & 5.7 \% & 10.3 \% & 16.3 \% & 24.2 \% \\ \hline
    \hline
    \hline
     $\eta$ & 40-48 GeV & 48-56 GeV & 56-64 GeV & 64-72 GeV & 72-80 GeV \\ \hline  
    \hline
    $N_{raw}$ & 15120 & 3998 & 766 & 111 & 14 \\ \hline
    $N_{unfolded}$ & 15475 & 2883 & 452 & 51 & 3 \\ \hline
    Efficiency & 0.0437 & 0.0678 & 0.0885 & 0.1040 & 0.1163 \\ \hline 
    Eff. Error & 1.74 \% & 2.99 \% & 5.91 \% & 10.40 \% & 16.74 \% \\ \hline 
    Background & 0.8173 & 0.8989 & 0.9186 & 0.9196 & 0.9258 \\ \hline 
    Bg. Error & 0.52 \% & 0.21 \% & 0.35 \% & 0.59 \% & 1.5 \%  \\ \hline 
    Unfld. Error & 4.35 \% & 2.5 \% & 6.3 \% & 17.7 \% & 48.4 \% \\ \hline 
    Calib. Error & 1.48 \% & 4.58 \% & 7.85 \% & 11.3 \% & 15.2 \% \\ 
    \hline
\end{tabular}
\caption[$\eta$ correction factors and systematic uncertainties]{Correction factors and systematic uncertainties. The ``background correction" refers not to the actual signal-to-background ratio, but the ratio of the background estimates in data and simulation. In the case of $\eta$, this factor is applied to the efficiency correction that already includes the simulation estimate (figure \ref{sig}). There is no additional background correction for the $\pi^0$, since the background estimate in the simulation is nearly identical to the estimate in the data }
\label{etacorr}
\end{table}

The error bars include both statistical and systematic uncertainties. The statistical uncertainties are indicated by the colored error bars, which are visible only in the very last bins. The sources of the systematic errors are the following. Firstly, there is the absolute calibration uncertainty of 3 \%, which is the dominant systematics for the $\pi^0$. Secondly, there is the uncertainty from the efficiency correction, due to the limited simulation statistics. Thirdly, we have the uncertainty from background estimation, which mostly affects the low energy $\eta$'s. Fourthly, there is the uncertainty from unfolding the energy smearing, which dominates the $\eta$ systematics at high energies. Finally, there is also a conservatively estimated normalization uncertainty of 20 \%. The total integrated luminosity for this run period has been previously measured at 6.8 $pb^{-1}$ \cite{STARAN}. Table \ref{etacorr} lists correction factors and systematic uncertainties, along with the original and unfolded number of events in each bin.\\

\begin{figure} [t]
\centering
\includegraphics[width=1.\textwidth]{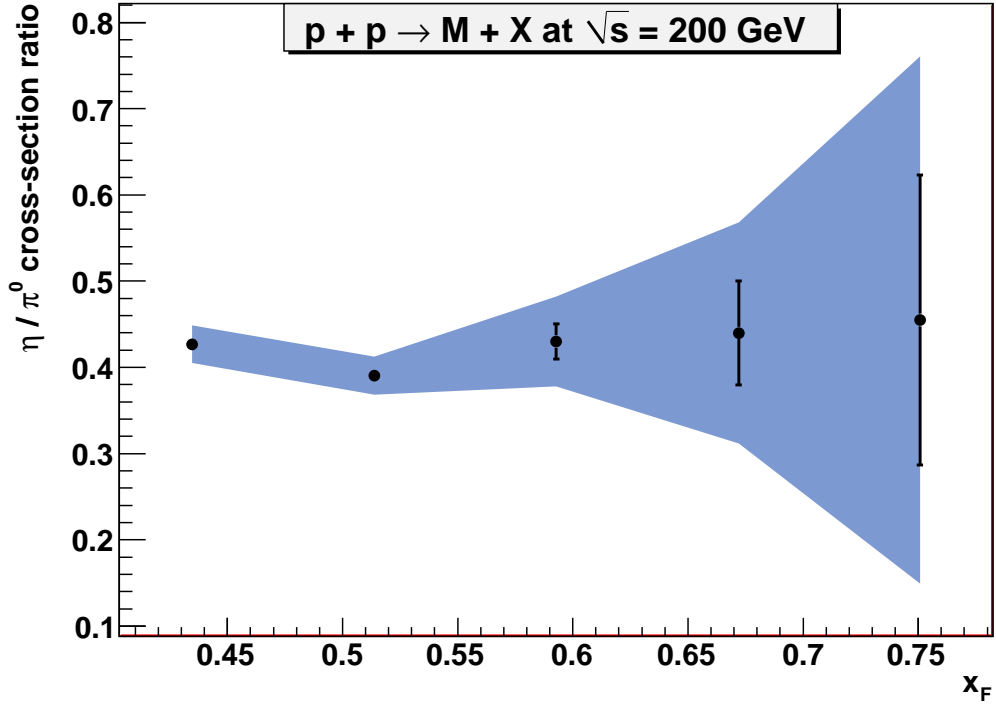}
\caption[$\eta$ to $\pi^0$ cross-section ratio]{$\eta$ to $\pi^0$ cross-section ratio at average pseudo-rapidity of 3.7, with center cut. $p + p \rightarrow M + X$  at $\sqrt{s}=200$ GeV. Error bars are for statistical uncertainties only. The error band indicates the combined statistical and systematic uncertainties.}
\label{XsecR}
\end{figure}  	

Figure \ref{XsecR} shows the $\eta$ to $\pi^0$ cross-section ratio. The error bars indicate the statistical uncertainty, while the blue error band includes both statistical and systematic uncertainties. The sources of systematic uncertainty are largely identical to those in the absolute cross-section measurement. The calibration uncertainty is reduced to 1.5 \%, as we only need to know the relative calibration, which is much better understood. The normalization uncertainty is not included, as the relative cross-section does not require absolute normalization. The $\eta$ fragmentation function in this kinematic region has not been measured, and we do not have a theory prediction at this point. \\

\section{Transverse Single Spin Asymmetries for Forward $\pi^0$ and $\eta$ Production}

\begin{figure} [b]
\centering
\includegraphics[width=1.\textwidth]{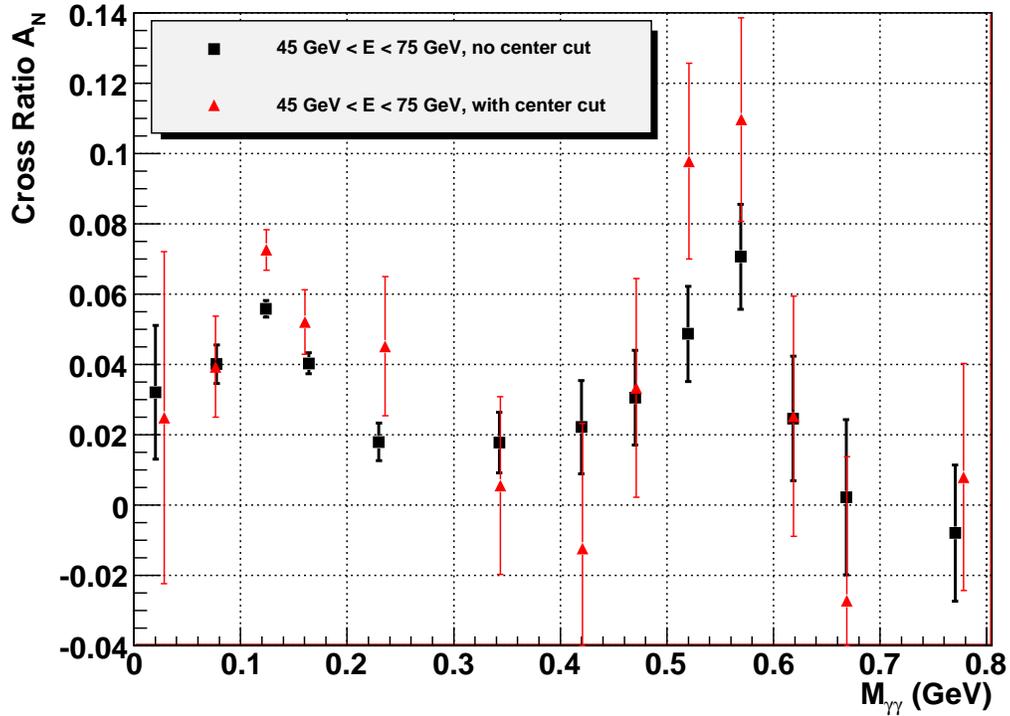}
\caption[$A_N$ vs $M_{\gamma\gamma}$]{Analyzing power ($A_N$) vs $M_{\gamma\gamma}$ at average pseudo-rapidity of 3.7. 56 \% beam polarization. $p^{\uparrow} + p \rightarrow M + X$  at $\sqrt{s}=200$. Center cut refers to equation \ref{eq:center}. The error bars are for the statistical uncertainty only.}
\label{ANM}
\end{figure}

Figure \ref{ANM} shows the analyzing power ($A_N$) as a function of the di-photon invariant mass ($M_{\gamma\gamma}$) for energy greater than 45 GeV. Two sets of data points are shown, with and without the center cut. The polarization of the beam was 56 $\pm$ 2.6 \%. \cite{polar} We see a clear suppression of $A_N$ in the continuum region, while the shape of the two mass peaks is recognizable in the structure of the asymmetry. Based on this result, we estimate the analyzing power of the continuum region to be around 0.02. Since we expect most of the events in this region to be related to $\pi^0$, it is conceivable that the asymmetry of the continuum may also grow with the energy. On the other hand, for those events to appear under the high energy $\eta$ mass peak, a substantial overestimation of energy is necessary. Combined with the kinematic limit, it is likely that the background under $\eta$ at high energy is related to low energy $\pi^0$'s, and therefore should have a very small analyzing power.\\ 

For the asymmetry measurement, the unfolding of the energy smearing has not been applied. This is due to the limited statistics and the relatively large smearing, which makes the unfolding difficult when splitting the data set into spin states and using 5 GeV binning. It is also noted that because the spin asymmetry varies only mildly with the energy (compared to the cross-section), it is not entirely necessary to unfold the smearing. \\

\begin{figure} [t]
\centering
\includegraphics[width=1.\textwidth]{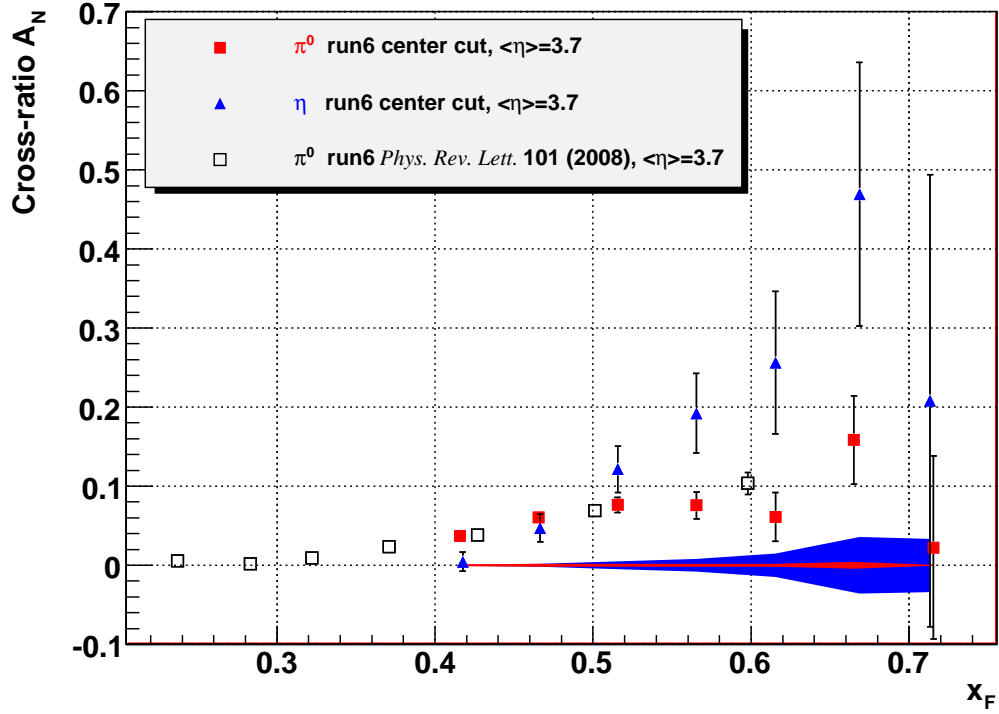}
\caption[$A_N$ vs $x_F$]{Analyzing power ($A_N$) vs $x_F$ at average pseudo-rapidity of 3.7 for $\pi^0$ and $\eta$. 56 \% beam polarization. $p^{\uparrow} + p \rightarrow M + X$  at $\sqrt{s}=200$. Also shown are the previously published results for the $\pi^0$ at lower $x_F$ based on the same data set, but without the center cut. The error bars are for the statistical uncertainty only. The red error band indicates the systematic uncertainty for the $\pi^0$, and the blue error band indicates the systematic uncertainty for the $\eta$.}
\label{ANxF}
\end{figure}  	

Finally, figure \ref{ANxF} shows the analyzing power as a function of $x_F$ for $\pi^0$ and $\eta$. Also shown are the published result for $\pi^0$ $A_N$ at lower $x_F$, which is based on the same data set as the current analysis. The previous result does not include the center cut. The background correction, which only significantly affects the $\eta$ asymmetry measurement, is done based on the assumed analyzing power of 0.02 for the background. Table \ref{ANBGcorr} lists the estimated signal fraction for each bin, along with the uncertainties. The error bars indicate statistical errors only, while the error bands indicate the systematic errors. The main source of the systematic uncertainty is the background correction. The polarization uncertainty plays a negligible role. \\

\begin{table}
\centering
\begin{tabular}{ | c || c | c | c | c | c | c | c |}
    \hline
      & 40 GeV & 45 GeV & 50 GeV & 55 GeV & 60 GeV & 65 GeV & 70 GeV \\ \hline  
    \hline
    $\pi^0$ & 0.944 & 0.955 & 0.96 & 0.967 & 0.972 & 0.967 & 0.955 \\ \hline
    $error$ & 0.32 \% & 0.14 \% & 0.19 \% & 0.28 \% & 0.42 \% & 0.78 \% & 1.7 \% \\ \hline
    \hline	
    $\eta$  & 0.506 & 0.641 & 0.7 & 0.778 & 0.786 & 0.851 & 0.737 \\ \hline
    $error$ & 9.8 \% & 2.6 \% & 2.9 \% & 3.3 \% & 5.2 \% & 7.1 \% & 15.9 \% \\ 
    \hline
\end{tabular}
\caption[Background corrections for $\pi^0$ and $\eta$]{Background corrections for $\pi^0$ and $\eta$ and associated uncertainties, from figure \ref{sig}.}
\label{ANBGcorr}
\end{table}

For all points between $x_F$ of 0.5 and 0.7, the $\eta$ meson exhibits an analyzing power that is significantly larger than that of the $\pi^0$. Furthermore, the size of the $\eta$ asymmetry may be larger than any transverse spin effects that have been reported at this energy. Currently, there are no definitive theoretical models that can explain the difference between the two neutral mesons, or the potentially very large $\eta$ asymmetry. \\

We emphasize that the $\pi^0$ cross-section measurement is consistent with the pQCD prediction in the same kinematic region in which the spin asymmetry measurement was made. Although there are no theoretical predictions for the $\eta$ cross-section in this kinematic region yet, we also note that the $\eta$-$\pi^0$ cross-section ratio is similar to what has been recently observed by PHENIX in mid-rapidity. \cite{phenix} \cite{phenix2} \\

As there had been a decades old concern regarding whether the transverse spin effects are in the realm of applicability of pQCD, as discussed in chapter \ref{ch:AN}, it is important to acknowledge the ability of the pQCD calculations to predict the unpolarized cross-section reliably. In addition, many pQCD based theoretical models have already been developed to describe the observed $\pi^0$ asymmetry. Nevertheless, it is currently unclear as to how these models can be extended to explain the difference between $\pi^0$ and $\eta$ asymmetries. Given the success of the pQCD framework in describing both the unpolarized cross-section and the $\pi^0$ asymmetry, the problems posed by the $\eta$ asymmetry should provide a fertile ground for future theoretical development.\\

\end{doublespace}
\chapter{Summary}\label{ch:summary}
\setlength{\parindent}{1cm}
\begin{doublespace}

In the past, there had been a considerable difficulty in understanding the large transverse spin effects within the framework of perturbative QCD. In fact, it was often true that where a large spin effect was observed, the unpolarized cross-section deviated significantly from the pQCD prediction. All of these results were at relatively low center of mass energy ($\sqrt{s}$), supporting the idea that the transverse spin effects were likely outside the regime in which pQCD was believed to be applicable.\\

However, we have seen that $A_N$ for forward $\pi^0$ production continues to be large at RHIC energy, where we can rely on pQCD to describe the unpolarized cross-section. There has been significant developments in the theory sector as well, which have produced multiple pQCD based models that turned out to be successful in describing most, if not all, aspects of the observed $\pi^0$ asymmetry. A related topic that was not covered in this thesis is the $p_T$ (transverse momentum) dependence of $A_N$, which remains an aspect of $\pi^0$ asymmetry that is yet to be fully understood. \cite{STARAN} \\

In order to make the asymmetry and cross-section measurement at higher energies than previously attempted with the FPD, we have improved upon a number of analysis procedures. We discovered that much of the previously observed discrepancy between simulation and data could be traced back to the inadequate simulation of the electromagnetic shower in the FPD. The charged particle energy loss scheme, which had been used for all past FPD analyses, significantly underestimates the energy smearing as well as lacking the energy dependent gain shift observed in data. Furthermore, the resulting shower shape differed in important ways from what is observed in data. We found dramatic improvements in all of these areas once we switched to the \v{C}erenkov effect based shower simulation that takes into account optical physics inside the Pb glass columns of the FPD. We have also implemented an incident angle dependent shower function, which reproduces the shower shape in data much more faithfully than the previous version based on charged particle energy loss. \\

Based on the improved shower simulation, we were able to generate off-line correction factors that are estimated to constrain the absolute calibration within 3 \%, and the relative calibration within 1.5 \%. While continuing with the $\pi^0$ mass based, iterative calibration procedure, we have found that the dependences of the invariant mass on a number of variables, such as the number of clusters, two photon separation, and the azimuthal decay angle, are now much better simulated. All of these factors are potentially problematic to the calibration process when left uncorrected. Our increased confidence in calibration at high energies is largely due to our improved ability to simulate these pathologies.\\

Another area of significant improvement is the $\pi^0 - \gamma$ separation. Initially suspected to be a considerable source of background to high energy $\pi^0$'s, the mis-reconstruction of a single photon into a $\pi^0$ event is now significantly more suppressed up to 70 GeV and above. This is largely due to the newly implemented, logarithm of energy weighted cluster width calculation, which allows for a surprisingly accurate distinction between one and two photon clusters at high energy. \\

As mentioned before, the energy smearing in data is now understood to be much larger than our original estimation. This finding necessitated a procedure to unfold the effects of energy smearing for the cross-section measurement. We implemented a relatively simple, smearing matrix based unfolding scheme that relies on a smoothness requirement to recover the true energy distribution from the apparent one. \\ 

Based on these improvements, we reported the first measurement of $A_N$ for forward $\eta$-meson production at RHIC energy, along with the cross-section measurements for $\pi^0$ and $\eta$ for $x_F>0.4$. Our result suggests that the $\eta$ asymmetry is significantly larger than that of the $\pi^0$ around a pseudo-rapidity of 3.7. At the same time, the $\pi^0$ cross-section is found to be consistent with the NLO pQCD prediction. Furthermore, while lacking a direct theoretical prediction, the $\eta$ to $\pi^0$ cross-section ratio is consistent with the results from other experiments. \cite{phenix} \cite{phenix2} These findings strongly suggest that the $\eta$ asymmetry can be understood within the pQCD framework, similar to $\pi^0$. However, currently no theoretical model exists that can satisfactorily explain either the difference between the $\pi^0$ and $\eta$ asymmetries, or the very large magnitude of the $\eta$ asymmetry.  \\

\end{doublespace}
%\include{Chapter-8/Chapter-8}
%%%%%%%%%%%%%%%%%%%%%%%%%%%%%%%%%%%%%%%%%%%%%%%%%%%%%%%%%%%%%%%
% Appendices
%
% Because of a quirk in LaTeX (see p. 48 of The LaTeX
% Companion, 2e), you cannot use \include along with
% \addtocontents if you want things to appear the proper
% sequence. Since the PSU Grad School requires 
%%%%%%%%%%%%%%%%%%%%%%%%%%%%%%%%%%%%%%%%%%%%%%%%%%%%%%%%%%%%%%%
\appendix
\Appendix{FMS Inner Calorimeter High Voltage System} \label{ch:CW}
\setlength{\parindent}{1cm}
\begin{doublespace}

\section{\sloppy Introduction}
Since RHIC run6, the west FPD has received significant upgrades, leading eventually to the commissioning of the Forward Meson Spectrometer (FMS) during RHIC run 8. The FMS is a Pb glass calorimeter employing a similar detector technology to the FPD, but it provides a vastly improved coverage both in azimuthal angle and pseudo-rapidity. The FMS consists of 1264 Pb glass cells (The FPD has 98), covering the full azimuth between a pseudo-rapidity of 2.5 and 4.0. Along with the existing Barrel and End-cap calorimeters, STAR now has an almost complete electromagnetic coverage in pseudo-rapidity from 1.0 to 4.0.\\

As a part of the effort to construct the FMS, the Penn State group has designed and built Cockroft-Walton type photo-multiplier tube (PMT) bases for the inner calorimeter of the FMS. Our design utilizes two Cockroft-Walton voltage multipliers for each base, and employs $I^2C$ serial communication protocol for the on-board intelligence. \cite{I2C}\\

The base design can be divided conceptually into two sections, one producing the high voltage, and the other providing the control and read-back circuits. In the first part of this chapter, we will study the load response of the Cockroft-Walton multiplier by introducing a simple theoretical model. Results from SPICE simulation will be presented and compared to the model predictions. In the second part of this chapter, we will discuss selected elements of the base design that form the on-board control circuitry. Finally, we will present a design overview of the controller boards, which integrate two heterogeneous systems (Penn State and Yale) to form the complete inner calorimeter high voltage system.\\   

\section{\sloppy Cockroft-Walton Voltage Multiplier}
A common choice for a PMT base is a resistor divider chain. It is extremely simple, yet can produce the necessary voltage steps required by a PMT. However, it has a number of disadvantages that make it less suitable for applications with limited space and heat budget. The FMS was built in a metal enclosure with very limited space, and the heat management is important especially for the tubes near the center of the enclosure. \\

The main shortfall of the resistor divider is that the structure of its output impedance (the ratio of the output voltage to the load current at a particular frequency) is nearly the opposite of the characteristic load structure of a PMT. During normal operations, the amount of current drawn from the PMT photo-cathode is negligible. But this is where the output impedance of the resistor divider is the lowest, (voltage drop is the lowest per unit current drawn) as it is usually connected directly to the high voltage supply. As the signal gets amplified through the dynode stages, the current increases exponentially, leading to roughly six to seven orders of magnitude amplification at the anode for a typical PMT. However, this is the highest output impedance point (often on the order of mega-ohms) of a resistor divider, because the current has to go through all the resistors in the chain. In short, where it can provide the most current is where the PMT needs the least, and where the PMT needs the most current is where it can provide the least. \\

A good photo-tube base should minimize the rate dependence of the PMT response. In other words, it should suppress the variation of the number of photo-electrons per unit energy due to the varying event rate. In case of a resistor divider, the rate dependence is dominated by the output impedance at the anode end, which is equal to the total resistance of the chain. But decreasing the resistance to improve the rate dependence in turn increases the quiescent current, which results in increased power consumption and heat generation. The total resistance of the chain then is chosen to provide the best compromise between these two competing effects. In reality, however, the rate independence requirement tends to drive the quiescent current to a level that is significantly higher than what can be achieved by a well designed Cockroft-Walton system. \\

\subsection{\sloppy Cockroft-Walton PMT Base}
A Cockroft-Walton (CW) voltage multiplier is a multi-stage charge pump that amplifies a low voltage AC input into a high voltage DC output. In its most basic form, each ``stage" is made up of two capacitors and two diodes. In figure \ref{CWscheme}, the first stage consists of capacitors C0 and C1, and diodes D0 and D1. The amplification factor is equal to the total number of stages, up to corrections some of which are driven by current load. \\

\begin{figure} [t]
\centering
\includegraphics[width=1.\textwidth]{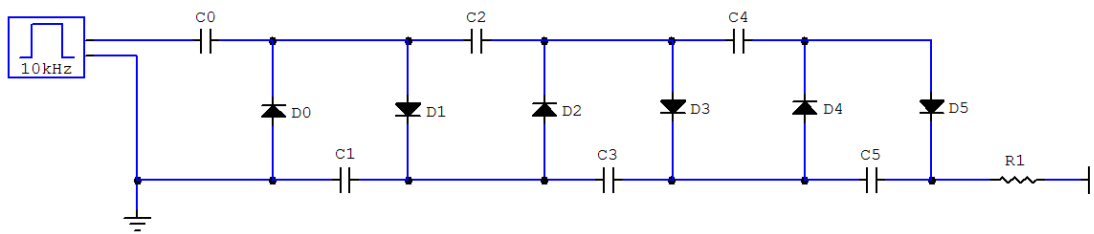}
\caption[Schematic of a simple three stage, half-wave Cockroft-Walton voltage multiplier chain]{Schematic of a simple three stage, half-wave Cockroft-Walton voltage multiplier chain. A load is represented by a resistive pathway to ground connected at the top of the chain.}
\label{CWscheme}
\end{figure}  	

A major advantage of the CW multiplier is that its output impedance structure is very similar to the load structure of a PMT. Because the charge is pumped from the anode up to the cathode, the output impedance is the lowest at the anode, and rapidly increases as the stage number increases. This pattern is well matched by a typical PMT, and allows for a much more efficient power and heat management compared to a resistor divider with a similar level of rate dependence. Furthermore, the CW multiplier itself has no quiescent current. If the PMT is not producing signals, the CW chain does not draw current. While the circuit necessary to drive the chain does have constant current load, it is on the low voltage lines. This allows for a design that is power efficient, while reducing the risk associated with high voltage lines. Finally, the on-board generation of the high voltage eliminates the need for external high voltage cables. Some examples of CW type photo-tube bases can be found in \cite{cw} \cite{cw1} \cite{cw2}\\

\subsection{Theoretical Model}

Figure \ref{CWscheme2} depicts a schematic of a 3-stage, half-wave CW chain in operation. The two panels correspond to the two half cycles of the driving pulse, which has peak to peak amplitude of $V_{input}$ and frequency of $f_{input}$. The voltage range of the driving pulse can be anywhere from $-V_{input} \sim 0$ V to 0 V$\sim V_{input}$ without changing the outcome. For convenience, we assume that it is $-V_{input}/2 \sim V_{input}/2$. The ``up"(``down") half cycle corresponds to the upper panel, where the driving pulse is high, and only the odd numbered diodes flow current. The active diodes are indicated in black, and the inactive one in grey. The ``down" half-cycle corresponds to the lower panel, when the driving pulse is low. Only the even numbered diodes are active in this case, and the charge flows are reversed from the previous half-cycle. The details of the charge flow, shown in red, will be derived later in this section. Naively, we would expect the output voltage to be simply $V_{input}$ times the total number of stages. The actual output voltage will differ from this expectation primarily due to two factors.\\

First, the load current introduces AC and DC voltage drops from the ideal output voltage. The amount of voltage drop per unit load has a quadratic or cubic dependence on the total number of stages in the CW chain, and to which stage the load is attached. This determines the output impedance structure of the CW chain, although it is important to note that the CW is not a linear circuit (one that conserves frequency between input and output), and a part of the voltage drop is an AC response to a DC load. In the next two sub-sections, we will derive the expression for the voltage drop as a function of load current, total number of stages, capacitance of the chain, and the frequency of the driving pulse.\\

Second, the diodes produce a voltage drop with a small but non-linear load current dependence. While the extreme non-linearity makes it difficult to predict the relationship between the load current and the voltage drop, the magnitude of the variation is small enough that it should be regarded as a constant for most high voltage applications. As such, we do not attempt to model the load dependence, but simply calculate the total voltage drop based on the nominal voltage drop across a diode. \\

\begin{figure} [t]
\centering
\includegraphics[width=1.\textwidth]{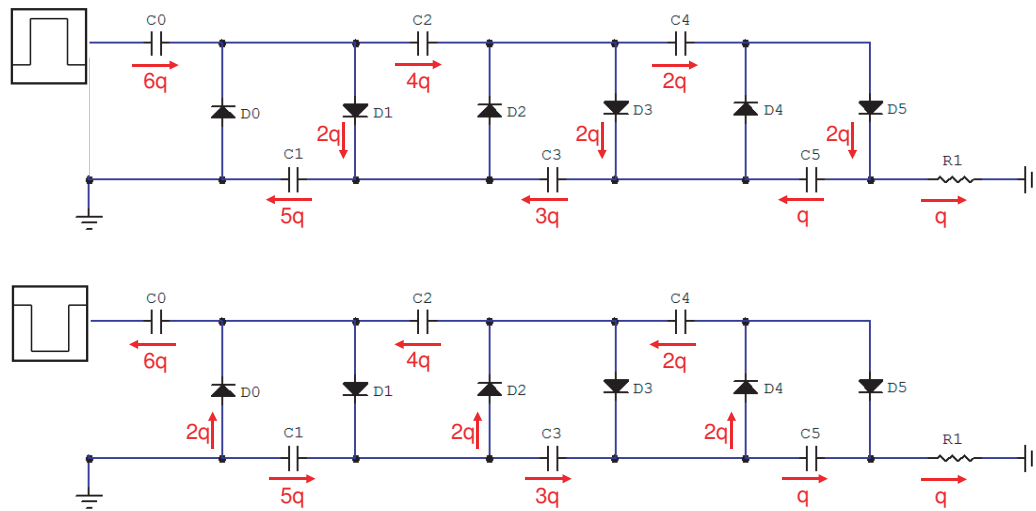}
\caption[Schematic of a three stage, half-wave Cockroft-Walton voltage multiplier chain in equilibrium]{Schematic of a three stage, half-wave Cockroft-Walton voltage multiplier chain in equilibrium. Upper panel: ``Up" half cycle in which the driving pulse is high. Lower panel: ``Down" half cycle in which the driving pulse is low. Indicated in red are the sum of charges that move across the capacitors and diodes during one half cycle. The equilibrium requirement implies that charge movements in the upper and lower panels should be symmetric.}
\label{CWscheme2}
\end{figure}  	

\subsubsection{\sloppy Charge Flow Pattern under Load}
For the analysis, the chain is assumed to be charged to the maximum possible voltage for the given load, and operating in equilibrium. (These two conditions imply that $f_{input}$ is slow enough to allow all the capacitors to complete the necessary charging and discharging during one half cycle.) There is a load current, $I_{load}$, which is drawn from the top of the chain where the output impedance is the highest. It is assumed that the amount of charge that leaves the system due to $I_{load}$ is the same for both half cycles. The load current may have finer time structure within a half cycle, but this is irrelevant as long as the integral is approximately the same for each half cycle. It is further assumed that the perturbation in the total output voltage $V_{out}$ due to $I_{load}$ is small compared to $V_{out}$. Under these assumptions, the load can be thought of as a large resistor attached in between the top of the chain and ground. \\

Figure \ref{CWscheme2} shows a total of three stages, with the charge flow indicated in red. In order to make the discussion more general, however, we will proceed by considering the case in which the total number of stages is equal to an arbitrary number N. The amount of charge drawn from the system at the top over one half cycle is $q$, same as shown in figure \ref{CWscheme2}. During the down half cycle, all the even numbered diodes are closed. Consequently, the load charge $q$ must come entirely from the final capacitor, C(N). (C3 in figure \ref{CWscheme2}) Conversely, the charge that leaves C(N) must also leave the system as it has nowhere else to go, and we conclude that over the down half cycle, C(N) is discharged by $q$. (Meaning $q$ flows through C(N))\\

Over the up half cycle, on the other hand, C(N) must be charged by the same amount, $q$, in order for the system to remain in equilibrium. But this charge can only come from C(N-1), as all the odd numbered diodes are closed. Furthermore, C(N-1) has to supply the load charge $q$ as well. Consequently, during the up half cycle, C(N-1) is discharged by $2q$. \\

Using similar arguments, we can easily see that over the down half cycle, C(N-2) is discharged by $3q$, and over the up half cycle, C(N-3) is discharged by $4q$, and so on and so forth. The pattern holds all the way down to the bottom, and we conclude that the amount of charge that moves in and out of the capacitor C(i) is equal to $(2N-i)q$. The charging bank gets charged during the down half cycle, and the output bank gets charged during the up half cycle. Every diode passes $2q$ over a half cycle in which it is active. It is clear that this is the only pattern of charge flow that satisfies both the equilibrium requirement and the charge conservation.\\

\subsubsection{\sloppy Calculating the Voltage Drop} 
Now that we have established the pattern of charge flow in equilibrium, we can derive the expressions for the non-diode part of the voltage drop, which is divided into two components. The DC component is called $V_{sag}$, and the AC component is called $V_{ripple}$.\\
$N_{stage} = \text{Number of stages in the CW chain}$\\
$V_{out} = \text{Output voltage of the chain at the top}$\\
$V_{input} = \text{Peak to peak amplitude of the input pulse}$\\
$V_{diode} = \text{Total voltage drop due to diodes}$\\
Using the above definitions, 
\begin{equation}\label{eq:Vsag}
V_{sag} = N_{stage} \cdot V_{input} - \langle V_{out} \rangle  + \langle V_{diode} \rangle
\end{equation}
\begin{equation}\label{eq:Vripple}
V_{ripple} = V_{out}(\text{up half cycle}) - V_{out}(\text{down half cycle})
\end{equation}
Brackets indicate time average over one full cycle of the driving pulse.\\

To get the expressions for $V_{sag}$ and $V_{ripple}$, we calculate the voltage drop for the two half cycles separately while ignoring the effects of the diode. The average of the two is equal to $V_{sag}$, while the difference is equal to $V_{ripple}$. To do so, it is useful to express the change in voltage over a capacitor when its stored charge increases or decreases by $q$. $q$ is defined above to be the amount of charge that leaves the system during a half cycle due to the load. From here on we assume that the load current $I_{load}$ is approximately constant, which would be true if the load was a simple resistor. The driving pulse has frequency of $f_{input}$ and period of $T$. 
\begin{equation}
q \approx I_{load} \cdot \dfrac{T}{2} = \dfrac{I_{load}}{2 \cdot f_{input}} 
\end{equation}
Using the capacitor equation $q=C \cdot V $,
\begin{equation}\label{eq:deltaV}
\delta V = \dfrac{I_{load}}{2 \cdot f_{input} \cdot C} 
\end{equation}

From figure \ref{CWscheme2}, we see that at the end of the down half cycle, the first capacitor C0 gets charged to $V_{input}/2$ as it is connected between ground and the driving pulse, which is at $-V_{input}/2$. A half cycle later, the charging bank is now maximally discharged, and the voltage across C0 is down by $2N \cdot \delta V$. We define the variable $V_{C(i)}(up/down)$ to be the voltage across C(i) at the end of the up/down half cycle.
\begin{equation}
V_{C0}(down) = \frac{1}{2} V_{input} 
\end{equation}
\begin{equation}
V_{C0}(up) = \frac{1}{2} V_{input} - 2N \cdot \delta V 
\end{equation}

The voltage relative to ground on the point between C0 and C2 is the sum of the voltage across C0 and the driving pulse voltage. During the up half cycle, it is therefore equal to $V_{input} - 2N \cdot \delta V$. In fact, the voltage at this point is independent of the choice of the driving pulse offset relative to ground, which is the reason that the change in offset does not alter the output. If the driving pulse offset is chosen such that the pulse range is from 0 V to $V_{input}$, C0 may be omitted with minimal change in the output. \\

The voltage between C0 and C2 is also the maximum charging voltage for C1 during the up half cycle, as it is connected between this point and ground through the active diode D1. 
\begin{equation}
V_{C1}(up) = V_{input} - 2N \cdot \delta V
\end{equation}
Since it is on the output bank, C1 goes through discharging during the down half cycle. The amount of charge lost is equal to $(2N-1)q$.
\begin{equation}\label{eq:cw1}
\begin{split}
V_{C1}(down) &= V_{C1}(up) - (2N-1) \cdot \delta V \\
&= V_{input} - (2 \cdot 2N-1) \cdot \delta V
\end{split}
\end{equation}
At the same time, C1 is connected to C2 on both ends via the active diodes D0 and D2. Therefore by the end of this half cycle, the voltage across the two capacitors should be equal. (As mentioned above, the diode effects are explicitly ignored during this derivation.)
\begin{equation}
V_{C2}(down) = V_{C1}(down) = V_{input} - (2 \cdot 2N-1) \cdot \delta V
\end{equation}
Since it is on the charging bank, C2 goes through discharging during the up half cycle. The amount of charge lost is equal to $(2N-2)q$.
\begin{equation}
\begin{split}
V_{C2}(up) &= V_{C2}(down) - (2N-2) \cdot \delta V \\&= V_{input} - (3 \cdot 2N-1-2) \cdot \delta V
\end{split}
\end{equation}
Similarly to the C1 and C2 pair in the down half cycle, C2 is connected to C3 through D1 and D3 in the up half cycle. 
\begin{equation}
V_{C3}(up) = V_{C2}(up) = V_{input} - (3 \cdot 2N-1-2) \cdot \delta V
\end{equation}
It is easy to see what the voltage across C3 is at the end of the down half cycle, given that the amount of charge that leaves C3 is equal to $(2N-3)q$.
\begin{equation}\label{eq:cw2}
\begin{split}
V_{C3}(down) &= V_{C3}(up) - (2N-3) \cdot \delta V \\&= V_{input} - (4 \cdot 2N-1-2-3) \cdot \delta V
\end{split}
\end{equation}
Looking at the equations from \ref{eq:cw1} to \ref{eq:cw2}, a pattern emerges. For an even numbered capacitor on the charging bank where $i=2k,(k>0)$,
\begin{align}
V_{C(i)}(down) &= V_{C(i-1)}(down) \\
V_{C(i)}(up) &= V_{C(i)}(down) - (2N-i) \cdot \delta V
\end{align}
For an odd numbered capacitor on the output bank where $i=2k-1,(k>0)$,
\begin{align}
V_{C(i)}(up) &= V_{C(i-1)}(up) \\
V_{C(i)}(down) &= V_{C(i)}(up) - (2N-i) \cdot \delta V
\end{align}

To summarize, the maximum voltage of any given capacitor is equal to the minimum voltage across the previous capacitor. The minimum voltage of the capacitor is obtained by subtracting $(2N-i) \cdot \delta V$ from its maximum voltage. The only exception to this rule is C0, which is a special case. It is clear how every capacitor has the ``floor" that becomes the ``ceiling" for the next capacitor, accumulating the effect of the load current as we move up along the chain. \\
%Figure \ref{} visualizes the continuous charging and discharging process among neighboring capacitors. 

In order to calculate the output voltage at the end of either half cycles, we simply have to add the voltages across all the odd numbered capacitors on the output bank. For the up half cycle, 
\begin{equation}
\begin{split}
V_{out}(up)=&V_{input} - 2N  \cdot \delta V + \\
&V_{input} - \lbrace 3 \cdot 2N - (1+2) \rbrace \cdot \delta V + \\
&V_{input} - \lbrace 5 \cdot 2N - (1+2+3+4) \rbrace \cdot \delta V + \\
&V_{input} - \lbrace 7 \cdot 2N - (1+2+3+4+5+6) \rbrace \cdot \delta V +  \\
&\ldots \\
=& N \cdot V_{input} - \sum_{i}^{N} \Big\lbrace (2i-1) \cdot 2N - \sum_{k}^{2i-2} k \Big\rbrace \cdot \delta V\\
=& N \cdot V_{input} - \sum_{i}^{N} \Big\lbrace -2i^2 + (4N+3)i -(2N+1) \Big\rbrace \cdot \delta V\\
=& N \cdot V_{input} - \frac{1}{6} \cdot \Big\lbrace 8N^3 + 3N^2 + N \Big\rbrace \cdot \delta V
\end{split}
\end{equation}
For the down half cycle, 
\begin{equation}
\begin{split}
V_{out}(down)=&V_{input} - \lbrace 2 \cdot 2N - 1\rbrace \cdot \delta V + \\
            &V_{input} - \lbrace 4 \cdot 2N - (1+2+3) \rbrace \cdot \delta V + \\
            &V_{input} - \lbrace 6 \cdot 2N - (1+2+3+4+5) \rbrace \cdot \delta V + \\
            &V_{input} - \lbrace 8 \cdot 2N - (1+2+3+4+5+6+7) \rbrace \cdot \delta V +  \\
            &\ldots \\
=& N \cdot V_{input} - \sum_{i}^{N} \Big\lbrace 2i \cdot 2N - \sum_{k}^{2i-1} k \Big\rbrace \cdot \delta V\\
=& N \cdot V_{input} - \sum_{i}^{N} \Big\lbrace -2i^2 + (4N+1)i \Big\rbrace \cdot \delta V\\
=& N \cdot V_{input} - \frac{1}{6} \cdot \Big\lbrace 8N^3 + 9N^2 + N \Big\rbrace \cdot \delta V
\end{split}
\end{equation}

The terms that include $\delta V$ are the load-dependent part of the output deviation from $N \cdot V_{input}$, which is the naive expectation. Referring back to equations \ref{eq:Vsag} and \ref{eq:Vripple}, $V_{sag}$ is obtained by taking the average of these two terms for up and down half cycles, 
\begin{equation}\label{eq:Vsag2}
V_{sag} = \frac{1}{6} \cdot \Big\lbrace 8N^3 + 6N^2 + N \Big\rbrace \cdot \delta V 
\end{equation}
and $V_{ripple}$ is obtained by taking the difference between $V_{out}(up)$ and $V_{out}(down)$.
\begin{equation}\label{eq:Vripple2}
V_{ripple} = N^2 \cdot \delta V 
\end{equation}

Finally, we note that the equation for $V_{ripple}$ can be obtained through an alternative method, by recognizing that it is simply the sum of voltage differences between two half cycles for all the capacitors in the output bank. Looking back at figure \ref{CWscheme2}, it is clear that $V_{ripple}$ is equal to the sum of all odd numbers from 1 to $2N-1$, times $\delta V$. Therefore we have,
\begin{equation}\label{eq:Vripple3}
V_{ripple} = \sum_{i}^{N} (2i-1) \cdot \delta V = N^2 \cdot \delta V 
\end{equation}
which gives the same result as equation \ref{eq:Vripple2}.  

\subsubsection{Full-Wave Cockroft-Walton Multiplier}
From equations \ref{eq:Vsag2} and \ref{eq:Vripple2}, we see that both types of load dependent deviations have a strong dependence on the total number of stages. For our purposes, the particular voltage tapering of the FEU-84 photo-tube required a twenty-two stage CW chain. For a capacitance of $0.5\: \mu$F and driving frequency of 20 kHz, we can calculate the sag and the ripple voltages per unit load current drawn at the top from equation \ref{eq:deltaV}; They are 1.46 V and 0.05 V per $1\: \mu$A, respectively. \\

Considering that the output voltage is typically on the order of 1500 V, these are fairly small numbers. On the other hand, the exponential dependence of the PMT gain on the high voltage means that a 1 V change in high voltage can cause a gain change on the order of 1 \%. \\

\begin{figure} [t]
\centering
\includegraphics[width=1.\textwidth]{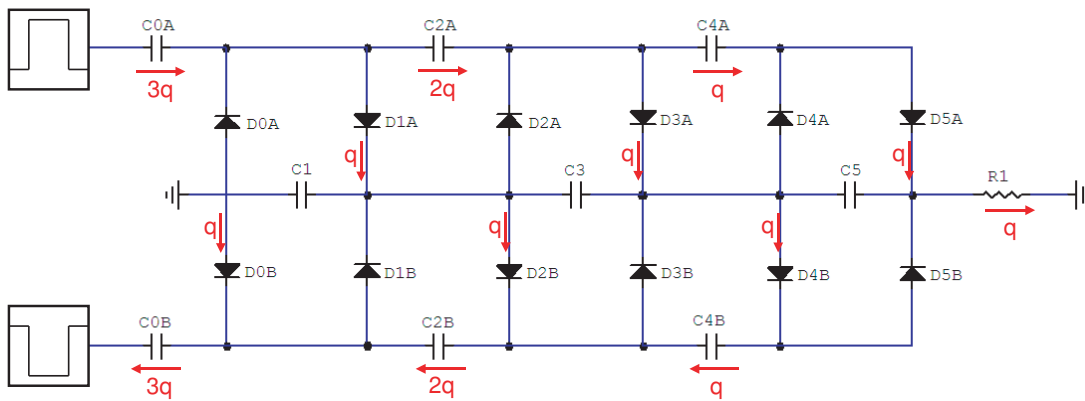}
\caption[Schematic of a three stage, full-wave Cockroft-Walton voltage multiplier chain in equilibrium]{Schematic of a three stage, full-wave Cockroft-Walton voltage multiplier chain in equilibrium. Indicated in red are the sum of charges that move across the capacitors and diodes during one half cycle. The equilibrium requirement implies that charge movements in the upper and lower banks should be symmetric.}
\label{CWscheme3}
\end{figure}  	

Figure \ref{CWscheme3} illustrates the schematic of a 3 stage, full-wave Cockroft-Walton chain. It is essentially two mirror imaged half-wave chains sharing a common output bank. The dual charging banks are driven by two pulses with their phases anti-aligned by 180 degrees. This ensures that at any given point, one of the charging banks is in the up half cycle, providing charges for the rest of the circuit. Consequently, there is no time at which the output bank has to replenish the charging bank capacitors, and this results in a significant reduction in both the sagging and the ripple voltages. \\

As before, the movement of the charges is indicated in the figure in red. Unlike figure \ref{CWscheme2}, there is no need to depict both half cycles, as the only difference between them is the two charging banks reversing their role. The argument for the charge movement is largely identical to the one made for the half-wave design, based on the equilibrium requirement and charge conservation. In fact, it is easy to see that the pattern shown in figure \ref{CWscheme3} is the only one that satisfies both conditions. The key point is to realize that the two charging banks are always charging each other. For example in figure \ref{CWscheme3}, C0A is charging C2B, C2A is charging C4B, and C4A is providing the load current.  \\

The most significant improvement is that now there is no movement of charge across output bank capacitors. Once brought to equilibrium, they do not charge or discharge. At any given half cycle, one of the charging banks is both replenishing the other charging bank, and providing the load current. Since there is no longer a difference between up and down half cycle for the output bank, the voltage ripple is limited to finer structures within a half cycle, which is a second order effect. \\
\begin{equation}\label{eq:Vripple4}
V_{ripple} \sim \sum_{i}^{N} 0 \cdot \delta V = 0 
\end{equation}

The equation for $V_{sag}$ can be obtained in a similar way as before, by recognizing the limit that each capacitor imposes on the maximum charging level of the subsequent capacitors. Once again, we derive the general formula based on the 3-stage schematic. Since the distinction between up and down half cycle no longer exists, we find expressions for the voltage across a capacitor in its maximum and minimum. From figure \ref{CWscheme3}, we see that the maximum voltage across C2 is equal to the minimum voltage across C0. (In the figure, C2B is at its maximum, and C0A at its minimum.) At the same time, this voltage is equal to the voltage across C1, as C2B and C1 are connected on both ends through D0B and D2B.
\begin{equation}
\begin{split}
V_{C2}(max) &= V_{C0}(min) = V_{C1}\\
            &= V_{input} - N \cdot \delta V 
\end{split}
\end{equation}
The minimum voltage across C2 is $(N-1) \cdot \delta V$ lower than its maximum, as is the case with C2A. 
\begin{equation}
\begin{split}
V_{C2}(min) &= V_{C2}(max) - (N-1) \cdot \delta V\\
            &= V_{input} - (2N-1) \cdot \delta V
\end{split}
\end{equation}
Similarly, for C3 and C4,
\begin{equation}
\begin{split}
V_{C4}(max) &= V_{C2}(min) = V_{C3}\\
            &= V_{input} - (2N-1) \cdot \delta V
\end{split}
\end{equation}
\begin{equation}
\begin{split}
V_{C4}(min) &= V_{C4}(max) - (N-2) \cdot \delta V\\
            &= V_{input} - (3N-1-2) \cdot \delta V
\end{split}
\end{equation}

The pattern is clear. We proceed to derive the equation for the output voltage by adding voltages across all odd numbered capacitors as before. 
\begin{equation}\label{eq:fullwave}
\begin{split}
V_{out}=&V_{input} - N  \cdot \delta V + \\
&V_{input} - \lbrace 2 N - 1 \rbrace \cdot \delta V + \\
&V_{input} - \lbrace 3 N - (1+2) \rbrace \cdot \delta V + \\
&V_{input} - \lbrace 4 N - (1+2+3) \rbrace \cdot \delta V +  \\
&\ldots \\
=& N \cdot V_{input} - \sum_{i}^{N} \Big\lbrace i \cdot N - \sum_{k}^{i-1} k \Big\rbrace \cdot \delta V\\
=& N \cdot V_{input} - \frac{1}{2}\sum_{i}^{N} \Big\lbrace -i^2 + (2N+1)i \Big\rbrace \cdot \delta V\\
=& N \cdot V_{input} - \frac{1}{6} \cdot \Big\lbrace 2N^3 + 3N^2 + N \Big\rbrace \cdot \delta V
\end{split}
\end{equation}
Since the output voltage no longer depends on the half cycle, the term that included $\delta V$ is equal to the voltage sag. 
\begin{equation}\label{eq:Vsag3}
V_{sag} = \frac{1}{6} \cdot \Big\lbrace 2N^3 + 3N^2 + N \Big\rbrace \cdot \delta V 
\end{equation}

Comparing equation \ref{eq:Vsag3} to equation \ref{eq:Vsag2}, we see that the full-wave design reduces $V_{sag}$ by a factor of 4 for N=22. Combined with the elimination $V_{ripple}$, it is clear that the full wave design can provide a much improved high voltage stability in both short and long time frames.

\subsection{SPICE Simulation Results}
In order to verify the model calculation, commercial circuit design software called $CircuitMaker^{tm}$ was used to perform SPICE simulation. Simulation Program with Integrated Circuit Emphasis (SPICE) is an analog circuit simulator widely used as a circuit design tool, while $CircuitMaker^{tm}$ provides an easy to use graphic interface to SPICE. \\

A full-wave Cockroft-Walton circuit with nine stages was simulated. The peak to peak amplitude of the driving pulse was set at 70 V, and its frequency at 10 kHz. ($N \cdot V_{input}$ = 630 V) The load was attached at the top of the chain, with a load current of $60 \: \mu$A. Using the more realistic load current of less than $1 \: \mu$A resulted in effects that were too small to be reliably read off from the simulation. \\

From the second to the last line from equation \ref{eq:fullwave}, we obtain the expression for the voltage drop corresponding to each stage of the output bank. It is given by,
\begin{equation}\label{eq:Vsagstage}
V_{drop}(i)=\frac{1}{2}\Big\lbrace -i^2 + (2N+1)i \Big\rbrace \cdot \delta V
\end{equation}
To make a direct comparison to this quantity, we sample the voltage at every stage of the charging bank at the beginning and the end of a half cycle in which it is being discharged. In figure \ref{CWscheme3}, this is equivalent to measuring voltages on the charging bank A at points to the right of the three capacitors. In this way, we can factor out the effects of diode related voltage drops, and only focus on the cumulative effects of charge flow. The difference between the beginning and the end of the half cycle for each of these points corresponds to one term in equation \ref{eq:Vsagstage}.\\

\begin{figure} [t]
\centering
\includegraphics[width=.8\textwidth]{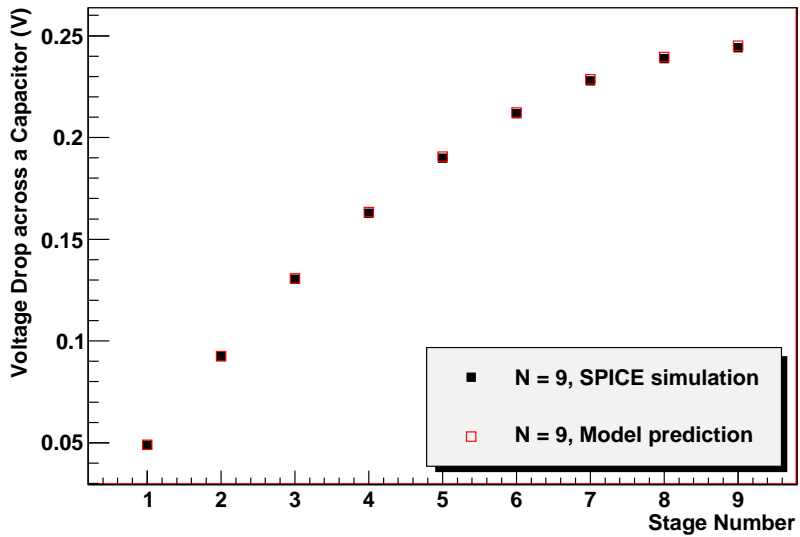}
\caption[$V_{drop}$ per stage vs Stage number for the SPICE simulation and model prediction]{$V_{drop}$ per stage vs Stage number for SPICE simulation and model prediction. The model prediction for $V_{drop}$ is defined in equation \ref{eq:Vsagstage}.}
\label{Vsag}
\end{figure}  	

The result is shown in figure \ref{Vsag}, where we see an excellent agreement between the model prediction and simulation. We conclude that our simple model of the Cockroft-Walton chain is adequate in describing the linear part of the voltage drop in the limits of small load current. \\

The piece of the voltage drop equation that has not been covered so far is the drop due to diodes. Looking back at figure \ref{CWscheme3}, there is a charge flow through diodes, which is equal to $q$ per half cycle. But unlike the capacitors, it is very difficult to predict how much voltage will be dropped due to this current. As is well known, the relationship between current and voltage across a diode is highly exponential. While it is common to approximate the voltage drop for silicon diodes to be around 0.6 V, we do not a priori know if this estimation is accurate enough for our purpose. What is relevant to us is the variation from this nominal value as a function of load current, as a constant drop does not affect the stability. Instead of trying to model this highly non-linear behavior, we will simply characterize it based on the SPICE simulation.\\

\begin{figure} [t]
\centering
\includegraphics[width=.8\textwidth]{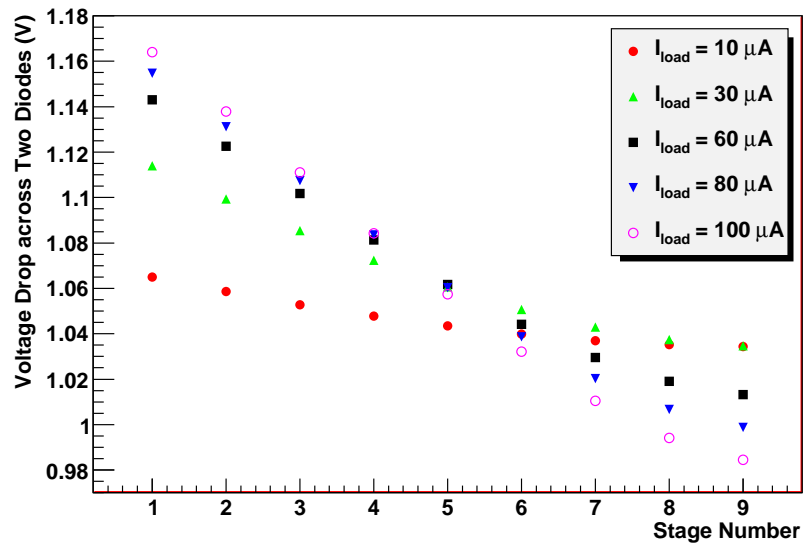}
\caption[Voltage drop across a pair of diodes in each stage vs. Stage number for the SPICE simulation]{Voltage drop across a pair of diodes in each stage vs. Stage number for SPICE simulation. For fixed frequency and amplitude of the driving frequency, the size of the load resistor connected to the top of the chain was varied to obtain a range of load current.}
\label{Vdiode}
\end{figure}

Each stage of a Cockroft-Walton chain has two diodes, and both of them contribute to the diode drop. Figure \ref{Vdiode} shows the diode drop per stage as a function of the stage number, for various load currents. The result is based on a SPICE simulation of a nine stage Cockroft-Walton chain. As the load current increases from ~$10 \:\mu$A to ~$100 \: \mu$A, the diode drop per stage changes by at most around 0.1 V. The maximum change occurs at the lowest stage, and curiously, the data points tend to merge when the stage number is equal to 6. Considering that the values of load currents used in this simulation are likely orders of magnitude larger than what is expected in reality, we conclude that the load current dependence of the diode drop, while highly non-linear, is very small compared to the scale of the high voltage. It can safely be considered a constant in the scale of the high voltage.\\

\section{\sloppy PMT Base Design Overview}
In designing the PMT base for the FMS inner calorimeter, there were several areas of concern that motivated our approach. The lack of space and cooling within the metal enclosure became one of the main constraints, raising a number of concerns. Because the inner calorimeter is located at the center of the enclosure, limiting the power consumption and heat generation was considered essential. There was a strict limit for the size of the base, and the ease of installation in tight, blind areas was also a concern. \\

On the logistics side, the main limitation was the time. The project went from concept study to mass production in less than a year, which forced us to focus on designs that were simple and robust. Any design that would have required extensive tuning was deemed unsuitable. It was also useful to put as much control circuits as we could on the base itself, so as to avoid building any elaborate control electronics later. Along with low power consumption, keeping the supply voltages under 50 V was necessary to reduce safety related concerns. The fabrication process had to be simple as well, much of which was done in-house by summer undergraduate students. \\

Figure \ref{CWscheme4} shows the simplified schematic of the Penn State photo-tube base. In this section, we will discuss selected elements of the base design. \\

\begin{figure} [p]
\centering
\includegraphics[width=1.\textwidth]{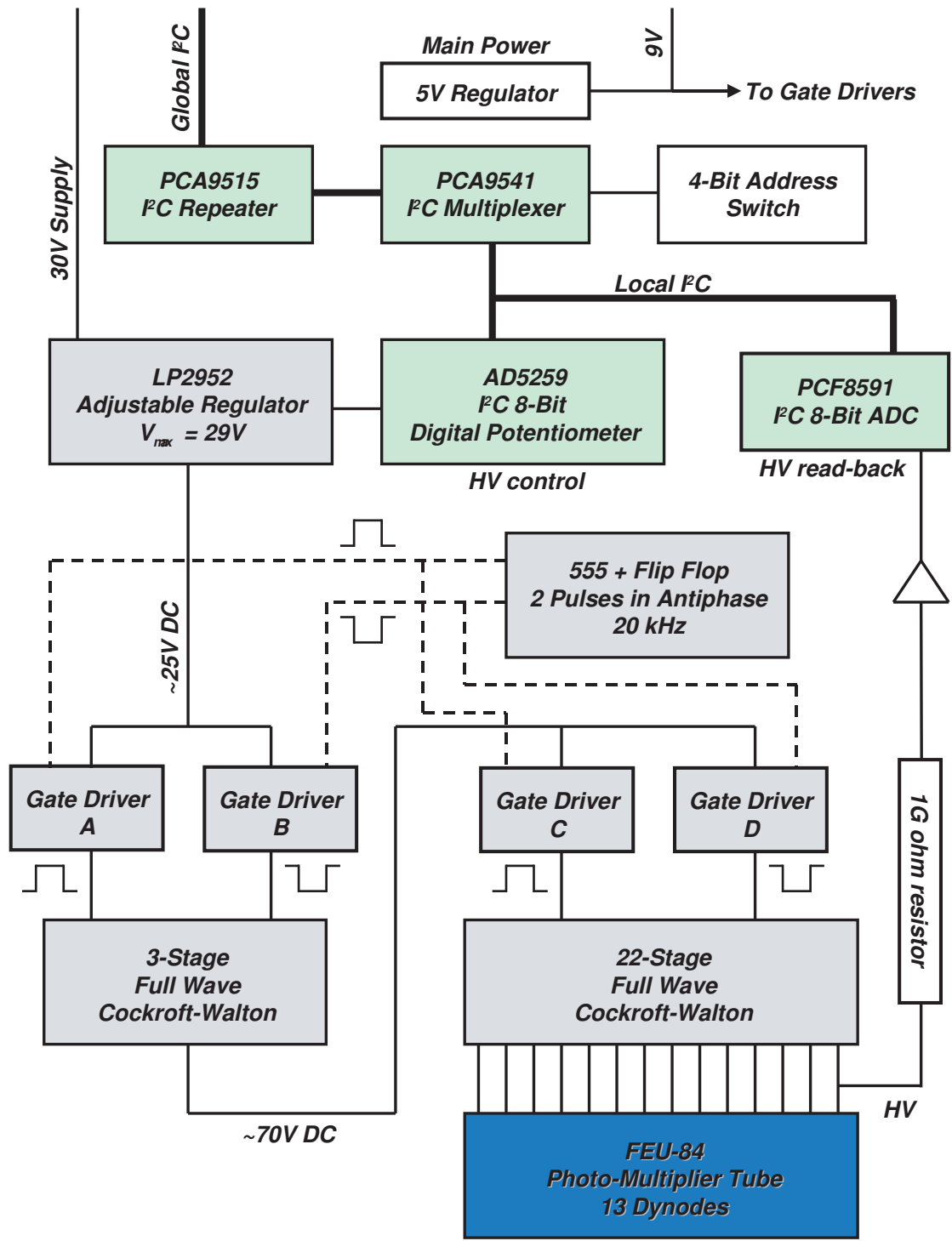}
\caption[Simplified schematic of the Penn State Cockroft-Walton PMT base]{Simplified schematic of Penn State Cockroft-Walton photo-multiplier tube base. $I^2C$ serial bus is indicated in double lines.}
\label{CWscheme4}
\end{figure}  

\subsection{Feedback-less Design}
Commonly, Cockroft-Walton systems employ a feedback circuit to counteract the time varying load, and stabilize the output voltage. \cite{cw1} This can be achieved by sampling the high voltage at the top of the chain (cathode) through a high impedance voltage divider, and adjusting the amplitude or frequency of the input pulse accordingly. The amplitude adjustment counteracts the change in voltage sag by changing the target voltage $N \cdot V_{input}$, whereas the frequency adjustment works by changing the magnitude of $\delta V$, which has a frequency dependence as given by equation \ref{eq:deltaV}. While this is the most direct way to ensure stability of high voltage, there are a number of drawbacks as well.\\

First, the feedback requires an accurate measurement of the high voltage, which involves drawing current at the top of the chain where the output impedance is the highest. The amount of current draw cannot be made arbitrarily small, as in general a higher precision measurement requires a larger amount of current. It is easy to make the feedback circuit the dominant load on the Cockroft-Walton chain, eclipsing the load from actual operation of the PMT. This is not necessarily a bad strategy in terms of stabilizing the rate dependence, as the change in event rate will result in a small fractional change in total load. However, the ripple can become a serious issue when running with a large constant load, especially for half-wave designs. \\

Secondly, the Cockroft-Walton chain is slow to react to the change in input pulse. The natural time scale is determined by the driving frequency, and any feedback-adjustment with finer time structure becomes irrelevant. (For instance, a feedback cannot cancel out the ripple, which has the same frequency as the driving pulse.) Furthermore, any change made to the input pulse takes many cycles to propagate through the chain, since the charges on the capacitors have to be adjusted sequentially. This is especially true when lowering the voltage, as the only way that the stored charges can leave the system is in the form of load current. This time delay can make stabilizing the feedback circuit very difficult. The time scale for the operating amplifiers would have to be carefully tuned to match that of the Cockroft-Walton chain.\\

Finally, the relative step sizes of the dynode voltages may not be preserved by the feedback mechanism. For instance, if the high voltage drops due to increased rate, the feedback circuit may counteract by increasing the amplitude of the driving pulse to maintain the high voltage at the target value. However, because we have increased $V_{input}$ to cancel out increased $\delta V$ in equation \ref{eq:fullwave}, the tapering of the voltages would change, potentially altering the gain and linearity of the PMT. In this sense, varying the frequency of the driving pulse is a better way to do feedback, as it directly cancels out the change in $I_{load}$ within $\delta V$. \\

While it is certainly possible to build a robust feedback circuit for a Cockroft-Walton system, we chose to forgo the feedback altogether in order to save development time. Based on the model study of Cockroft-Walton as described in previous sections, we determined that the full-wave design was inherently stable enough to make active corrections unnecessary. This was especially true in the absence of a feedback circuit, which puts significant load at the top of the chain. Another advantage of building a robust system that minimizes voltage drop was that the tapering was better preserved under load. \\

\subsection{Dual Cockroft-Walton Amplifiers}
With the required 22 stages in the main Cockroft-Walton chain, the amplitude of the driving pulse needed to be upward of 80 V to reach the maximum required output voltage of 1800 V. In order to keep the supply voltage levels low, it was necessary to derive this 80 V from a lower voltage on-board. Utilizing much of the circuitry needed to operate the main Cockroft-Walton, we chose to add an additional three stage, high capacitance full-wave chain to serve as the first level amplifier. With the amplification factor close to three, the supply voltage of no more than 30 V was needed to produce the 80 V required by the main Cockroft-Walton. \\

The obvious downside to this approach is the increased load current dependence due to voltage sag. However, as described by equation \ref{eq:fullwave}, the sagging has a strong dependence on the total number of stages. During normal operation, the total current draw from a PMT is on the order of 1 mA, which sets the scale on the potential rate dependent load variations. With N=3 and capacitance of $10 \: \mu$F, 1 mA of load current only introduces 0.035 V of voltage sag. After amplification by the main Cockroft-Walton, this is still well under 1 V off the output high voltage.\\

\subsection{$I^2C$ Serial Bus}
$I^2C$ is a two line serial bus created by Philips. The typical bus speed is 100 kbits/s, which is slow by modern standards. However, it is commonly used for household electronics and computer peripherals that do not require fast communication. In addition, there is no licensing fee. Consequently, there is an abundance of affordable integrated circuits from multiple manufacturers that conform to $I^2C$ protocol. It is also relatively power efficient, and the low line count makes it very simple to design. In short, it is well suited for the high voltage control circuit.\\

In figure \ref{CWscheme4}, there are four integrated circuits that are connected to the $I^2C$ serial bus, shown in double lines. The first two ICs, the repeater (PCA9515) and the multiplexer (PCA9541), handle the connection between global and local $I^2C$ buses. The address space for most $I^2C$ compatible chips is limited to two or three bits, meaning that we can only connect a handful of the same types of ICs to the bus before we run out of address space. To get around this issue, each base is given a local $I^2C$ bus that can be disconnected from the global bus. The multiplexer, which has a four bit address space, serves as the gate between the two buses, making the local bus visible to the master only when it is called upon. A user configurable 16 position address switch is connected to it, giving each base an address from 1 to 16. \\

One downside of this multiplexer is that it does not isolate the capacitance of the local bus. Due to the maximum bus capacitance limit of 400 pF, much of which is taken up by the capacitance of the cables, it is essential that the capacitance of the base be disconnected from the global bus. This is the job of the repeater, which physically separates the two buses by introducing a pair of buffers for each of the two lines. The added benefit of the local and global bus separation is that when a failure occurs in the local bus, it does not bring down the entire communication system. \\ 

\subsection{High Voltage Generation and Read-Out}
High voltage generation starts with the regulation of the 30 V analog supply voltage. The output voltage is controlled via the $I^2C$ serial bus, which is fed into a pair of gate drivers (drivers A and B in figure \ref{CWscheme4}) that produce two phase anti-aligned pulses. These pulses drive the first level Cockroft-Walton chain, the output of which is fed into another pair of gate drivers (drivers C and D in figure \ref{CWscheme4}). They in turn drive the main 22 stage amplifier, producing the final high voltage. Consequently, the output voltage of the PMT base is directly proportional to the output of the 30 V regulator. \\

LP2952 is a low drop out, high precision adjustable regulator with a maximum input voltage rating of 32 V. With a 30 V supply, it is capable of producing an output up to 29 V. As with most adjustable regulators, the output is controlled by two resistors connected to the output and feedback pins, the ratio of whose resistances determines the output voltage. \\

One part of this resistor assembly consists of a 50 k$\Omega$ resistor and a 50 k$\Omega$ digital potentiometer (AD5259) in series. In this configuration, the output can be adjusted down to 50 \% of its maximum value. The digital pot has an 8-bit volatile register that controls the resistance between two output pins from 0 to 50 k$\Omega$ in 256 steps. In addition, it has a non-volatile register that stores the start-up value. Read and write operations to these registers are carried out via the $I^2C$ serial bus, enabling digital control of the high voltage. The advantage of digital control is that there is no need for an external ``reference" signal. Usually, the reference signal needs to be very carefully regulated, as it is directly proportional to the high voltage, and sending it through a long cable can become an issue. \\ 

There are two pairs of phase anti-aligned pulses that are needed to drive the pair of Cockroft-Walton chains. The first pair has the maximum amplitude of around 30 V, and the second pair around 80 V. We use a 555 timer to generate a 20 kHz clock signal, which is fed into a flip-flop. The two complementary outputs of the flip-flip produce a pair of TTL signals that are off-phase by 180 degrees. These pulses control the gate driver (LM5000A), which chops a DC input of up to 100 V into a square pulse of identical frequency and duty cycle to the TTL driving signal. The only downside of the gate drivers is that they require a somewhat high supply voltage of 9 V. This was the reason for choosing 9 V as the second supply voltage.    \\

While there is no feedback circuit, it is still useful to read out the high voltage for diagnostic purposes. As discussed earlier, putting a constant load of reasonable size can also improve the overall stability. For these reasons, a 1 G$\Omega$ resistor was attached to the top of the main Cockroft-Walton, through which a high input impedance, unity gain operational amplifier samples $1/1000$ of the high voltage. The output of the op-amp is routed to the input of the PCF8591, an $I^2C$ compatible, four channel 8-bit ADC. Since the entire range of the output has to be covered by 8 bits, the accuracy of the read out is limited to $\sim \pm 5$ V. This means that the resolution is insufficient to provide meaningful information about photo-tube gains. Nevertheless, it is useful for monitoring the operation of the bases, as the common modes of failure tend to produce large change in high voltage.\\

\subsection{Power Consumption and Cabling}
There are two supply voltages in total, 30 V and 9 V. 30 V is used exclusively to drive the photo-tube, whereas 9 V is used mostly to power the on-board electronics. The average current draws are $\sim$ 3 mA for the 30 V line, and $\sim$ 10 mA for the 9 V line during normal operation. This results in total power consumption of around 200 mW, which is about half as much as that of the resistive divider that it replaces. \\

In addition to the two supply voltage lines, the $I^2C$ bus has two lines for data and clock, making the total line count equal to four. (The photo-tube signal is routed through the usual 50 $\Omega$ BNC cables.) The supply voltages carry no more than 10 mA per line, while the $I^2C$ lines have a maximum voltage of 3.3 V and minimal current. The lack of high voltage, high current, and analog signal allowed us to choose the CAT5E cable, commonly used for Ethernet connections. It has eight conductors, which are used to form four pairs of signal and ground. It comes with a locking and polarized connectors, alleviating the difficulties with the installation within the enclosure. The low price and wide availability were also useful.  \\

\section{\sloppy Controller Boards}
The initial plan was to use the FEU-84 photo-tube and the Penn State base combination for the entire FMS inner calorimeter. However, a last minute logistics problem prevented us from acquiring enough FEU-84's to cover the full detector. Instead, we had to find a replacement photo-tub and base for roughly half of the inner calorimeter. Fortunately, the STAR collaborators from Yale University donated to the FMS project XP2972 photo-tubes and Nanometric bases that were used in the AGS E864 experiment. \cite{E864} As a result, the inner calorimeter high voltage system was built as a hybrid unit consisting of two different photo-tube and base combinations.\\

\subsection{\sloppy Penn State Controller}    
Because the Penn State base has all the intelligence it needs on-board, the Penn State controller is simply a CAT5e hub that allows 16 bases to be connected to a single CAT5e cable. The only additional components on the board are the poly-fuses (thermisters) for current limiting $+30$ V and $+9$ V lines, and an EEPROM to store the controller ID.

\subsection{\sloppy ``Yale" Controller}  
The Nanometric base is also of Cockroft-Walton type, with similar power consumption to that of the Penn State Base. It is a half-wave design utilizing feedback circuitry that controls the frequency of the driving pulse. Unlike the Penn State base, it requires an external analog input to set the high voltage, and produces an analog output that is proportional to the high voltage. It also has a discriminator output with its threshold set externally. (This functionality is not used in our system) It needs a set of supply voltages that are different from the PSU base, including a negative voltage line. Twelve conductor ribbon cables are used for connections. In order to interface the ``Yale" system to the existing $I^2C$ framework, a control board was designed to operate the Nanometric base via $I^2C$ communication. \\

\begin{figure} [p]
\centering
\includegraphics[width=1.\textwidth]{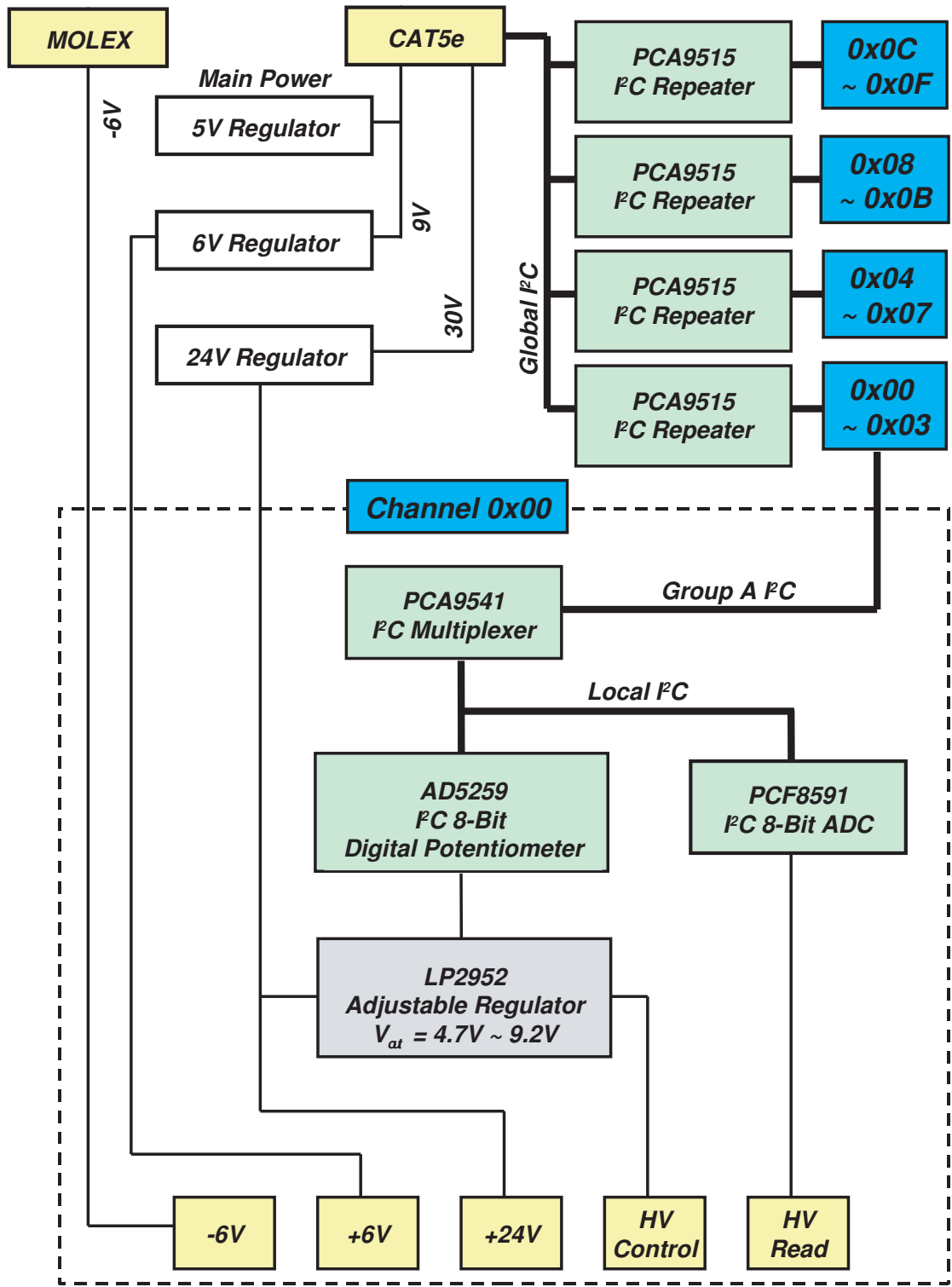}
\caption[Simplified schematic of the ``Yale" controller]{Simplified schematic of the ``Yale" controller. The Yale controller has 16 channels, one of which is shown in detail within the dashed line. A 12 conductor ribbon cable connects to each channel, which is shown in light yellow boxes at the bottom. The $I^2C$ serial bus is indicated in double lines.}
\label{CWscheme5}
\end{figure}  

Figure \ref{CWscheme5} shows the simplified schematic of the ``Yale" controller board. The goal was to make the integration as transparent as possible by using a modified version of the circuits that are found in the Penn State base. One Yale controller board has 16 channels, and each of these channels receives an $I^2C$ circuit that is very similar to that of one Penn State base. For instance, the addressing and capacitance management are handled by the same two IC's, PCA9515 and PCA9541. The differences are minor, such as the fact that there are 4 repeaters (PCA9515) per controller, each forming an $I^2C$ subnet that talks to four channels. In the Penn State system, each channel has its own repeater. (Sharing repeaters does increase the risk of multiple channel communication failure.) Another difference is that while there is an address switch in every Penn State base that allows the user to choose a 4-bit address, the addresses for the Yale system are built into the controller.\\

The Nanometric base uses three types of supply voltages, which are $+24$ V, $+6$ V, and $-6$ V. Of these, $+24$ V and $+6$ V are derived from the existing $+30$ V and $+9$ V, respectively, on the controller board. Only $-6$ V is unique to the Yale system. The two positive voltages ($+30$ V and $+9$ V) arrive at the Yale board through a CAT5e connection, which also carries the $I^2C$ lines. In this way, we have only one type of CAT5e connection in the system, which is shared by the Penn State and Yale sub-systems. Additional two pin Molex connectors are used for distributing $-6$ V to the Yale controllers. \\

As mentioned before, high voltage control for the Nanometric base requires an external analog signal between 0 V and 10 V. The constant of proportionality is around 200, which means that a 10 V signal corresponds roughly to 2000 V output. We use the same precision regulator (LP2952) and the 50 k$\Omega$ digital potentiometer (AD5259) circuit used for high voltage control in the Penn State base to generate this control voltage. The only change is the size of the resistors in the digital pot circuit, since the output for the Yale system needs to be between 4.7 V and 9.2 V. (This corresponds to roughly 900 V to 1800 V range in output.) High voltage read-out is done similarly as well, utilizing the same 8-bit ADC (PCF8591) used in the Penn State base. The analog output from the Nanometric base has a 1000:1 reduction factor from the actual high voltage, which is the same as the read-back signal in the Penn State base.\\

Because each Yale channel uses the same addressing, high voltage control and read-out scheme, it looks almost identical to a Penn State base as an $I^2C$ device. Even the high voltage read-out is independent of the base type. The only significant difference is the interpretation of the set high voltage bits, which is determined automatically by the in-house control software based on the controller board ID. \\

\subsection{\sloppy Master Controller}  
The master controller provides the highest level $I^2C$ multiplexing and supply voltage distribution. The master controller can operate up to 16 daughter controllers, which may be any mixture of Penn State or Yale type. This allows it to control up to 256 bases, enough to cover one half of the FMS inner calorimeter that has 238 active cells. \\

\begin{figure} [p]
\centering
\includegraphics[width=1.\textwidth]{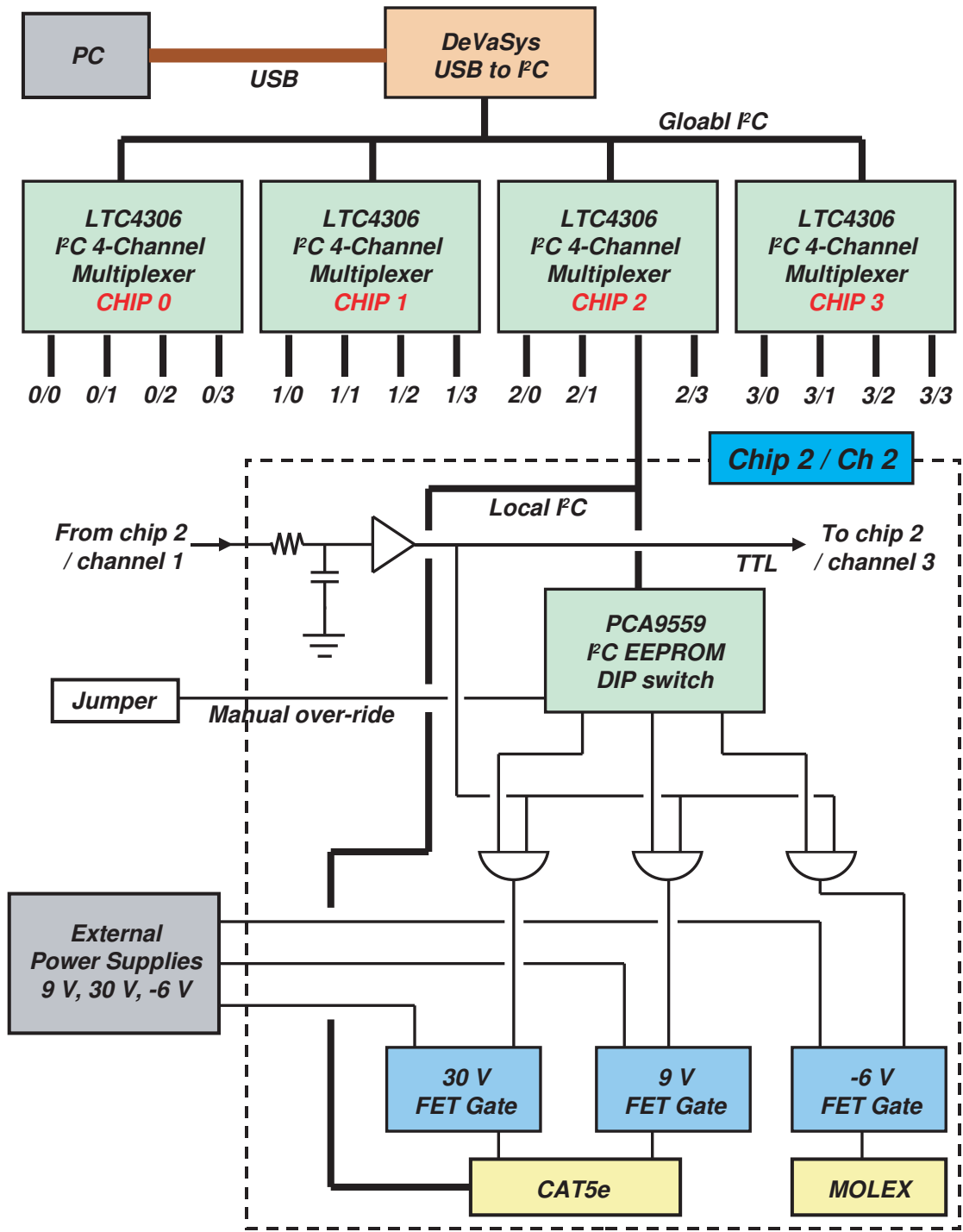}
\caption[Simplified schematic of the master controller]{Simplified schematic of the master controller. The master controller has four master multiplexers, each connected to 4 downstream channels. The master multiplexers are named chip 0, 1, 2, and 3. The 16 downstream channels are identified by their chip and channel number, which is shown in ``$N_{chip} / N_{channel}$" format. Chip 2 channel 2 is shown in detail within the dashed line. The $I^2C$ serial bus is indicated in double lines. The USB line is indicated by the brown line. }
\label{CWscheme6}
\end{figure}  

Figure \ref{CWscheme6} shows a simplified schematic of the master controller. Its primary function is $I^2C$ multiplexing. The address space for each Penn State or Yale channel is limited to 4 bits, which means that at any given time, no more than 16 bases should be visible to the $I^2C$ master. The multiplexer chip used in the bases (PCA9541) cannot be reused at a higher level, since its $I^2C$ address will conflict with one of the down-stream channels. The master controller employs four 4-channel ``master" multiplexers (LTC4306) to route the communication to only one of the 16 daughter controllers at a time. Each master multiplexer provides multiplexing and capacitance decoupling for four downstream channels. A NOR gate circuit (not shown in figure \ref{CWscheme6}) connecting the enable pins of the four master multiplexers ensures that only one of them is active at any given time. (If more than one downstream channels become visible to the master, it requires a reboot of the communication system to restore normal operation.)\\

The downside of the LTC4306 master multiplexer is that it is in principle incompatible with the PCA9515 repeater. Both devices decouple the capacitance of the upstream and downstream buses by inserting two buffers for each communication line. Two buffers are needed to drive the communication in both up-to-downstream and down-to-upstream directions. Since one buffer's output is connected to the other buffer's input, a lock-up can occur where the output of one buffer in turn drives the buffer in the opposite direction. To avoid this condition, a ``low" level voltage that is slightly higher than the legal level is used to distinguish the buffer driven low from the actual lows driven by other components. Because both LTC4306 and PCA9515 employ this scheme to avoid lock-up, and the illegal-low for LTC4306 is lower than that of PCA9515, the lows created by the PCA9515 are not recognized by LTC4306 as such, blocking the down-to-upstream communication. We got around this issue by inserting a simple RC circuit into the two $I^2C$ lines in between LTC4306 and PCA9515 (not shown in figure \ref{CWscheme6}) to manually lower the voltage level on the LTC4306 side, so that the PCA9515 generated low comes upstream as a legal low level. \\

The secondary function of the master controller is to distribute supply voltages to the daughter controllers. For each of the 16 channels, there are three MOSFET voltage switches for $+30$ V, $+9$ V and $-6$ V. They are controlled by an $I^2C$ EEPROM DIP switch (PCA9559) that provides channel by channel switching of the three supply voltages. PCA9559 is the only $I^2C$ device located between the master multiplexer and the repeater. The voltage control bits are stored in the EEPROM, allowing any combination of supply voltages to become a start-up value for each channel. An emergency manual override function is implemented, which opens all three switches for all 16 channels by forcing the DIP switches to produce high level outputs regardless of the stored values in the register. It is done by simply removing a jumper on the master controller board. This can be useful if the communication circuit is damaged, and the stored control bits leave some voltage gates closed.\\

The rush current at startup for both Penn State and Yale system can be an order of magnitude greater than the normal operating current. Having all 16 channels go through the start-up process can put unnecessary load on the power supply. In order to reduce this effect, a sequential turn-on function was employed. There is a daisy chain of 16 digital buffers, one for each channel. In between the neighboring buffers is an RC circuit with time constant on the order of 30 seconds. Once a buffer flip its state to high, it has to charge up the capacitor before the next buffer can flips the state. The output of a buffer is routed into three AND gates, whose inputs are the three voltage control bits from the DIP switch. This allows the channels to receive the supply voltage in sequence at startup, with around 30 seconds delay from one channel to the next. All of the supply voltages are current and voltage limited on the master controller before distribution, and current limited on the daughter controller channel by channel. \\

The communication to the master controller is done through a small commercial circuit board that translates USB into $I^2C$, called DeVaSys. This device allows a PC running the control software to operate the entire inner calorimeter high voltage system through two USB ports, one for each half of the detector. As of 2010, the control system has been integrated into the STAR main slow control system.\\

\end{doublespace}

%\include{Appendix-B/Appendix-B}
%%%%%%%%%%%%%%%%%%%%%%%%%%%%%%%%%%%%%%%%%%%%%%%%%%%%%%%%%%%%%%%
% ESM students need to include a Nontechnical Abstract as the %
% last appendix.                                              %
%%%%%%%%%%%%%%%%%%%%%%%%%%%%%%%%%%%%%%%%%%%%%%%%%%%%%%%%%%%%%%%
% This \include command should point to the file containing
% that abstract.
%\include{nontechnical-abstract}
%%%%%%%%%%%%%%%%%%%%%%%%%%%%%%%%%%%%%%%%%%%
% End of the \allowdisplaybreak command %
%%%%%%%%%%%%%%%%%%%%%%%%%%%%%%%%%%%%%%%%%%%

%%%%%%%%%%%%%%%%
% BIBLIOGRAPHY %
%%%%%%%%%%%%%%%%
% You can use BibTeX or other bibliography facility for your
% bibliography. LaTeX's standard stuff is shown below. If you
% bibtex, then this section should look something like:
%   \begin{singlespace}
%   \bibliographystyle{FG-bibstyle}
%   \addcontentsline{toc}{chapter}{Bibliography}
%   \bibliography{Biblio-Database}

\begin{thebibliography}{99}
\addcontentsline{toc}{chapter}{Bibliography}
\frenchspacing

\bibitem{asympt} D. J. Gross and F. Wilczek, Phys. Rev. Lett. {\bf 30}, 1343 (1973).
\bibitem{star} K. H. Ackermann {\it et al.}, Nucl. Instrum. Meth. A {\bf 499}, 624 (2003).
\bibitem{STARAN} I. Abelev {\it et al.}, Phys. Rev. Lett. {\bf 101} 222001 (2008).
\bibitem{paper704} D. L. Adams {\it et al.}, Phys. Lett. B {\bf 261} 201 (1991).
\bibitem{paper704b} D. L. Adams {\it et al.}, Phys. Lett. B {\bf 264} 462 (1991).
\bibitem{KPR} G. L. Kane, J. Pumplin, and W. Repko, Phys. Rev. Lett. {\bf 41} 1689 (1978).
\bibitem{oldAN1} R. D. Klem {\it et al.}, Phys. Rev. Lett. {\bf 36} 929 (1976). 
\bibitem{oldAN2} W. H. Dragoset {\it et al.}, Phys. Rev. D {\bf 18} 3939 (1978).
\bibitem{oldAN3} S. Saroff {\it et al.}, Phys. Rev. Lett. {\bf 64} 995 (1990).
\bibitem{paper704c} D. L. Adams {\it et al.}, Nucl. Phys. B {\bf 519} 3 (1998).
\bibitem{paper704d} B. E. Bonner {\it et al.}, Phys. Rev. Lett. {\bf 61} 1918 (1988).
\bibitem{STARX} J. Adams {\it et al.}, Phys. Rev. Lett. {\bf 97} 152302 (2006).
\bibitem{STARX2} J. Adams {\it et al.}, Phys. Rev. Lett. {\bf 92} 171801 (2004).
\bibitem{facto} J. C. Collins, D. E. Soper, G. and Sterman, ``Factorization of hard processes in QCD", in ``Perturbative
QCD", (ed. A.H. Mueller) (World Scientific, Singapore, 1989) and references therein.
\bibitem{collins} J. C. Collins, S. Heppelmann, and G. Ladinsky, Nucl. Phys. B {\bf 420} 565 (1994). 
\bibitem{barone} V. Barone, A. Drago, and P. G. Ratcliffe, arXiv:hep-ph/0104283v3.
\bibitem{nofacto2} J. Collins and J. Qiu, Phys. Rev. D {\bf 75} 114014 (2007).
\bibitem{nofacto} T. C. Rogers, and P. J. Mulders, Phys. Rev. D {\bf 81} (2010). 
\bibitem{collins2} J. C. Collins, Nucl. Phys. B {\bf 396} 161 (1993).
\bibitem{sivers} D. Sivers, Phys. Rev. D {\bf 41} 83 (1990).
\bibitem{sivers2} D. Sivers, Phys. Rev. D {\bf 43} 261 (1991).
\bibitem{sivers3} U. D'Alesio and F. Murgia, Phys. Rev. D {\bf 70} 074009 (2004).
\bibitem{boer} D. Boer and P. J. Mulders, Phys. Rev. D {\bf 57} 5780 (1998). 
\bibitem{boer2} D. Boer, Phys. Rev. D {\bf 60} 014012 (1999). 
\bibitem{twist3} J. W. Qiu and G. Sterman, Phys. Rev. Lett. {\bf 67} 2264 (1991).
\bibitem{Jaffe} R. Jaffe, arXiv:hep-ph/9602236v1.
\bibitem{twist3a} C. Kouvaris, J. W. Qiu, W. Vogelsang, and F. Yuan, Phys. Rev. D {\bf 74} 114013 (2006).
\bibitem{sivt3} X. D. Ji, J. W. Qiu, W. Vogelsang, and F. Yuan, Phys. Lett. B {\bf 638} 178 (2006).
\bibitem{white} J. Adams {\it et al.}, Nucl. Phys. A {\bf 757} 102 (2005).
\bibitem{deltaG}  B. I. Abelev {\it et al.}, Phys. Rev. Lett. {\bf 97} 252001 (2006).
\bibitem{deltaG2} B. I. Abelev {\it et al.}, Phys. Rev. Lett. {\bf 100} 232003 (2008).
\bibitem{deltaG3} M. M. Aggarwal {\it et al.}, Phys. Rev. Lett. {\bf 106} 62002 (2011).
\bibitem{nuclinst} I. Alekseev {\it et al.}, Nucl. Instr. and Meth. Phys. Res. A {\bf 499} 392 (2003).
\bibitem{rhicprl} M. Bai {\it et al.}, Phys. Rev. Lett. {\bf 96} 174801 (2006).
\bibitem{snake} S. R. Mane, Yu. M. Shatunov, and K. Yokoya, J. Phys. G: Nucl. Part. Phys. {\bf 31} R151 (2005). 
\bibitem{originsnake} Ya. S. Derbenev, A. M. Kondratenko, S. I. Serednyakov, A. N. Skrinsky, G. M. Tumaikin, and Yu. M. Shatunov Yu, Part. Accel. {\bf 8} 115 (1978).
\bibitem{fpd} L. C. Bland, arXiv:hep-ex/0602012.
\bibitem{NETshower} L. C. Bland {\it et al.}, Nucl. Exp. Tech. {\bf 51} 3 (2008).
\bibitem{fms} L. K. Eun, J. Phys. Conf. Ser. {\bf 230} 012041 (2010).
\bibitem{yiqun} Yiqun Wang, Ph.D. Thesis, University of Texas at Austin, (2004).
\bibitem{pythia} T. Sj\"{o}strand {\it et al.}, Comput. Phys. Commun. {\bf 135} 238 (2001).
\bibitem{vogel} W. Vogelsang, private communication, (2011).
\bibitem{CTEQ} W. K. Tung, H. L. Lai, A. Belyaev, J. Pumplin, D. Stump, and C. P. Yuan, JHEP {\bf 0702} 053 (2007).
\bibitem{DSS} D. de Florian, R. Sassot, and M. Stratmann, Phys. Rev. D {\bf 75} 114010 (2007).
\bibitem{polar} A. Bazilevsky {\it et al.}, RHIC/CAD Accelerator Physics Note 298 (2007).
\bibitem{phenix} S. S. Adler {\it et al.}, Phys. Rev. C {\bf 75} 024909 (2007).
\bibitem{phenix2} C. Aidala, F. Ellinghaus, R. Sassot, J. Seele, and M. Stratmann, Phys. Rev. D {\bf 83} 034002 (2011).
\bibitem{I2C} $I^2C$ specification and user manual \url{http://www.nxp.com/documents/user_manual/UM10204.pdf}
\bibitem{cw} A. Brunner {\it et al.}, Nucl. Instr. and Meth. Phys. Res. A {\bf 414} 466 (1998).
\bibitem{cw1} I. D'Atone, M. Lolli, and M. Zanotti, Nucl. Instr. and Meth. Phys. Res. A {\bf 480} 555 (2002).
\bibitem{cw2} Yu. Gilitsky {\it et al.}, Nucl. Instr. and Meth. Phys. Res. A {\bf 571} 294 (2007).
\bibitem{E864} T. A. Armstrong {\it et al.}, Phys. Rev. Lett. {\bf 79} 3351 (1997).

\end{thebibliography}
%   \end{singlespace}

\begin{singlespace}

\end{singlespace}

\backmatter

% Vita
% Place your vita below.

\begin{itemize}
\item Pennsylvania State University, PhD candidate in Physics, Aug, 2004 to present

\item Pennsylvania State University, B.S in Physics, May, 2004

\item Military Service, Korean Augmentation to the United States Army, Jan. 2001 to Mar. 2003

\item Seoul National University, English Literature Department, (no degree) Mar. 1996 to Dec. 1998

\item Myungduk Foreign Language High School, Seoul, Korea, HS Diploma, Mar. 1996 

\end{itemize}

\end{document}